\documentclass{article}
\usepackage{graphicx} 
\usepackage{amsmath} 
\usepackage{xcolor}
\usepackage{amssymb}
\usepackage{commath}
\usepackage{braket}
\usepackage{authblk}
\usepackage{titlesec}

\titleformat{\paragraph}
{\normalfont\normalsize\bfseries}{\theparagraph}{1em}{}
\titlespacing*{\paragraph}
{0pt}{3.25ex plus 1ex minus .2ex}{1.5ex plus .2ex}

\begin{document}


\title{Two dimensional semiconductors:\\ optical and electronic properties}
\author[1]{Roberto Rosati}
\author[2]{Ioannis Paradisanos}
\author[1]{Ermin Malic}
\author[3]{Bernhard Urbaszek}
\affil[1]{Department of Physics, Philipps-Universit\"at Marburg, Marburg, Germany.}
\affil[2]{Institute of Electronic Structure and Laser (IESL), Foundation for Research and Technology (FORTH), Heraklion, Greece}
\affil[3]{Institute of Condensed Matter Physics (IPKM), Technische Universit\"at Darmstadt, Germany}

\date{}

\maketitle

\begin{abstract}
    In the last decade atomically thin 2D materials have emerged as a perfect platform for studying and tuning light-matter interaction and electronic properties in nanostructures. The optoelectronic properties in layered materials such as transition-metal-dichalcogenides (TMDs) are governed by excitons, Coulomb bound electron-hole pairs, even at room temperature. The energy, wave function extension, spin and valley properties of optically excited conduction electrons and valence holes are controllable via multiple experimentally accessible knobs, such as lattice strain, varying atomic registries, dielectric engineering as well as electric and magnetic fields. This results in a multitude of fascinating physical phenomena in optics and transport linked to excitons with very specific properties, such as bright and dark excitons, interlayer and charge transfer excitons as well as hybrid and moir\'e excitons.
In this book chapter we introduce general optoelectronic properties of 2D materials and energy landscapes  in TMD monolayers as well as their vertical and lateral heterostructures, including twisted TMD hetero- and homobilayer bilayers with moir\'e excitons and lattice recombination effects. We review the recently gained insights and open questions on exciton diffusion, strain- and field-induced exciton drift. We discuss intriguing non-linear many-particle effects, such as exciton halo formation, negative and anomalous diffusion, the surprising anti-funneling of dark excitons. 
\end{abstract}

\newpage

\tableofcontents

\newpage

\section{Introduction}\label{sec:intro}

Some school students can create great things with just a scotch tape and a pencil. Some physicists can win a Nobel prize this way. In 2010, Andrey Geim and Konstantin Novoselov received the Nobel Price of Physics for their work on graphene, which  is just graphite, the black material in our pencils, brought down to its ultimate low thickness of only one atom \cite{Novoselov04,Novoselov11}. To do this, Geim and Novoselov exfoliated graphite with an adhesive tape, repeating the procedure until a monolayer flakes appeared.
Being only one atom thick, graphene is the ultimate two-dimensional (2D) material, the closest we can get in physics to the concept of planes in mathematics. 
Importantly, the properties of 2D graphene are drastically different compared to  the bulk/3D graphite, 
with electrons behaving similarly to photons exhibiting a linear dispersion relation (the relation between energy and momentum of electrons) \cite{Malic13}.

The success of graphene has triggered the search and discovery of many other two-dimensional  materials \cite{Novoselov05}. These originate from bulk layered materials, which are formed by atoms interacting strongly with other atoms in the same plane (covalent bonds), while interacting very weakly with those belonging to different layers via weak van der Waals interactions.
It is the latter property allowing for the exfoliation down to the monolayer -- so don't try this at home, unless you know that your material is a layered one! Interestingly, most of  2D monolayers share with graphene the hexagonal structure, because this is mechanically more stable than other structures \cite{Ding17}.
Despite these similarities, different 2D materials show an extremely broad range of properties, ranging from metals and semimetals (e.g. NbSe$_2$ and graphene) to insulators (e.g. hexagonal boron nitride (hBN)). 
Here, we focus on the specific class of 2D semiconductors, hence showing a Fermi energy lying in the bandgap between the conduction and the valence band. Such bandgap is in the range of the visible light. 
Besides having revolutionized electronics - think about the silicon-valley - semiconductors have remarkable optical properties.
When light of suitable wavelength shines on these materials, it excites an electron into the conduction band. This leaves a lack of negative charge in the valence band which can be interpreted as a positively charged quasi-particle called \textit{hole} \cite{Grundmann16,Rossi02,Haug09,Kira06}. Following the excitation, the  electron can return to the valence band, a process called electron-hole recombination. Such a process is of crucial relevance for fundamental as well as application reasons, because it results in emission of light with an energy dictated by the bandgap due to energy conservation. The 2D semiconductors are extremely interesting for their optics as well as transport properties, which are the two focus areas of this chapter.
\begin{figure}[t!]
    \centering
    \includegraphics[width=0.8\linewidth]{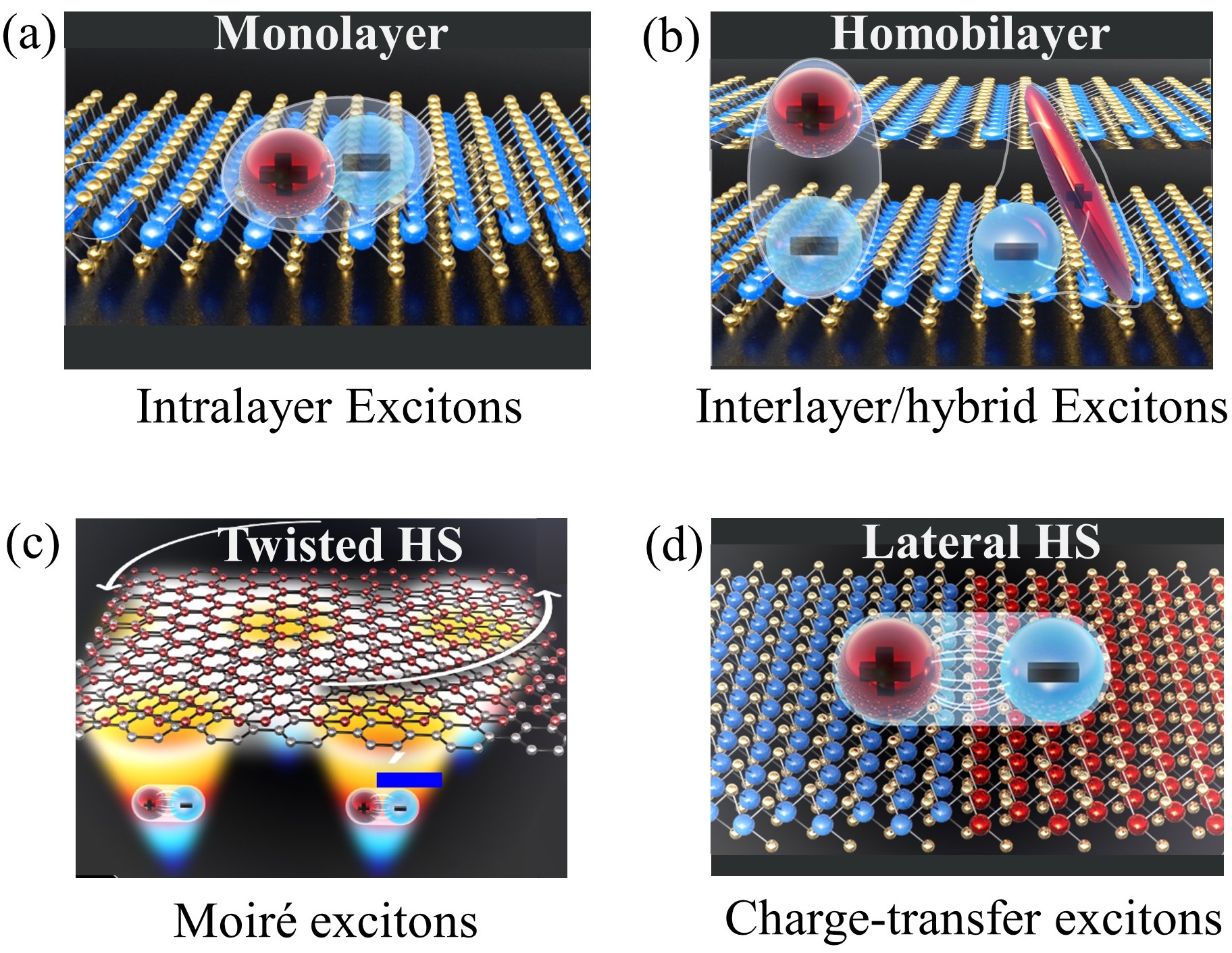}
    \caption{(a) Sketch of (a) TMD monolayers, (b) homobilayers as well as their (c) twisted vertical heterostructures (HS), and (d) lateral heterostructures. Thanks to their two-dimensional nature these materials host deeply-bound excitons with binding energies of several hundreds of meV. In addition, new quasi-particles appear in TMD bilayers including interlayer and hybrid excitons as well as localized moir\'e excitons in twisted HS or charge-transfer excitons in lateral HS.}
    \label{intro1}
\end{figure}

More specifically, we focus on  transition metal dichalcogenides (TMDs), which are 2D semiconductors formed by a layer of  transition metal atoms (Mo or W, cf. the blue spheres in Fig. \ref{intro1}(a)) sandwiched between two layers of chalcogen atoms (S or Se, cf. gold spheres in Fig. \ref{intro1}(a)). Similarly to graphene and many other 2D materials, also TMD monolayers form a hexagonal lattice, with transition-metal and chalcogen atoms alternating on the hexagon corners. These materials are relatively common in the bulk version, for example used as excellent lubricant - also this property stems from the weak van der Waals interaction between their layers \cite{Irving17}. Importantly, they are very stable in the exfoliated version, hence making their experimental realization particularly affordable. 
In addition they are very flexible and their 2D nature makes it particularly easy to integrate them with different materials \cite{Estrada23},  as these can be directly deposited on top of them resulting in  vertical heterostructures (HS), cf. Fig. \ref{intro1}(b), held together by van der Waals bonding. This allows an additional control via stacking and twisting the two layers, resulting in the formation of an in-plane moir\'e potential \cite{Tran19}, cf. Fig. \ref{intro1}(c).  Growing two different monolayers in plane results in lateral heterostructures, where Coulomb-bound electrons and holes can be separated along the interface, cf. Fig. \ref{intro1}(d) \cite{Najafidehaghani21,Rosati23}. While flexibility and integrability are properties shared by almost all 2D materials,  TMD-based nanomaterials show  fascinating many-particle phenomena reflected by the remarkably versatile exciton landscape \cite{Wang18,Mueller18}. As it will be further addressed later, excitons are bound pairs of a negatively charged electron with a positively-charged hole (light blue and red in Fig. \ref{intro1}). These quasi-particles show remarkable properties in TMD-based nanomaterials, forming a landscape including bright and dark states, spatially separated interlayer, hybrid and charge transfer excitons as well as moire excitons (Fig. \ref{intro1}) that will be discussed in the next section. 

\subsection{Excitons in layered semiconductors}\label{sec:intro}

\begin{figure}[t!]
    \centering
    \includegraphics[width=1\linewidth]{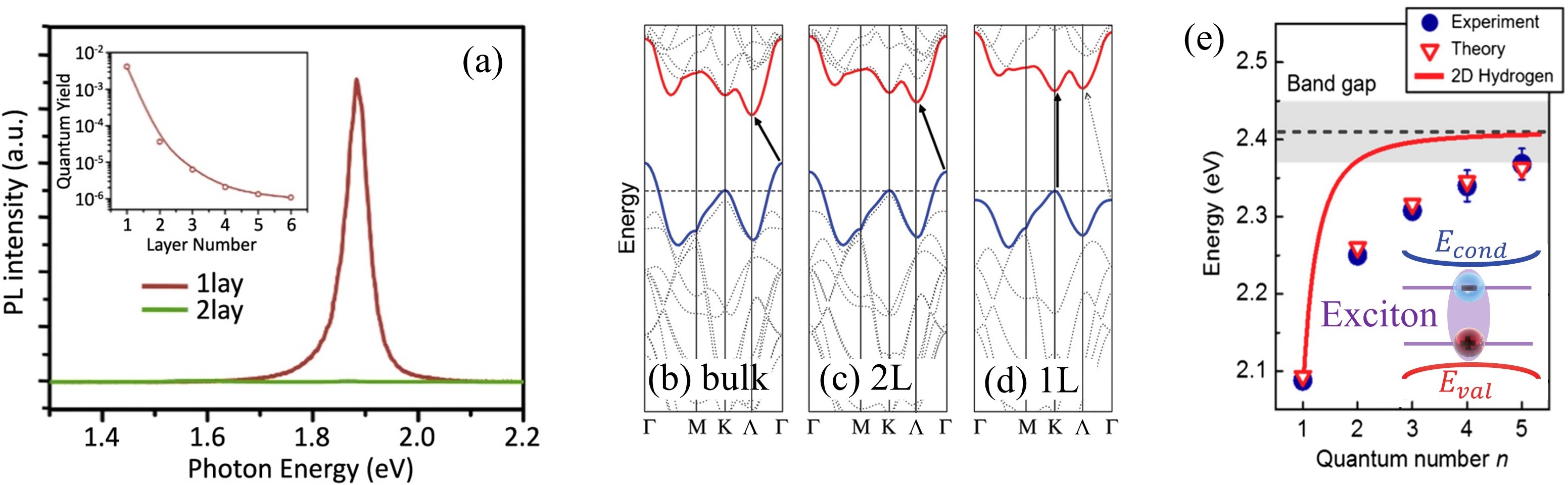}
    \caption{(a) Photoluminescence spectra in mono- (red) and bilayer MoS$_2$ (green), showing a much more intense emission for the monolayer case as reflected by the obtained quantum yield as a function of sample thickness (inset). (b-d) First-principle predictions of the single-particle dispersion in (b) bulk, (c) bilayer and (d) monolayer MoS$_2$, showing the transition from indirect to direct semiconductors, cf. arrows between the energetically-favourable valence- and conduction-band extreme. (e) Exciton energies in WS$_2$ on SiO$_2$ revealed by reflectance measurements: All peaks below the bandgap (dashed line) correspond to bound excitons, cf. sketch in the inset.  Figs. (a), (b-d) and (e) adapted respectively from Ref. \cite{Mak10}, \cite{Splendiani10} and \cite{Chernikov14}.}
    \label{intro2}
\end{figure}
Two identical lamps emit twice as much light as one. This does not apply to TMD heterostructures: A bilayer of MoS$_2$ emits much less light than one single monolayer \cite{Mak10}, cf. Fig. \ref{intro2}(a). This remarkable and counter-intuitive observation originates in first place from their single-particle bandstructure. This changes abruptly when reducing the thickness to the single-layer case \cite{Splendiani10,Kuc11}, as shown in Fig. \ref{intro2}(b-d) respectively for bulk, bilayer and monolayer MoS$_2$. 
The bulk and bilayer material are indirect semiconductors, i.e. the maximum of the valence band is not vertically aligned with the minimum of the conduction band, cf. arrows in Figs \ref{intro2}(b,c). 
 Electron and holes will 
 occupy mostly these two energetically favourable valleys. From here they can not recombine and emit light because the light-matter interaction preserves both energy and momentum. Since photons carry only a negligible momentum, only direct vertical transitions are allowed. 
This momentum condition is fulfilled for the monolayer case, cf. solid arrow in Fig. \ref{intro2}(d),  explaining the drastically  different emission intensity between bi- and monolayer MoS$_2$ in Fig. \ref{intro2}(a). 
Besides the intensity, also the energy position of resonances in optical spectra reveals a characteristic property of each single TMDs, cf.  Fig. \ref{intro2}(e) for the case of WS$_2$ deposited on SiO$_2$. Interestingly, most of these peaks appear well below the single-particle bandgap, hence can not originate from an electron-hole recombination. The resonance rather stem from excitons. It is the remarkably versatile excitonic landscape which makes TMD nanomaterials so unique, as we discuss in the following.

Positive and negative charges attract each other due to the Coulomb interaction: This triggers the formation of hydrogen atoms, where an electron is Coulomb-bound to a proton, forming a set of orbitals with finite energies (1s, 2s, etc.). This general rule of physics applies to semiconductors as well: Here an exciton is formed by a Coulomb-bound electron-hole pair. 
The energy of such an exciton is given by the bandgap minus the binding energy, cf. sketch in Fig. \ref{intro2}(e). 
Similarly to the case of the hydrogen atom, also here multiple orbitals with discrete energies are present, cf. dots in Fig. \ref{intro2}(e). These states can hence be interpreted as 1s, 2s, etc. exciton states, with additional orbitals analogous to 2p, 3p etc. being present, as revealed by infrared or second-harmonic-generation studies \cite{Berghauser18,Wang15}. Higher quantum indices imply an increase of transition energy via a decrease of the binding energy. The energy separation between the exciton states in TMD monolayers however does not entirely follow the Rydberg series of the hydrogen atom \cite{Chernikov14}, as shown respectively by dots and solid line in Fig. \ref{intro2}(e). This happens because in solid state  excitons do not live in empty space, but rather are formed in a material which hosts multiple atoms, and these dielectrically screen the Coulomb interaction, leading to the Keldysh-Rytova potential \cite{Keldysh79,Rytova67} and a Rydberg series modified by the TMD polarizability \cite{Molas19}.

On a qualitative level, excitons are a common feature of all semiconductors. However, excitons in TMDs are unique for multiple reasons. Here, electrons and holes are bound together in a stronger way than in bulk TMD materials \cite{Lynch22}, in typical III-V \cite{Moore90}, or II-VI quantum wells \cite{Broser82}. This is reflected by huge binding energies of several hundreds of meV, cf. the energy difference between bandgap and 1s state in Fig. \ref{intro2}(e). These values are much larger than the thermal energy of 25 meV at room temperature, hence excitons dominate the response of TMDs even at room temperature. In real space this strong Coulomb interaction implies that for 1s excitons the electron and hole are placed within a few nanometers, the so-called Bohr radius \cite{Zipfel18}. This increases the chances of electrons and holes being in the same position, which is an additional necessary condition for an efficient light-matter interaction (together with the energy and momentum conservation discussed above). As a consequence, 1s states are much brighter than higher-order states \cite{Chernikov14,Wang18}, and can absorb up to  20\% of light \cite{Mak10,Splendiani10,Wang18}.

\begin{figure}[t!]
    \centering
    \includegraphics[width=1\linewidth]{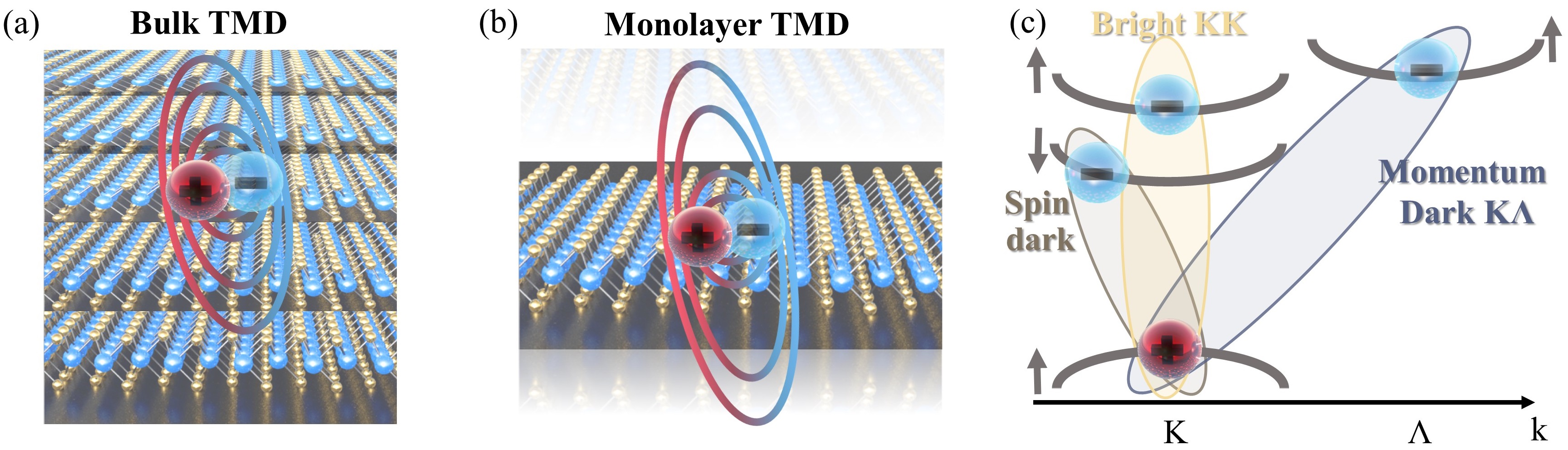}
    \caption{Sketch of excitons in (a) bulk  and (b) monolayer TMDs. In the first case the field lines between electron and holes remain within the material, where the atoms can dielectrically screen the Coulomb interaction, hence considerably weakening it. In contrast in the monolayer case, the lines to a large extent live outside of the monolayer, leading to a reduced screening. (c) Sketch of the single-particle valleys in two crucial high-symmetry points of the Brillouin zone, with bright as well as spin- and momentum-dark excitons.}
    \label{intro3}
\end{figure}
Why is the Coulomb interaction so strong for TMD monolayers? Besides benefiting from high single-particle effective masses \cite{Kormanyos15}, these materials show huge excitonic binding energies thanks to their two-dimensional nature, leading in principle to a four-fold increase compared to the bulk case \cite{Bastard82}. 
Nevertheless, the electrical field lines connecting the negatively charged electron and the positively charged hole could lead to a dielectric screening \cite{Hueser13,Latini15,Laturia18} which weakens the Coulomb interaction.
As a consequence the binding energy depends inversely and quadratically on the optical dielectric constant of the host material \cite{Chernikov14}, which for TMD is unfortunately particularly large \cite{Laturia18,Latini15}. Nevertheless, the atomically thin nature of monolayers allows a reduced screening.
In bulk materials, the electrical lines remain within the system, Fig. \ref{intro3}(a), resulting in a efficient screening of Coulomb interactions. In contrast, in the monolayer case most of the field lines live outside of the monolayer, Fig. \ref{intro3}(b), where the dielectric constant is typically (much) smaller than in TMD materials. 
This leads to a drastically reduced screening and  results in  stronger Coulomb interactions and hence larger exciton binding energies. This applies particularly well to  free-standing TMD monolayers, as in this case the field lines will extend to the vacuum, resulting in binding energies one order of magnitude larger for free-standing monolayers \cite{Kumar23} compared to the corresponding bulk \cite{Lynch22}. More involved is the case of TMDs on substrates: In this case the binding energy depends strongly on the dielectric constant of the surrounding, allowing for dielectric engineering of exciton properties \cite{Raja19}.

Besides exhibiting huge  binding energies, excitons in TMD monolayers form a remarkable energy landscape with different excitonic species. This stems primarily from their involved single-particle dispersion, which shows multiple interesting high-symmetry points in the Brillouin zone (cf. Fig. \ref{intro2}(d)) and tens to hundreds meV spin-splitting in conduction and valence band, respectively, due to the spin-orbit coupling \cite{Xiao12,Roldan14,Kormanyos15}.
The presence of several relevant high-symmetry points can be traced back to the multi-atomic composition of TMDs. The orbitals of different states are centered around different atoms, e.g. transition metal or chalcogen atoms for respectively K and $\Lambda$ states in the conduction band \cite{Brem20b}. 
As a consequence, we have a variety of different exciton species including bright KK excitons with electrons and holes being located at the K point. Additionally, momentum-dark or momentum-indirect K$\Lambda$ can be formed with Coulomb-bound electrons and holes being located at different valley, cf. the blue oval in Fig. \ref{intro3}(c). Here, the first (K) and the second letter ($\Lambda$) indicate the valley of the hole and the electron, respectively. 
The excitonic states form valleys in the center-of-mass momentum $\textbf{Q}$. From the transport perspective, this momentum triggers the excitonic mobility, opening the way to  remarkable transport features discussed in Sec. \ref{sec:transport}. From the optics perspective, excitons can directly recombine into light only when both energy and momentum are conserved, hence only KK states with momentum $Q\equiv |\textbf{Q}|$ in the light-cone interact directly with light. Nevertheless, momentum-dark excitons can also become optically active via phonon-assisted mechanisms, when their occupation is high enough to compensate for the much weaker efficiency of higher-order phonon-assisted emission compared to the direct recombination of bright excitons \cite{Brem20}.   Here, a crucial distinction arises between molybdenum- and tungsten-based TMD monolayers, as the energetically-lowest states are bright KK and  dark KK$^ \prime$ as well as K$\Lambda$ excitons in Mo- and W-based materials, respectively \cite{Selig18,Deilmann19}. This difference stems mostly from the single-particle dispersion: While the energy position at K$^\prime$ reflects the spin-splitting of the conduction band at K point (see below), the energy at $\Lambda$ point is much higher than the one at the K point in Mo-based TMDs, compared to the W-based ones \cite{Kormanyos15}. In addition, thanks to larger effective masses the binding energies of dark valleys are typically larger than those of  bright excitons. In tungsten-based TMDs, this results in dark excitons being energetically below the bright states, cf. the sketch in Fig. \ref{intro3}(c).

The spin-splitting stems from the spin-orbit coupling and the broken symmetry displayed by TMD monolayers\cite{Kormanyos15}. Interestingly, the spin-splitting is very different for conduction and valence bands, being respectively of the order of few tens and hundreds of meV at the K point. In optics, mainly the spin-splitting at the valence band induces the energy separation between  A and B excitons, which constitute the two most intense peaks in absorption spectra and observed also in early studies on bulk-like TMD \cite{Wilson69}. In photoluminescence or transport measurements, B excitons have only negligible effects, because 
they are much less occupied than  A excitons in view of their large energy separation. In contrast, the small spin-split at the conduction band has a crucial impact, in particular in tungsten-based materials. Here, the spin-dark states composed by electron and hole with opposite spins have an energy smaller than the one of bright KK states, cf. brown oval in Fig. \ref{intro3}(c). Before exchange interaction \cite{Qiu15,Echeverry16,Deilmann19} these states are degenerate with the momentum-dark KK$^\prime$ states \cite{Yang22} and show a peculiar interaction with light.
The bright excitons interact with light propagating out-of-plane, as this can couple only with the orbital angular momentum hence preserving the spin \cite{Xiao12}. In contrast, these states composed by electron and hole with opposite spin couple with  photons propagating in-plane \cite{Wang17,Robert17}, hence requiring a finite aperture of the microscope to be detected.
Such an interaction is, however, considerably weaker and as a consequence we refer to these states as spin-dark states. Nevertheless this emission can be detected via a finite aperture of the microscope and at low temperatures, where compared to the bright states  their higher occupation somewhat compensates the smaller exciton-light coupling \cite{Wang17,Robert17}. 

The excitonic landscape is further enriched in the presence of TMD heterostructures (HS), cf. Figs. \ref{intro1}(b-d). The vertical HS, also called van-der Waals HS, are composed by monolayers stacked on top of each other (Fig. \ref{intro1}(b)). While homobilayers can be formed also directly by exfoliation of bulk materials, the stacking of individually exfoliated monolayers opens the way to the formation of hetero- or twisted bilayers (Fig. \ref{intro1}(c)), as composed  by two monolayers  of different composition or twisted one respect to each other. Even without twisting, the stacking of different monolayers has a drastic impact on the excitonic properties. 
First, electrons and holes from different layers can bound to form interlayer excitons, cf. Fig. \ref{intro1}(b), whose electrons and holes are spatially separated in two different layers. These are particularly relevant in heterobilayers, as these typically form type-II heterostructures. Here, due to a band-offset between the conduction and valence bands of different layers interlayer excitons become the energetically lowest states.
Second, electrons in one layer experience an electrostatic potential coming also from the atoms in the other layer, resulting in a new contribution to the Hamiltonian, the so-called moir\'e potential. 
Third, the wavefunctions of states living in one layer can now overlap with those of the other layer, resulting in a new tunneling contribution to the Hamiltonian. Such an interlayer hopping will be stronger for the states with orbitals mostly around the chalcogen atoms, as these are closer to the other layer. At the single-particle level, this induces drastic changes at the $\Gamma$ and $\Lambda$ points in the Brillouin zone, cf. Figs. \ref{intro2}(c,d), while the K point in the conduction band is less affected because here the orbitals are mostly located at the central transition metal atoms \cite{Brem20b}. At the exciton level this has even more remarkable effects. First of all, it allows the formation of hybrid excitons, where either the electron or the hole lives in a quantum superposition between the two layers, cf. Fig. \ref{intro1}(b). This leads to crucial variations of excitonic energies also in homobilayers, while in heterobilayers it crucially triggers the sub-picosecond formation of interlayer excitons, which are formed via phonon-mediated scattering into hybrid excitons after an optical excitation of intralayer states \cite{Schmitt22,Meneghini23}. In general, the interlayer tunneling leads to an energetically larger separation between bright and dark exciton states compared to the monolayer case \cite{Hagel21,Kumar23}. In addition, more dark states become relevant, including e.g. $\Gamma$K excitons \cite{Meneghini23} and the recently-observed $\Gamma\Lambda$ states \cite{Kumar23}. Interestingly, momentum-dark states are crucial also for homobilayers based in molybdenum-based TMDs \cite{Meneghini23}, contrary to their monolayer counterpart \cite{Cadiz17b,Rosati21a}.

So far we have introduced excitons which can be localized in their relative motion, but not in their center-of-mass motion, i.e. Coulomb-bound electrons and holes must be close to each other, but they can propagate freely as a quasi-particle. The situation differs significantly in the presence of twisted or lateral heterostructures, cf. Figs. \ref{intro1}(c) and \ref{intro1}(d). In the presence of a rigid twist, a superlattice is formed in the heterostructure (Fig. \ref{fig:B1}(a)). The electrostatic interaction results in an energy potential within the plane, showing the smooth formation of potential minima where excitons can be localized, cf. Fig. \ref{intro1}(c). The shape and periodicity of this moir\'e potential depend on the angle of the rigid twist, cf. Eq. (\ref{eq:moireperiod}) below. In particular, the spatial width of the potential minima is crucial, as the localization is possible only for widths larger than the excitonic Bohr radius \cite{Brem20c}. As it will be further discussed below, the situation becomes more involved with small twist angles (typically smaller than 1$^\circ$). In this case, the approximation of a rigid rotation is not valid anymore, because atoms rearrange themselves resulting in a peculiar atomic reconstruction \cite{Zhao23,Arnold23,Hagel24}, cf. Fig. \ref{fig:B1}(b).

In lateral HS two different monolayers are grown directly one beside each other and bind each other covalently at an interface,  cf. Fig. \ref{intro1}(d). The materials at the two sides of the junction have different single-particle energies, resulting in a type-II lateral heterostructure qualitatively similar to the one of van der Waals heterobilayers. As a consequence one expects the formation of charged-transfer (CT) excitons, constituted by electron and hole living in different monolayer materials forming the HS, in analogy to interlayer excitons in vertical HS, cf. Figs. \ref{intro1}(b) and \ref{intro1}(d). However, while the first observations of interlayer excitons in vertical heterostructures appeared already in 2015 \cite{Rivera15}, only very recently they have been observed in lateral HS \cite{Rosati23,Yuan23}. This is because of the required high quality of the samples, which has recently improved leading to narrow interfaces with a small residual alloying \cite{Chu18,Najafidehaghani21,Herbig21,Pielic21,Beret22,Shimasaki22,Rosati23,Yuan23}. 
Recently it has been shown microscopically how the shrinking of interface width toward the Bohr radius opens up the way for new physics, in particular allowing the stable formation of CT and their optical observation \cite{Rosati23}. These lateral HS also hosts a remarkable transport both across \cite{Beret22,Lamsaadi2023} and along the interface \cite{Yuan23}, as it will be discussed in Sec. \ref{sec:transport}.

In this book chapter, we will address crucial features in both exciton optics and transport in TMD nanomaterials including TMD monolayers as well as bilayers, twisted van der Waals and lateral heterostructures. In Sec. \ref{sec:optics} we will focus on exciton optics in TMD bilayers, in particular addressing effects induced by moir\'e potentials. In Sec. \ref{sec:transport} we will discuss exciton transport, in particular addressing what rules exciton diffusion and exciton drift and what impact dark exciton states have.

\section{Exciton optics}\label{sec:optics}
In the introduction, we have already mentioned the strong light-matter coupling in TMDs as well as the drastic suppression of photoluminescence when going from a TMD monolayer to bilayers. TMDs show a variety of intriguing optical properties including  specific valley- and spin-dependent optical selection rules. In TMD homo- and heterobilayers, we find even richer optics phenomena thanks to the formation of hybrid and interlayer excitons (cf. sketch in Fig. \ref{intro1}(b)). As briefly introduced in Fig. \ref{intro1}(c), twisting one layer compared to the other results in the formation of moir\'e potentials and associated moir\'e excitons that can be, dependent on the twist angle, trapped resulting in a peculiar polarization, energy, linewidth and even single-photon character of the emitted light. The optical response of these materials can be tuned by external electric and magnetic fields. In the following,  we review these remarkable properties of TMD-based vertical heterostructures.

\subsection {Moir\'e effects in relaxed and rigid superlattices} 
Fabricating van der Waals bilayers by stacking TMD monolayers of the same or different composition offers a distinct physical system with novel functionalities. Stacking, twisting, and electrically gating van der Waals bilayers open up a unique avenue for systematically investigating the physics across a wide spectrum from weak to strong correlations within many-body frameworks  \cite{he2021moire}. A relative, rigid rotation between the constituent monolayers creates a new artificial lattice with a much larger periodicity than that of the individual layers, named \textit{moir\'e} superlattice. The long-wavelength periodicity in real-space introduces new electronic properties in reciprocal space, such as the formation of flat electronic bands. These refer to regions in the electronic band structure where the energy does not vary significantly with momentum. Here, electrons are effectively trapped and exhibit enhanced electron-electron interactions, resulting in correlated states, such as unconventional superconductivity \cite{cao2018unconventional} ($i.e.$ vanishing electrical resistance) and Mott insulator states \cite{devakul2021magic}. A Mott insulator is a material that is expected to behave like a metal according to conventional band theory, however it exhibits insulating properties due to strong carrier-carrier interactions. Additionally, phenomena such as exciton condensation, where excitons transition into a coherent state at low temperatures, displaying macroscopic quantum behavior \cite{wang2019evidence}, and the emergence of non-trivial phases of matter, can occur due to the modified electronic band structure of the twisted bilayer system. Consequently, the \textit{twist-angle}, allowed by the weak van der Waals interaction between the monolayers, is a decisive parameter in the study of two-dimensional quantum materials and driving research into the emerging field of \textit{twistronics}.
\begin{figure}[t!]
  \centering
  \includegraphics[width=1\linewidth]{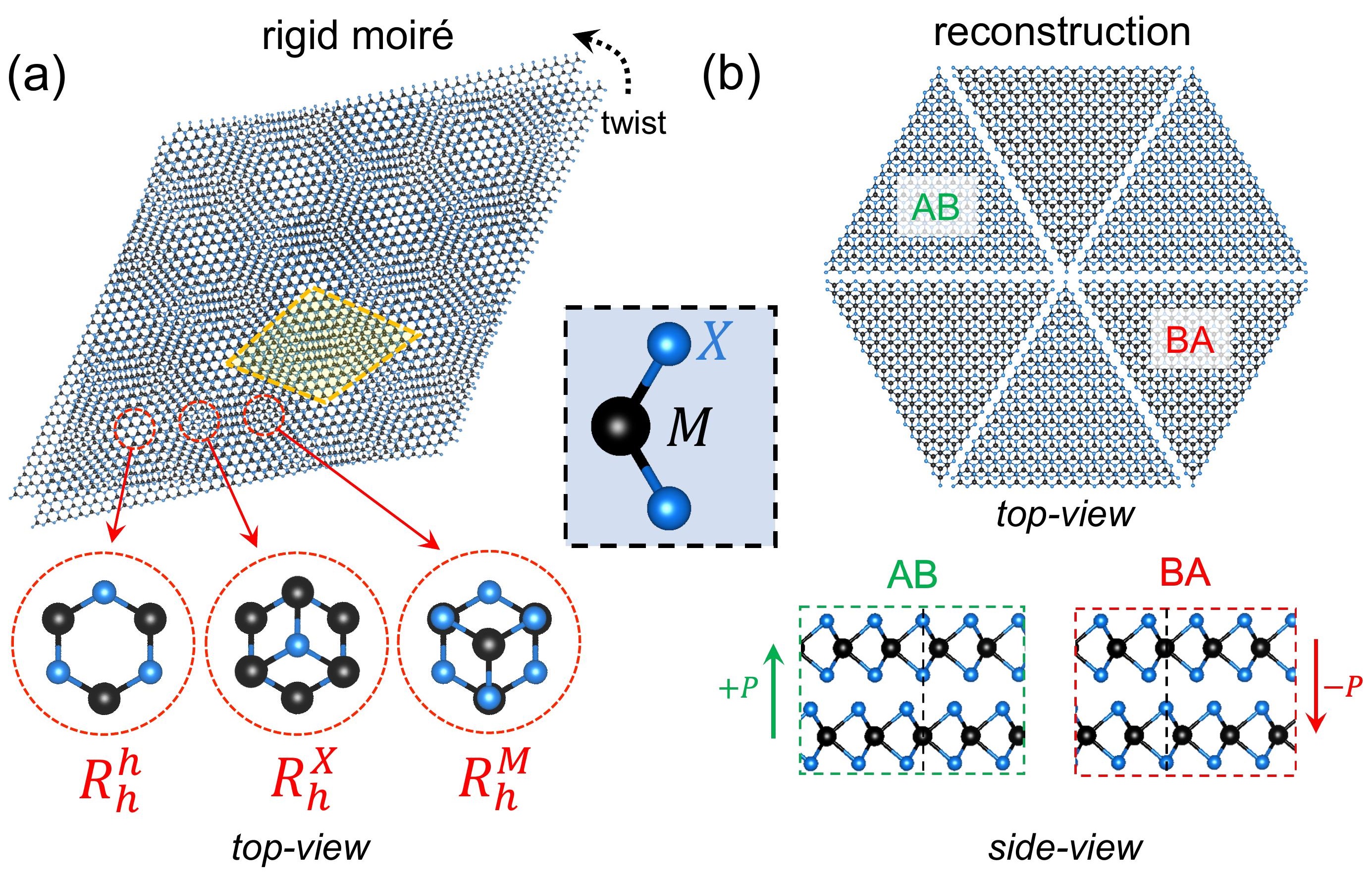} 
  \caption{(a) Schematic illustration of a slightly twisted TMD homobilayer showing the rigid moir\'e superlattice. Three distinct atomic registries are highlighted, with the superlattice unit cell depicted as a yellow diamond.(b) Lattice reconstruction in a TMD homobilayer, featuring two domains: AB and BA. These domains arise from a spatial translation of the top layer relative to the bottom layer. In the AB domain, the top chalcogen atoms are positioned above the bottom transition metal atom, while in the BA domain, the top transition metal atom aligns directly above the bottom chalcogen atoms (vertical dashed lines), resulting in an opposite ferroelectric polarity compared to the AB domain.}
  \label{fig:B1}
\end{figure}

TMD bilayers can consist of monolayers with either identical (matching lattice constants) or different (varying lattice constants) compositions. The emergence of two distinct scenarios upon twisting depends on the lattice mismatch and the twist angle between the monolayers: (a) rigid moir\'e superlattice and (b) lattice reconstruction  \cite{Zhao23,Arnold23,Hagel24} (Figure \ref{fig:B1}). The distinction between a rigid moir\'e superlattice and lattice reconstruction lies in the nature of the interlayer interaction and the resulting structural behavior. In bilayers with a slight twist angle (near to 0$^{o}$ or 60$^{o}$), the superlattices can experience a significant structural reconstruction (atomic relaxation) to minimize the additional energy due to the small misalignment  \cite{carr2018relaxation,enaldiev2020stacking,weston2020atomic}. This atomic rearrangement modifies the optoelectronic properties of the twisted bilayers, mainly due to interlayer hybridization of electronic states, leading to the emergence of multiple flat bands with distinct energy levels, as well as ferroelectricity ($i.e.$, spontaneous electric polarization that can be switched by the application of an external electric field) with domains of opposite out-of-plane polarities  (Figure \ref{fig:B1}(b)). The precise value of the twist angle at which reconstruction occurs is influenced by the lattice mismatch and van der Waals interactions between the layers and it is larger in homobilayers compared to heterobilayers. This value, known as the critical angle, represents the upper limit of the twist angle where reconstruction occurs and it is still under investigation for different TMD monolayer combinations. 

\begin{figure}[b!]
  \centering
  \includegraphics[width=0.9\linewidth]{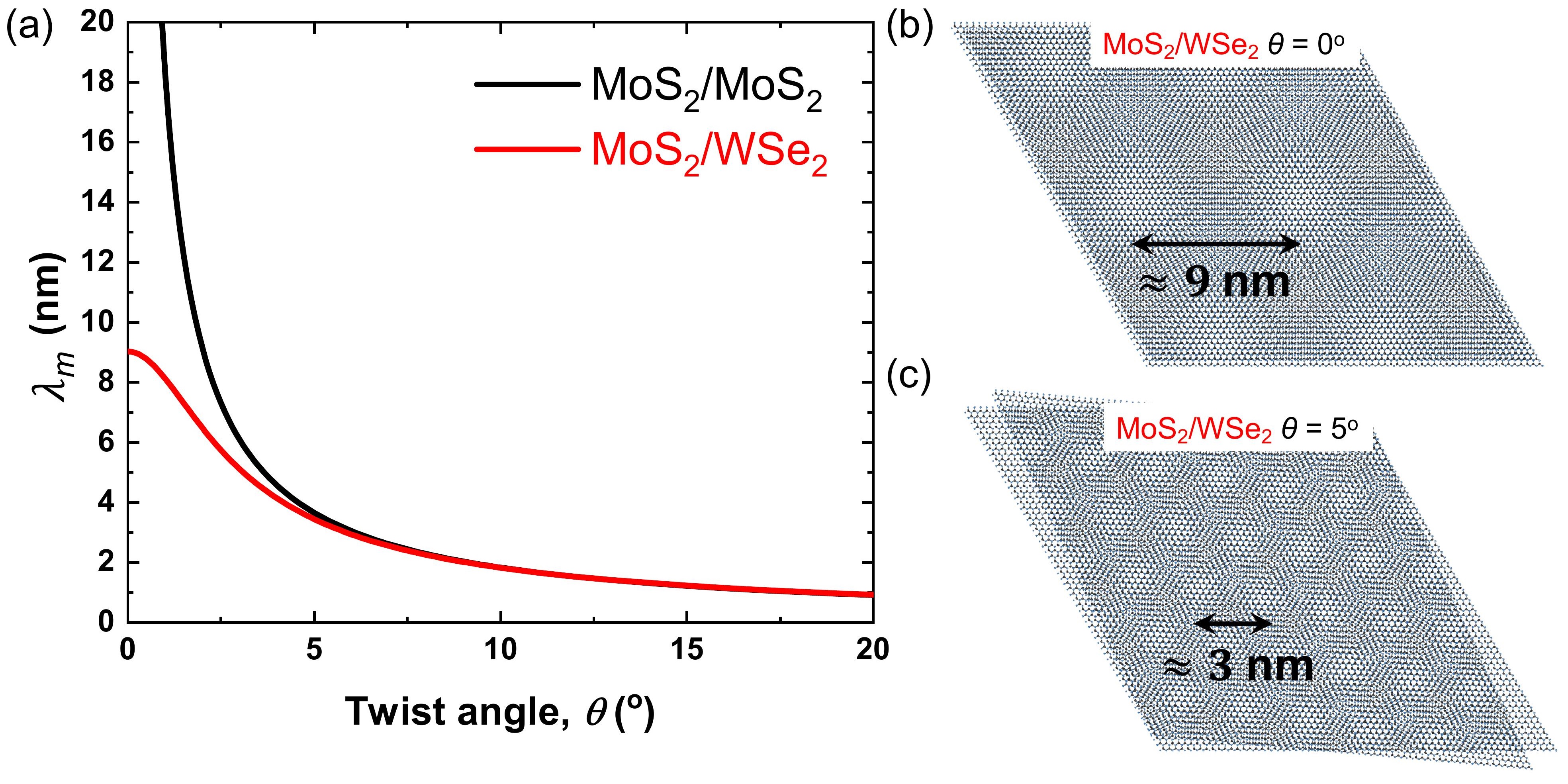} 
  \caption{(a) Calculated moir\'e wavelength as a function of the twist angle, using Eq. \ref{eq:moireperiod}, for a homobilayer MoS$_2$ (black) and a heterobilayer MoS$_2$/WSe$_2$ (red). (b) Illustration of the 9nm moir\'e periodicity in a zero and (c) 5$^{o}$ twisted MoS$_2$/WSe$_2$ heterobilayer, considering their lattice constants. }
  \label{fig:B2}
\end{figure}

In contrast, in a rigid moir\'e superlattice the twist angle between two lattices creates a periodic pattern resulting from a simple interference of the original lattices while preventing significant reconstruction of the atomic arrangement (Figure \ref{fig:B1}(a)). In this case, a new supercell emerges with a larger unit cell in real space (yellow diamond in Figure \ref{fig:B1}(a)) that creates a smaller,  moir\'e-Brillouin zone in reciprocal space.  The name \textit{moir\'e} suggests a large-scale interference pattern that emerges upon superposition of two structures with similar periodicities. The period of the moir\'e superlattice can be determined via   \cite{ribeiro2018twistable,tran2020moire}:
\begin{equation}
\label{eq:moireperiod}
\lambda_m = \frac{{(1+\Delta_l)a_0}}{{\sqrt{2(1+\Delta_l)(1-\cos\theta)+\Delta_l^2}}},
\end{equation}
where $\Delta_l$ represents the lattice mismatch, calculated as $\frac{a'_0 - a_0}{a_0}$, with $a'_0$ and $a_0$ being the lattice constant of the monolayer with the larger and smaller value, respectively. In Figure \ref{fig:B2}(a), two examples of moir\'e periodicity resulting from varying twist angles, $\theta$, are plotted using Eq. \ref{eq:moireperiod},  neglecting atomic relaxation.  Here, the lattice constants for MoS$_2$ and WSe$_2$ are 0.319~nm and 0.331~nm, respectively. Comparing a homobilayer MoS$_2$/MoS$_2$ (black line, with identical lattice constants, $\Delta_l$=0) to a heterobilayer MoS$_2$/WSe$_2$ (red line, with $\approx3.7\%$ lattice mismatch), a significant deviation in moir\'e periodicity is observed for twist angles below 5$^{o}$. Unlike heterobilayers, homobilayers at zero twist angles have an infinite moir\'e periodicity, indicating the absence of a superlattice. Figures \ref{fig:B2}(b,c) illustrate a schematic representation of the long-wavelength, periodic moir\'e pattern formed by overlaying two monolayers with differing lattice constants (MoS$_2$ and WSe$_2$). The moir\'e periodicity of MoS$_2$/WSe$_2$ decreases from 9~nm to 3~nm as twist angles vary from 0$^{o}$ (Figure \ref{fig:B2}(b)) to 5$^{o}$ (Figure \ref{fig:B2}(c)), respectively. Consequently,  combining TMD monolayers with varying lattice constants  \cite{kormanyos2015k} and twist angles enables precise control over moir\'e superlattice periodicity. The resulting moir\'e supercell extension can be substantial,  typically ranging from a few to tens of nanometers.\\

Notably,  there are two energetically favorable stacking orders that offer distinct physics in hexagonal bilayers, \textit{i.e.} 0 and 60 degrees twist.  In the case of homobilayers,  the stacking order of 0 degrees twist is named \textit{AB} and for the 60 degrees, \textit{AA$^{\prime}$}. A side-view example of the \textit{AB} stacking is shown in Figure \ref{fig:B1}(b).  The interlayer distance in these two cases is very similar, on the order of  6.17 \AA. In contrast,  for 0 and 60 degrees twisted heterobilayers the local symmetry and interlayer distance periodically varies (because each high symmetry stacking has a preferred interlayer distance) following the moir\'e period due to the lattice constant difference between the constituent monolayers (Figure \ref{fig:B2}(b)).  Various terminologies have been employed in different studies to describe the stacking orders of bilayers, leading to unnecessary confusion within the community. To eliminate this confusion, the stacking arrangement in both homo- and heterobilayers is now commonly denoted as R-type for parallel (0$^{o}$) and H-type for anti-parallel (60$^{o}$) configurations. The letters H and R represent $hexagonal$ and $rhombohedral$ respectively, mirroring the bulk \textit{2H} and \textit{3R} stacking polytypes  \cite{van2014tailoring}. It is noteworthy that H-type bilayers possess an inversion center, while R-type bilayers do not, resulting in significant differences in their respective nonlinear properties \cite{wen2019nonlinear,shree2021interlayer}. \\ 

\begin{figure}[t!]
  \centering
  \includegraphics[width=0.8\linewidth]{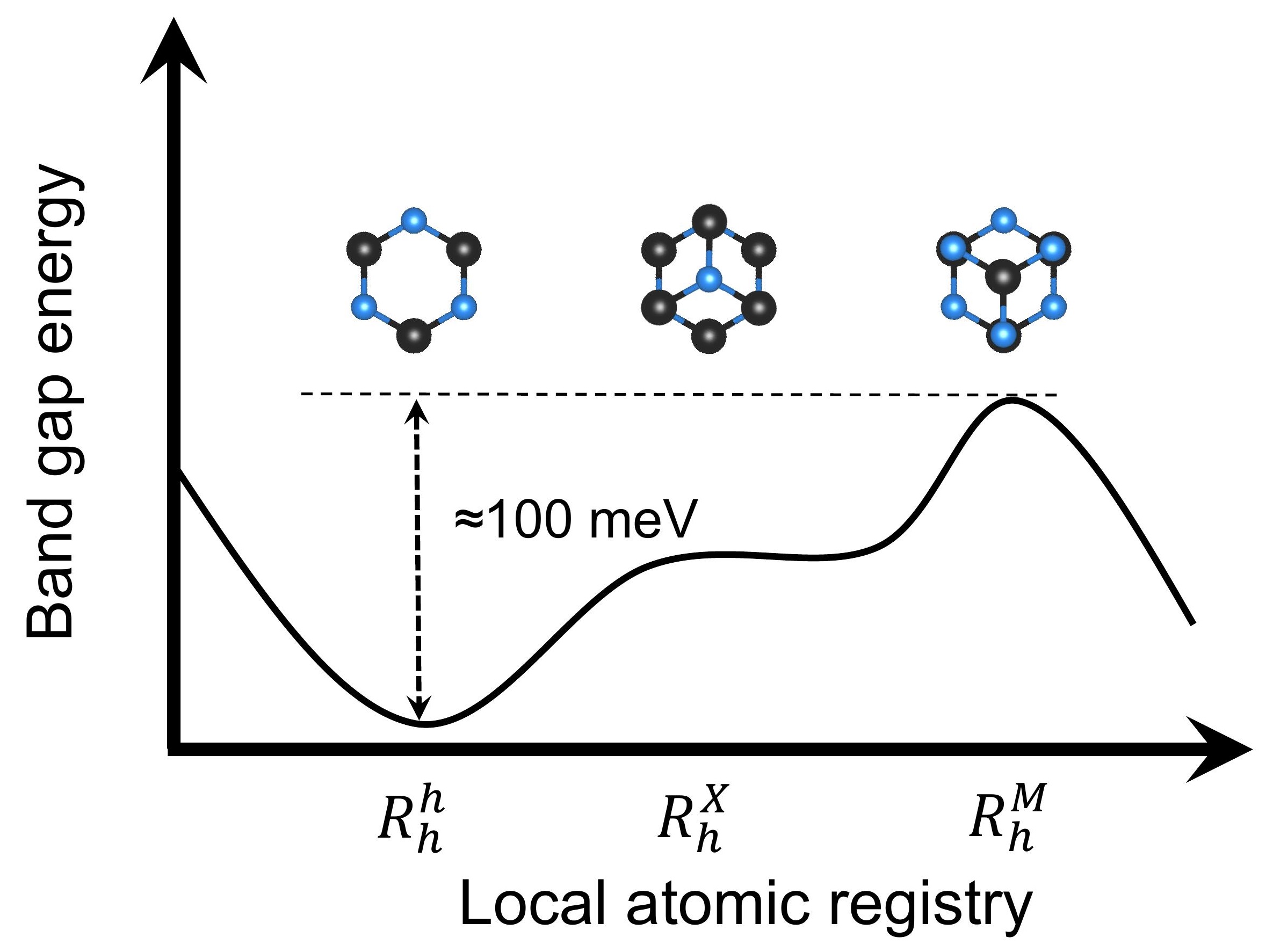} 
  \caption{Band gap energy landscape of a bilayer for parallel (R-type) atomic registries. The potential difference between the highest and lowest energy is approximately 100 meV for R-type stacking, while similar -but weaker- modulation is expected for H-type stacking. The values are taken from Refs  \cite{gillen2018interlayer,wu2018theory}.}
  \label{fig:B3}
\end{figure}

Within the moir\'e supercell, three locations with a three-fold rotational symmetry are identified: R$^{h}_{h}$,  R$^{X}_{h}$,  and R$^{M}_{h}$ (inset of Figure \ref{fig:B1}(a)).  Here,  R$^{\mu}_{h}$,  denotes R-type stacking with the $\mu$ site (h, corresponding to the hexagonal center,  X the chalcogen atom, and M the metal atom) of the top layer aligning with the hexagonal center (\textit{h}) of the bottom layer. Correspondingly, for anti-parallel stacking orders the notations switch to H$^{h}_{h}$,  H$^{X}_{h}$,  and H$^{M}_{h}$ (not shown in Figure \ref{fig:B1}(a)).  These high-symmetry areas, often referred to as \textit{local atomic registries}, play a significant role in determining the local optoelectronic properties and optical selection rules of the moir\'e superlattice. The interactions between neighboring atoms within each atomic registry, together with the symmetry and periodicity of the superlattice, lead to modifications in the electronic band structure of the homo- or heterobilayer. The possibility of engineering such a superlattice presents exciting opportunities for band gap engineering, the formation of moir\'e minibands in the Brillouin zone, and the realization of topological states of matter. The moir\'e superlattice induces a periodic spatial variation in the band gap, closely linked to the local atomic registry. Notably, the band gap can undergo modulation with a potential depth of $ \approx $ 100 meV in R-stacking, in contrast to a much weaker modulation of $ \approx $ 20 meV in H-stacking  \cite{yu2017moire, tran2020moire}. Figure \ref{fig:B3} illustrates an example of the lateral modulation of the band gap as a function of direction across the three rotational symmetries of the superlattice in R-stacking. 

Now, we briefly introduce experimental methods to distinguish between H- and R-stacking orders and to elucidate information on rigid and reconstructed moir\'e superlattices in TMD bilayers.

\subsubsection {Experimental techniques for investigating moir\'e superlattices}
 
We categorize three distinct experimental techniques for examining moir\'e superlattices, each with its own advantages and drawbacks: scanning and transmission electron imaging, tip-based methods, and optical spectroscopy and imaging. Traditionally, imaging methods used to experimentally investigate twisted, rigid, and reconstructed moir\'e superlattices rely on the high spatial resolution of electron imaging techniques  \cite{he2021moire}, such as tunneling electron microscopy (TEM) \cite{weston2020atomic,rosenberger2020twist}, scanning tunneling electron microscopy (STEM) \cite{huang2018topologically}, and secondary electron microscopy (SEM)  \cite{sushko2019high}. While these methods offer powerful real-space imaging with atomic resolution and they can easily offer imaging of nanometer-scale moir\'e superlattices, they can also be destructive and impractical for certain device geometries. Kelvin probe and piezo force microscopy (KFM and PFM) also provide direct imaging of moir\'e superlattices and offer quantitative insights into the surface potential of both reconstructed and rigid moir\'e configurations. These techniques can also reveal opposing polarities within ferroelectric reconstructed domains  \cite{stern2021interfacial}.  However, tip-based techniques require the tip to be brought extremely close to the sample's surface. Consequently, the top surface must remain bare and not covered by an hBN layer. This lack of coverage limits the sample's protection from its environment. 
Non-destructive optical methods, such as polarization-resolved microphotoluminescence ($\mu$PL), second harmonic generation (SHG), and microRaman spectroscopy, are routinely used to extract the twist angle in bilayers and investigate moir\'e superlattices, though their spatial resolution is given by light diffraction limit (hundreds of nanometers) \cite{shree2021guide}.\\

In the case of homobilayers, differential reflectivity experiments can distinguish between H$^{h}_{h}$ and R$^{X}_{h}$ stacking orders by measuring the energy difference between A and B excitons \cite{paradisanos2020controlling}. The spatial resolution of optical techniques can be enhanced by integrating optical spectroscopy/imaging methods with the near-field enhancement provided by a metallic tip, such as tip-enhanced PL and Raman spectroscopy. Furthermore, recent advances in scanning near-field optical microscopy methods (SNOM)  \cite{luo2020situ} provide information based on light scattering phase and dynamics of quasiparticles such as phonon polaritons (coupling between phonons and photons) and plasmon polaritons (coupling between collective oscillations of electrons and photons) in the moir\'e domains, with spatial resolution limited only by the tip radius (approximately 10 nm), comparable to the moir\'e supercell. In optical spectroscopy, it is important to emphasize that, with its ability to probe the electronic and vibrational properties of materials, it offers a non-invasive and highly sensitive approach to characterizing the microscopic details of these structures. By utilizing advanced optical techniques, such as ultra-low and high frequency Raman spectroscopy, as well as low-temperature photoluminescence spectroscopy,  it is possible to distinguish subtle variations in lattice configurations and explore the manifestation of moir\'e patterns. 
In addition, TMD monolayers can serve as efficient sensors to optically probe correlated states of adjacent rigid moir\'e structures \cite{gu2024remote} or as optical probes for ferroelectricity in the underlying reconstructed structure. The doping density and type that is induced in the monolayer by the underlying ferroelectric structure can be controlled by adjusting the position of the monolayer semiconductor in relation to the ferroelectric interface. For example, recent results highlight the potential of the ferroelectric hBN/WSe$_2$ van der Waals stacking as a promising optoelectronic structure \cite{fraunie2023electron}. 

In the context of rigid moir\'e structures, the periodic potential induces exciton localization, which can be investigated through low-temperature, polarization-resolved photoluminescence spectroscopy. This allows to record the emergence of spectrally-sharp emission from excitons trapped in the moir\'e potential, exhibiting anti-bunching photon emission statistics \cite{baek2020highly}. This analysis follows specific optical selection rules based on the corresponding atomic registry \cite{yu2017moire}.  The alignment and twist angle of the layered structure can be systematically monitored before, during, and after fabrication of bilayer superlattices, using polarization-resolved second harmonic generation \cite{paradisanos2022second}. For the lattice reconstruction, resonant second-harmonic generation spectroscopy recently served as a non-destructive method to characterize the lattice structure of small-angle twisted bilayer samples.  \cite{xu2023nonlinear}. Notably, ultra-low and high-frequency Raman spectroscopy identifies the relative intensity ratio of various signature lattice vibrations \cite{quan2021phonon}. This approach, applied as a function of twist angles, reduces the stacking order and rotation uncertainty and provides valuable insights into the nature of the moir\'e superlattice, distinguishing between relaxed and rigid configurations. 

\begin{figure}[t!]
  \centering
  \includegraphics[width=1\linewidth]{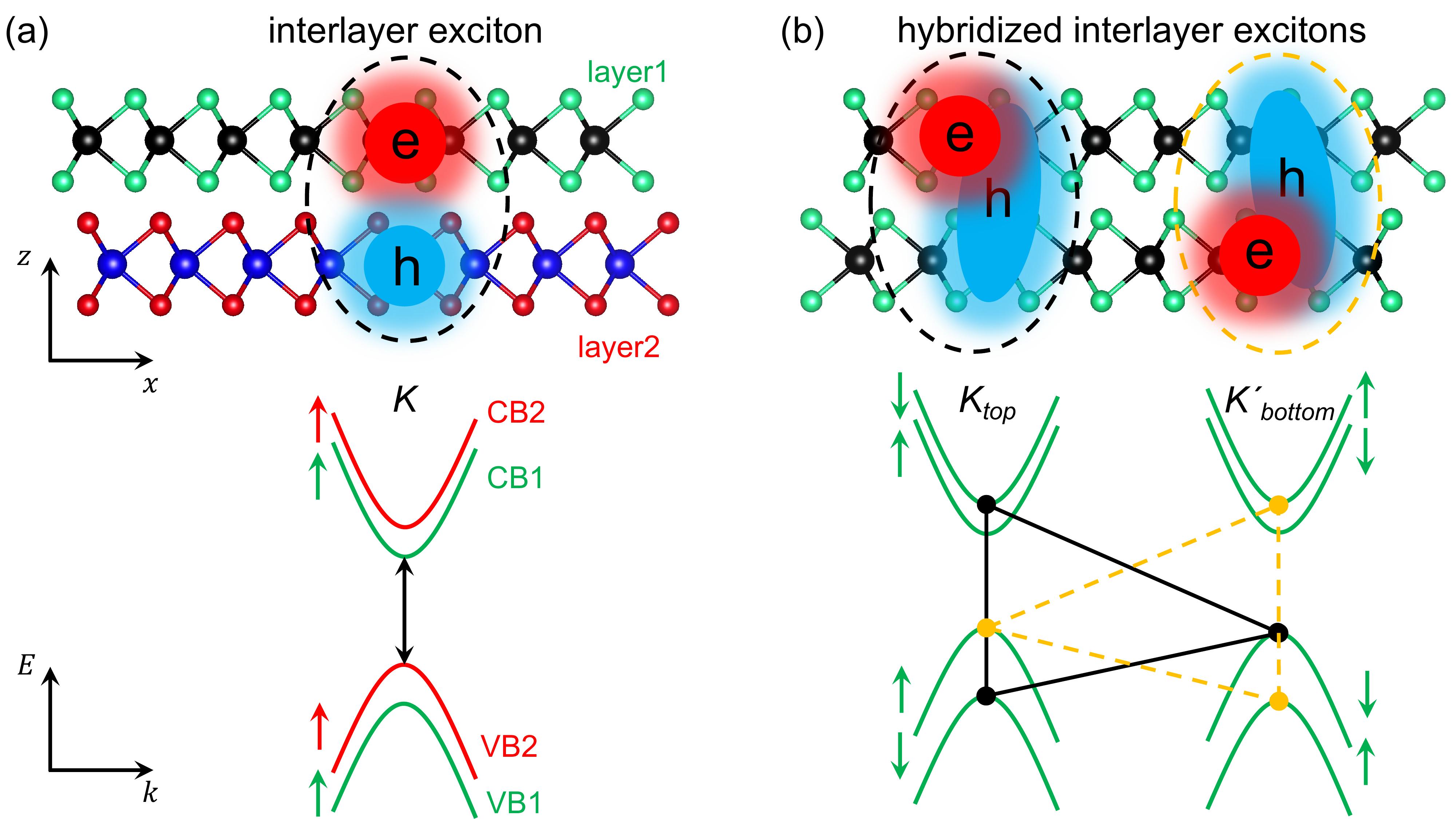} 
  \caption{Formation of interlayer excitons in (a) heterobilayers and (b) hybridized interlayer excitons in  homobilayers. In heterobilayers, electrons and holes are confined to different layers. Radiative recombination involves spin and momentum conserving transitions between the bottom conduction band of the upper layer and the top valence band of the bottom layer (indicated by black arrow). In H$^{h}_{h}$ homobilayers, the electron is localized in one of the two layers, while the hole is delocalized over both layers due to mixing with B-excitons. This hybridization arises from spin-conserving hole tunneling processes that involve the lower and upper valence bands of different layers. This unique process in this stacking order results in the formation of two symmetric hybridized interlayer excitons with opposite out-of-plane dipole moments. The states involved in the formation of each hybridized exciton are depicted in black and orange colors. For clarity, the K valley of the top layer and the K$^{\prime}$ valley of the bottom layer are presented separately in momentum space.}
  \label{fig:B4}
\end{figure}      

\subsection {Interlayer excitons}\label{subsec:interlayer-excitons}
In this section, we focus on  Coulomb-bound electron-hole pairs in TMD bilayers, known as interlayer excitons, where the constituent electron and hole can be located in different layers, separated by the van der Waals gap. These interlayer excitons have garnered significant scientific interest due to their unique and tunable properties for applications in optoelectronics and quantum optics \cite{ciarrocchi2022excitonic}.  Interlayer excitons in TMD bilayers significantly enhance the capabilities initially offered by interlayer excitons in III-V coupled quantum wells \cite{butov2017excitonic}. The advantages of TMDs include their remarkably thin layers (less than 1 nm), which can be assembled with sharp interfaces and transferred to any photonic circuit. The substantial binding energies of interlayer excitons in TMDs, typically around 150 meV, compared to just a few meV for interlayer excitons in III-V systems, allows their observation even at room temperature. Furthermore, the ability to control the interlayer distance by adding thin hBN layers at the heterobilayer interface, along with twist angle degree of freedom between the two layers, enables new functionalities for controlling the absorption, emission, transport, and localization of interlayer excitons. We  distinguish between two types of interlayer excitons: those that form in heterobilayers and those that form in homobilayers. However, we will only discuss direct transitions in momentum space. Therefore, for TMD heterobilayers (Figure\ref{fig:B4}(a)) and homobilayers (Figure\ref{fig:B4}(b)), only transitions between conduction and valence bands at the K-points of the Brillouin zone will be introduced.

\subsubsection {Hybrid interlayer excitons in homobilayers}
Vertical stacking of monolayers forming bilayers, allows the formations of interlayer and hybrid excitons, Figure \ref{intro1}(b). Hybrid excitons are crucial in the dynamics, where they provide a key intermediate step between the formation of intra- and interlayer excitons in both hetero- and homobilayers  \cite{Schmitt22,Meneghini22,Meneghini23}. In addition, their degree of hybridization and resulting spatial dipole can be electrically tuned, providing a crucial transport control which will be discussed in Figure \ref{tr2-conventional}  \cite{Tagarelli23}. Momentum-dark interlayer and hybrid excitons can have particularly low energy, resulting in emission of photoluminescence via phonon-assisted mechanisms  \cite{Hagel21,Kumar23}.
We emphasize that there are striking differences in absorption and emission energy and strength when comparing interlayer excitons in heterobilayers and hybridized interlayer excitons in homobilayers. Considering the hybridized interlayer exciton shown in Figure\ref{fig:B4}(b) (black lines) post-DFT calculations predict for MoS$_2$ bilayers an unusually high oscillator strength on the order of 20\% compared to the corresponding intralayer exciton transition \cite{deilmann2018interlayer,gerber2019interlayer}. 
\begin{figure}[t!]
  \centering
  \includegraphics[width=1\linewidth]{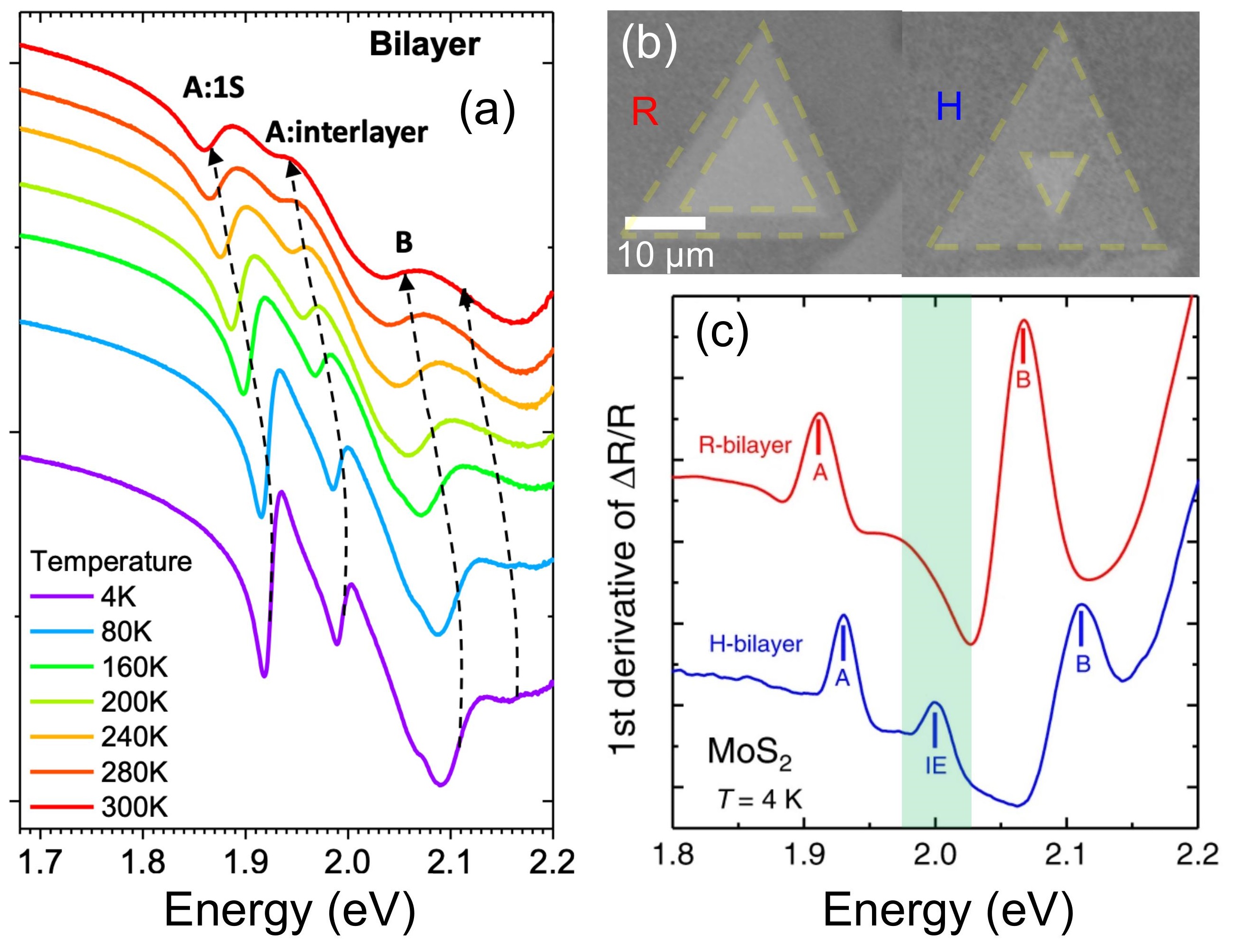} 
  \caption{(a) Temperature dependent reflectivity experiments of H$^{h}_{h}$ MoS$_2$ homobilayers. Hybridized interlayer excitons have sufficient oscillator strength and binding energy, making them detectable even at room temperature.  (b) Optical microscope images of as-grown R (left) and H$^{h}_{h}$ MoS$_2$ homobilayers (right) on silicon oxide/silicon substrate.(c) Twist angle-dependent reflectivity experiments demonstrating that interlayer hopping of holes for the formation of hybridized excitons (highlighted in green color) is allowed in H$^{h}_{h}$ MoS$_2$ homobilayers but not in R-stacking order ($ i.e. $, zero twist angle). Figs. (a) and (b), (c) are adapted from Refs. \cite{gerber2019interlayer} and  \cite{paradisanos2020controlling}, respectively. }
  \label{fig:B5}
\end{figure}
This is surprising for an interlayer exciton state where usually the reduced wavefunction overlap between the spatially separated electron and hole results in low oscillator strengths.  Optical spectroscopy experiments verify the prediction, showing a transition with high oscillator strength in reflectivity spectra of MoS$_2$ bilayers (similar to MoSe$_2$ bilayers) that persists even at room temperature (Figure \ref{fig:B5}(a)).The primary contributions to this transition arise from states corresponding to the spin-down bottom valence band of the top layer, partially hybridized with the spin down upper valence band of the bottom layer, along with a well-localized electron at the second lowest conduction band states of the top layer. The symmetric counterpart of this hybridized exciton, where the localized electron resides in the bottom layer, is illustrated by yellow dashed lines in Figure \ref{fig:B4}(b). This symmetric arrangement of two degenerate hybridized excitons is not feasible in heterobilayers, where the electron and hole are distinctly localized in separate layers. The remarkable oscillator strength observed for this spatially indirect transition in MoS$_2$ bilayers arises from its strong mixing with intralayer B-exciton states. It is important to note that the symmetry of the first valence bands in K is mainly characterized by $d_{x^{2}-y^{2}}$ and $d_{xy}$ orbitals, mixed with $ p_{x,y} $ orbitals of the chalcogen atom, while the first conduction bands are composed of $d_{z^{2}}$ orbitals around the metal atom, hence resulting in a negligible overlap with the analogous orbitals from the other layer. As a result, hybridization (or interlayer hopping) of holes in the valence bands is possible, but remains forbidden for electrons in the conduction bands.  The stacking order symmetry of the homobilayer also influences the formation of hybridized excitons, primarily occurring in the H$^{h}_{h}$ stacking order, as can been seen in Figure \ref{fig:B5}(b,c) where the interlayer exciton absorption 
is visible for H-type but not R-type stacking \cite{gerber2019interlayer,Hagel23}. Therefore, the ability to engineer the formation of hybridized excitons via the twist angle and stacking order presents an exciting opportunity for producing photonic structures using homobilayers, as recently showcased using large-scale chemical vapor deposition MoS$_2$ samples (see Figure \ref{fig:B5}(b,c)) \cite{paradisanos2020controlling}.

We emphasize that homobilayers are indirect gap semiconductors with the lowest energy transitions involving states away from the K-points of the Brillouin zone (e.g. $\Lambda$ conduction band and $\Gamma$ valence band \cite{Hagel21,Kumar23}). Consequently, hybridized interlayer excitons in homobilayers are not strong light emitters, however recent reports demonstrate light emission from these exotic complexes \cite{zhao2022interlayer,Kumar23}. Notably, the interlayer coupling for hole states is also influenced by the the magnitude of the spin-orbit splitting. As the spin-orbit splitting increases, the probability of hole tunneling decreases. This is because a larger energy barrier must be overcome for holes to hybridize between the two monolayers. Bilayer MoS$_2$ \cite{gerber2019interlayer,paradisanos2020controlling} and MoSe$_2$ \cite{horng2018observation,Hagel23} offer advantageous conditions for observing interlayer excitons, as the interlayer coupling of valence bands is allowed and the spin-orbit splitting is smaller ($ \approx $180 and $ \approx $200 meV, respectively) compared to MoTe$_2$, WSe$_2$, and WS$_2$. By measuring the energy splitting between A and B excitons in bilayers, it is possible to experimentally measure the interlayer hopping term, $t_\perp$, linked to the interlayer interaction strength, in very good agreement with predictions based on k$\cdot$p models of bilayers \cite{gong2013magnetoelectric}. Later, we will explore how hybridized excitons can be widely tuned in the presence of electric and magnetic fields.

\begin{figure}[t!]
  \centering
  \includegraphics[width=0.9\linewidth]{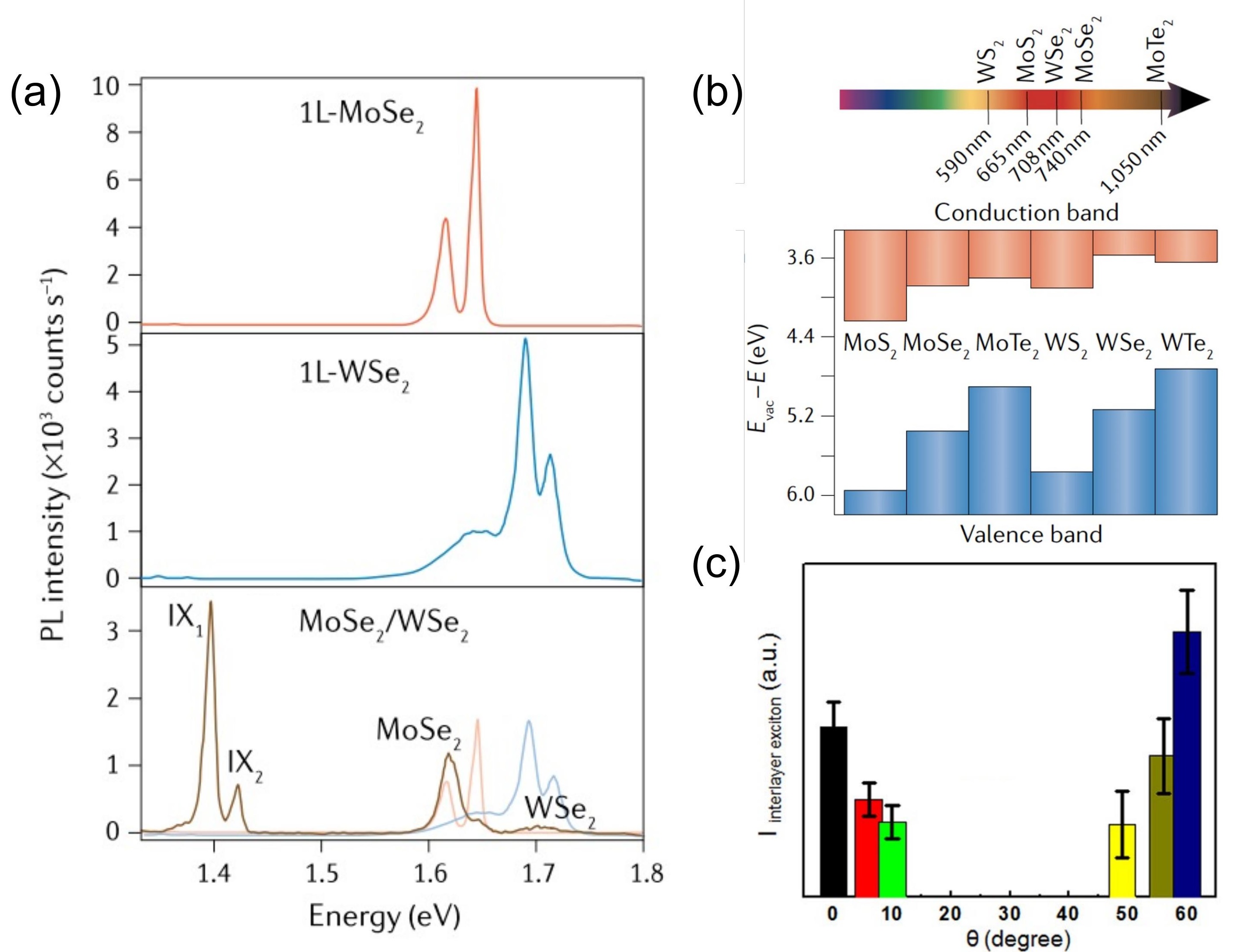} 
  \caption{(a) PL at T = 4.2 K from hBN-encapsulated MoSe$_2$ monolayer (orange), WSe$_2$ monolayer (blue), and MoSe$_2$/WSe$_2$ heterostructure (brown). Intralayer exciton emission is observed from the MoSe$_2$ and WSe$_2$ monolayers. Interlayer exciton emission (IX1 and IX2) appears at lower energies compared to the intralayer resonances in the heterostructure. (b) Band alignment between 2D semiconductor monolayers as calculated in Ref.  \cite{liu2016van}. The arrow indicates the wavelength of the optical gap for various materials ranging from visible to near-infrared. (c) Intensity of the interlayer exciton emission in MoSe$_2$/WSe$_2$ as a function of the twist angle. For twist angles away from lattice alignment (0 or 60 degrees), the interlayer exciton intensity drops dramatically due to momentum mismatch between the K valleys involved. Figs. (a), (b) and (c) are adapted respectively from Refs. \cite{shree2021guide,ciarrocchi2019polarization}, \cite{shree2021guide,liu2016van} and \cite{nayak2017probing}. 
  }
  \label{fig:B6}
\end{figure}

\subsubsection {Interlayer excitons in heterobilayers}
Combining monolayers of different composition, such as MoSe$_2$ and WSe$_2$, can favor the formation of excitonic states where the electron and hole are spatially separated in adjacent monolayers, due to the different band-edge energies. This spatial separation results from the type-II (staggered) band alignment commonly found in TMD heterobilayers, where the conduction and valence band edges reside in different layers (Figure \ref{fig:B4}(a)). They exhibit differences compared to hybridized interlayer excitons in homobilayers. Their transition energy lies below the intralayer excitons in both layers, resulting in strong light emission in PL, also due to the direct gap nature of aligned heterobilayers (Figure \ref{fig:B6}(a)).  Typical calculated valence and conduction band energy values and relative band alignment between different TMD monolayers are presented in Figure \ref{fig:B6}(b) \cite{liu2016van}. The spatial separation of charge carriers results in a reduced overlap between the electron and hole wavefunctions that impacts the physical properties of interlayer exciton, such as their lifetime, dipole moment and transport, as we will see in the transport section. Additionally, factors such as the strength of interlayer coupling, dielectric screening, and spin-orbit splitting can influence the formation and properties of interlayer excitons in TMD heterobilayers. The weak interfacial van der Waals forces in heterobilayers allow for the stacking of constituent layers with lattice mismatch, a crucial aspect for transitions at the K-points. Here, the conduction and valence band edges at the Brillouin zone corners of different TMD monolayers are displaced in momentum space due to lattice mismatch and twist angle (see also Figure \ref{fig:B8}(c)). For instance, twisting monolayers in a heterobilayer significantly impacts the radiative strength of interlayer excitons due to the introduced momentum mismatch between the K bands (Figure \ref{fig:B6}(c)). However, for small twist angles, momentum conservation enabled by the moir\'e Brillouin zone overcomes the momentum mismatch caused by the twist \cite{rivera2018interlayer}. It is important to highlight that for interlayer excitons involving transitions at the $\Gamma$-point (the center of the Brillouin zone), as observed in InSe-based heterostructures, no dependence on twist angle is anticipated, thereby avoiding any momentum mismatch  \cite{ubrig2020design}. However, significant hole tunneling at this point could result in an efficient twist-induced dehybridization  \cite{Sokolowski23}.
   
\subsubsection {Optical selection rules}\label{sec:optSelRule}

Beyond conservation of energy, momentum and electron-hole overlapping in real-space, there are fundamental rules that dictate the allowed and forbidden optical transitions in semiconductors. They are known as \textit{optical selection rules} and they are associated with the transition dipole moment and the specific symmetry of the states involved in the transition. The transition dipole moment (or optical transition dipole) differs from the permanent electric dipole, which arises from the electron-hole spatial separation in bilayers and can be measured in experiments using out-of-plane electric fields (see Stark effect \cite{leisgang2020giant}) or laser power dependent experiments \cite{Rosati23}. Optical selection rules play a crucial role in spectroscopic analyses and studies of electronic band structures and quantify the strength of interaction between light and matter. Knowing the wave functions of initial and final states, one can evaluate the dipole matrix element, a term that describes the strength of the interaction, linked to the transition dipole moment. Consequently, it is possible to determine the probability of a transition and whether it is dipole allowed or forbidden for a given polarization of light. The primary interaction stems from the electric dipole moment, where the electric field of light interacts with the electric charge distribution of matter. Typically, the electric dipole approximation is employed to describe this interaction, assuming that the light wavelength (and hence the amplitude of the electric field) does not vary significantly over the spatial distribution of carriers. This interaction dominates in most optical processes and forms the basis for many optical selection rules. Additionally, magnetic interactions may also play a role, particularly when considering the interaction between the electronic magnetic moment with the oscillating magnetic field of light. Beyond the dipole interaction, higher-order interactions such as quadrupole and octupole moments become relevant, however they are orders of magnitude weaker for the systems that we describe here. In the following, we discuss the coupling between light polarization and interlayer optical transitions at the K valleys. 

\begin{figure}[t!]
  \centering
  \includegraphics[width=1\linewidth]{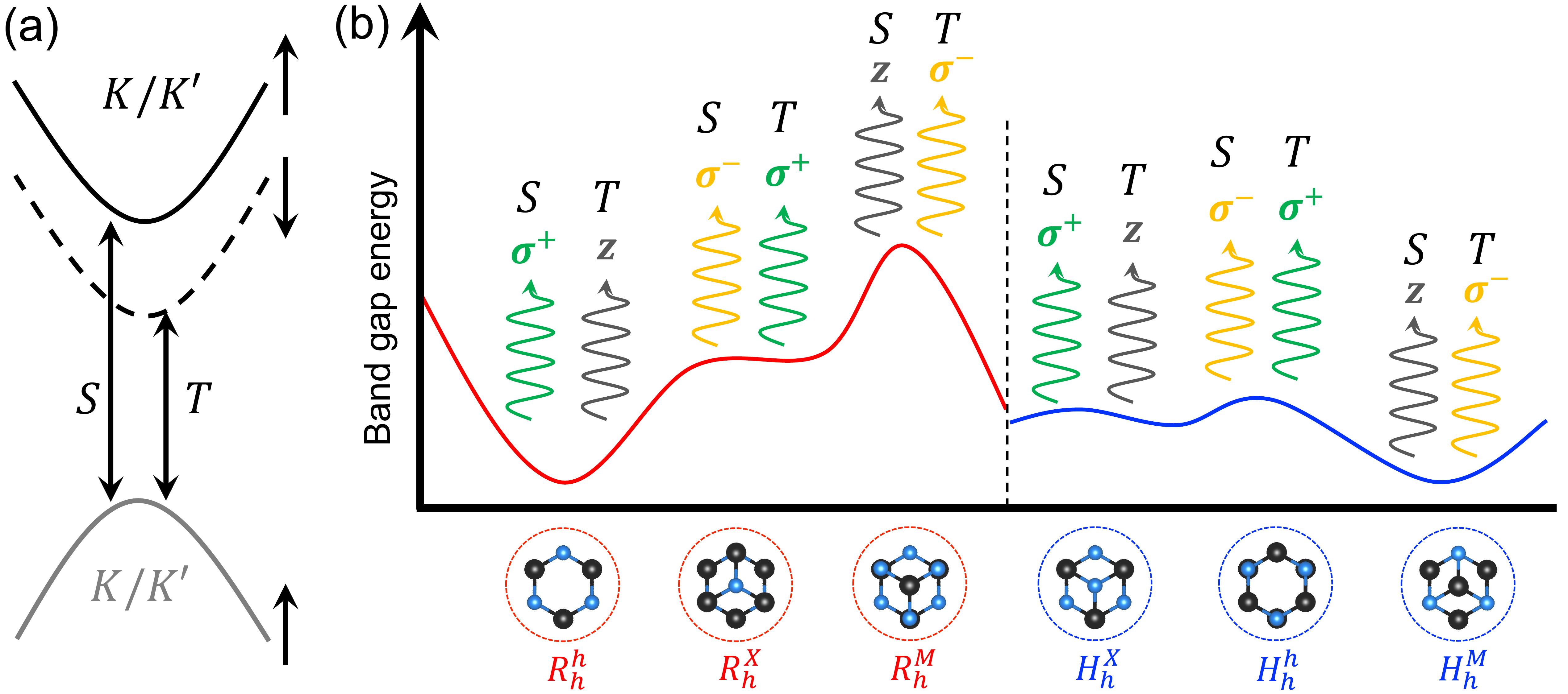} 
  \caption{(a)Schematic representation of a heterobilayer's band structure near the corners of the Brillouin zone. For $R$-type bilayers, interlayer transitions occur between K (K$^{\prime}$) conduction band of the top layer (black color) with K (K$^{\prime}$) valence band of the bottom layer (grey color). In contrast,  for $H$-type bilayers, interlayer transitions occur between K (K$^{\prime}$) conduction band of the top layer (black color) with K$^{\prime}$ (K) valence band of the bottom layer (grey color). Spin-conserved (singlet) and spin-flipped (triplet) transitions are indicated by black arrows.  Note that the spin order of the lower energy conduction band can be also opposite, depending on the specific combination of TMD monolayers. (b) Optical selection rules for singlet and triplet transitions, indicating light polarization for every high-symmetry atomic registry in $R$ and $H$ type bilayers. An example of the band gap modulation is shown with red and blue color for $R$ and $H$ stacking orders, respectively, to emphasize on the expected variation of the exciton emission energies at each atomic registry. Green color corresponds to $\sigma^{+}$ polarization, yellow color corresponds to $\sigma^{-}$ polarization and grey color corresponds to out-of-plane (z) polarization.}
  \label{fig:B7}
\end{figure}

Before delving into TMD bilayers, we emphasize that the crystal symmetry together with the orbital character of the bands and spin-orbit interaction are responsible for the optical selection rules in TMD monolayers. Interband transitions at the K$^{\pm}$ in monolayers are chiral due to the Bloch functions of valence and conduction bands, each possessing different orbital magnetic quantum numbers.  This chirality is linked to the threefold rotational symmetry, $\hat{C}_{3}$ \cite{cao2012valley} and broken inversion symmetry of monolayers \cite{xiao2012coupled}. Considering transitions between valence and conduction bands at K and K$^{\prime}$, the angular momentum components must change as $\Delta\textit{m}=\pm1$. Thus, the optical transition matrix element gives right ($\sigma^{+}$) and left ($\sigma^{-}$) handed polarized light coupling with K and K$^{\prime}$ valleys, respectively, resulting in a valley-selective circular dichroism \cite{mak2012control}. 
Similar symmetry arguments can be applied in TMD bilayers.  For natural $H$-stacked Mo-based homobilayers, circular dichroism should vanish for $K, K^{\prime}$ transitions due to restoration of inversion symmetry \cite{liu2015electronic}. Applying out-of-plane electric fields in $H$-stacked natural Mo-based homobilayers \cite{wu2013electrical} or introducing a twist angle \cite{scuri2020electrically}, can break inversion symmetry and result in a net circular dichroism. Interestingly, experiments using polarization-resolved photoluminescence in the presence of vertical electric fields have recently reported hybridized interlayer excitons in natural MoS$_2$ homobilayers demonstrating near 100\% negative circular polarization \cite{zhao2022interlayer}. For W-based, naturally stacked $H$ homobilayers, despite the restoration of inversion symmetry, an extra degree of freedom exists: the layer polarization. The large spin-orbit splitting can suppress interlayer hopping of electrons and holes, localizing the carriers in either the upper or lower layer depending on the valley and spin state. This can result in a spin optical selection rule with circularly polarized exciton emission in homobilayers WSe$_2$ \cite{jones2014spin,wang2014exciton} and WS$_2$ \cite{zhu2014anomalously} due to the spin-layer locking effect \cite{gong2013magnetoelectric}.

For TMD heterobilayers, out-of-plane mirror symmetry is always broken partly due to the difference in the lattice constant between the top and bottom layer. However, this also applies in lattice-matched heterobilayers either prepared by chemical vapor deposition (CVD) \cite{hsu2018negative} or thermally-induced reconstructed heterostructures \cite{baek2023thermally} where the two monolayers may have the same lattice constant but they differ in composition. The broken mirror symmetry in TMD heterobilayers has important consequences in the coupling between in-plane or out-of-plane polarized light with singlet (spin-conserving) and triplet (spin-flip) interlayer transitions (see Figure \ref{fig:B7}(a)). Specifically, the transition dipole moment is calculated to be comparable between singlet and triple interlayer excitons \cite{yu2018brightened} and can be experimentally evaluated by polarization-resolved optical spectroscopy. The local optical selection rules for interlayer excitons in the moir\'e potential are governed by the $\hat{C}_{3}$ operator (symmetry operation that represents a rotation of 120$^o$ about a threefold symmetry axis) for each atomic registry and the corresponding participation of orbitals, around a common rotation center of the two layers (Figure \ref{fig:B7}(b))  \cite{yu2018brightened}. 

In Figure \ref{fig:B7}(b), the spatial modulation of the band gap in a moir\'e heterobilayer is illustrated together with the light polarization emission expected in each atomic registry with three-fold symmetry. It is evident that the transition dipole moment can be either in-plane ($\sigma^{+}$, $\sigma^{-}$ polarization) or out-of-plane (z polarization) with a spatial variation that depends on the location of the interlayer exciton. Identifying the polarization and emission energy of interlayer excitons through low-temperature photoluminescence spectroscopy -where the linewidths are sufficiently sharp for spectral separation- can provide crucial insights into their physical origin within the moir\'e superlattice. A significant challenge lies in spatially distinguishing light emissions originating from different atomic registries, primarily due to the diffraction limit, (several hundreds of nanometers) which averages signals over thousands of moir\'e supercells. Local interface disorder due to strain and impurities can result into inhomogeneous broadening, thus limiting the spectral identification of interlayer excitons with different local origin. We emphasize that recent reports determine transition dipole moments of interlayer excitons to be 99\% in-plane, using low temperature back focal plane imaging \cite{sigl2022optical}. Numerous studies claim emission of multiple, broad interlayer excitons with opposite circular polarization, attributed either to ground and excited states \cite{tran2019evidence}, momentum indirect transitions separated by spin-orbit coupling \cite{hanbicki2018double} or spin-singlet and triplet excitons \cite{jin2019identification}. Despite the predicted significant oscillator strength of z-polarized interlayer excitons, their negligible contribution in the optical spectra remains unclear. In future experiments, tip-enhanced, low temperature PL spectroscopy with a spatial resolution of $ \approx20 $nm (similar to the size of the moir\'e unit cell) can probably enable the detection of z-polarized interlayer excitons at specific areas of the superlattice \cite{park2018radiative}. Moir\'e-trapped interlayer excitons with sharp emission and uniform polarization degrees will be discussed below.

\subsubsection {Localization of interlayer excitons for quantum optics}

\begin{figure}[t!]
  \centering
  \includegraphics[width=1\linewidth]{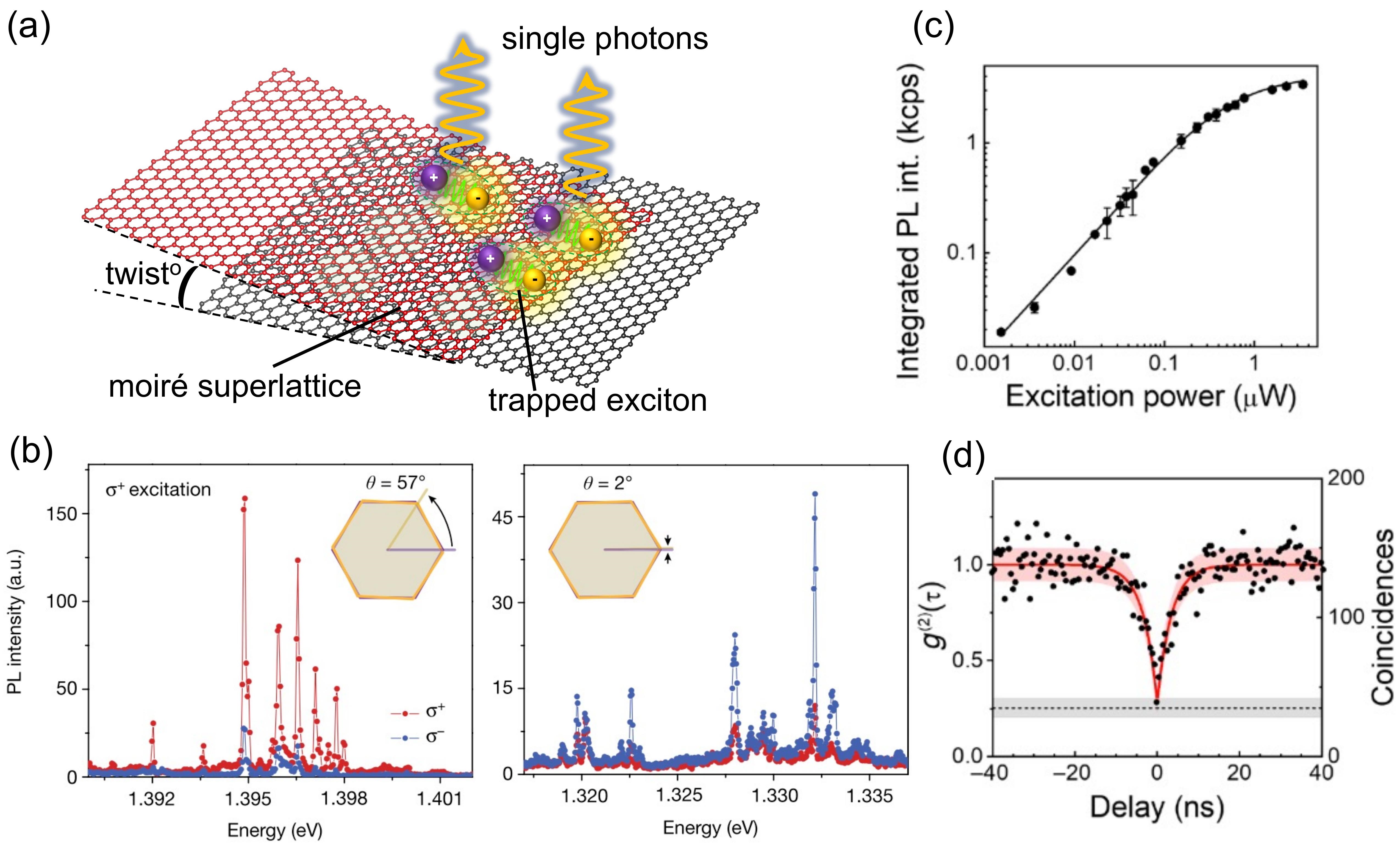} 
  \caption{(a) Illustration of a twisted heterostructure showing a moir\'e superlattice, where excitons can be trapped at local periodic potentials and create arrays of single photon emitters. (b)  Helicity-resolved photoluminescence spectra of moir\'e-trapped interlayer excitons of heterobilayers close to $H$-stacking (left) and $R$-stacking (right).  Insets show the precise twist angle and red (blue) spectra correspond to $\sigma^{+}$ ($\sigma^{-}$) light emission.  (c) Power dependent photoluminescence integrated intensity of a single emitter (taken from reference  \cite{baek2020highly}). (d) Second-order photon correlation statistics of a single emitter showing antibunching photon statistics with a $ g^{(2)}(0) $ value of 0.28 $ \pm $ 0.03 (also taken from reference  \cite{baek2020highly}). Figs. (b) and (c,d) adapted respectively from Ref.  \cite{seyler2019signatures} and  \cite{baek2020highly}. }
  \label{fig:B8}
\end{figure}

TMD heterostructures offer a versatile approach to creating quantum light sources by exploiting moir\'e trapping potentials for excitons. These potentials, as discussed in previous chapters, are periodic in space and can reach depths from several tens to a hundred meV, depending on the composition and stacking order of the monolayers. Within these potentials, excitons can become effectively localized, forming uniform arrays of quantum (single photon) emitters and unlocking new opportunities for quantum material design (Figure \ref{fig:B8}(a)). For an effective exciton localization, it is necessary that the Bohr radius ($ \approx $1 nm) is smaller than the moir\'e period (see Eq. \ref{eq:moireperiod}) \cite{Brem20c}. Through the strategic stacking of different atomic layers, it is possible to engineer novel quantum phenomena with unprecedented precision, leveraging various tuning parameters such as twist angle, interlayer spacing, strain, and external electric and magnetic fields. Many studies have observed multiple sharp lines in photoluminescence experiments with the energy separation between the peaks and the polarization of the emission to be dependent on the twist angle \cite{seyler2019signatures,tran2019evidence,forg2021moire}. Typical linewidths of these sharp lines are on the order of 100 $\mu$eV, similar to quantum emitters in WSe$_2$ monolayers \cite{koperski2015single,branny2017deterministic}. 

Moir\'e-trapped interlayer excitons exhibit strong circular polarization in photoluminescence experiments with values exceeding 70$\%$ at low temperatures, while the sign of the polarization is reversed between $H$ and $R$ stacking order (see Figure \ref{fig:B8}(b)). In contrast to quantum emitters in monolayers,  moir\'e-trapped interlayer excitons show negligible linear polarization emission. The similar circular polarization emission between the emitters shown in Figure \ref{fig:B8}(b) suggest a common interlayer atomic registry with a threefold rotational symmetry, $\hat{C}_{3}$. Experimental signatures of interlayer exciton localization can be obtained by laser power dependent photoluminescence experiments, where at very low powers (order of $ \mu $W) saturation of emission intensity is observed (Figure \ref{fig:B8}(c) \cite{baek2020highly}). Depending on the depth of the moir\'e potential and the emission rate of interlayer excitons, the power curve saturates because higher excitation leads to a delocalization of moir\'e excitonic states \cite{brem2023bosonic}. Remarkably, interlayer excitons can also exhibit dipolar interactions, leading to optical nonlinearities when a multitude is confined within a harmonic trapping potential \cite{li2020dipolar,kremser2020discrete}.  Another method to create periodic arrays of quantum emitters based on interlayer excitons involves transferring TMD heterostructures onto nanopatterned substrates containing sharp ($\approx$100 nm) nanopillars on their surface. This process introduces point-like strain perturbations, which locally modify the band gap and effectively trap excitonic species \cite{montblanch2021confinement,branny2017deterministic,palacios2017large,kern2016nanoscale,michaelis2022single}. 

However, until recently, there was no unambiguous demonstration of quantum light emission from moir\'e-confined interlayer excitons. Previous reports only offered signatures and indirect evidence of quantum emission (such as sharp linewidths and early power dependent saturation intensities), mainly due to weak emission intensities. Notably, second-order correlation function $ g^{(2)}(0) $ experiments with optimized collection efficiency in MoSe$_2$/WSe$_2$ heterostructures revealed photon antibunching and provided direct evidence of the quantum nature of moir\'e interlayer excitons (Figure \ref{fig:B8}(d)) \cite{baek2020highly}. The topic is still under intense investigation to understand the microscopic origin of the numerous quantum emitters and determine their association with the local atomic registry.
Signatures of moir\'e intralayer excitons (electrons and holes residing within the same layer) in heterostructures have also been reported in absorption experiments,  affected by the periodic moir\'e potential \cite{jin2019observation}. Intralayer excitons can also hybridize with interlayer excitons under the premise that the conduction or valence bands are delocalized over both layers, enhancing the impact of the moir\'e potential on the optical properties of the heterostructure \cite{alexeev2019resonantly}.

\subsubsection {Interlayer exciton lifetime}\label{sec:lifetime}

Spatially indirect excitons and their prolonged lifetimes have been a subject of active research since the 1980s following the discovery of double quantum well systems  \cite{charbonneau1988transformation,golub1990long}. There are similarities between indirect excitons in double quantum wells and those in TMD heterostructures, such as a static dipole moment and the reduced overlap of electron-hole wavefunctions, resulting in extended lifetimes. However, interlayer excitons in TMD heterostructures offer significant advantages that make them appealing for fundamental studies and potential optoelectronic applications. These advantages include stronger binding energies that surpass thermal dissociation, a small Bohr radius ($ \approx $ 1 nm), tunable optical selection rules and additional degrees of freedom to engineer lifetimes. Consequently, exotic phenomena such as excitonic condensation at elevated temperatures and superfluidity are anticipated with interlayer excitons in TMD heterostructures  \cite{fogler2014high}. 

\begin{figure}[t!]
  \centering
  \includegraphics[width=1\linewidth]{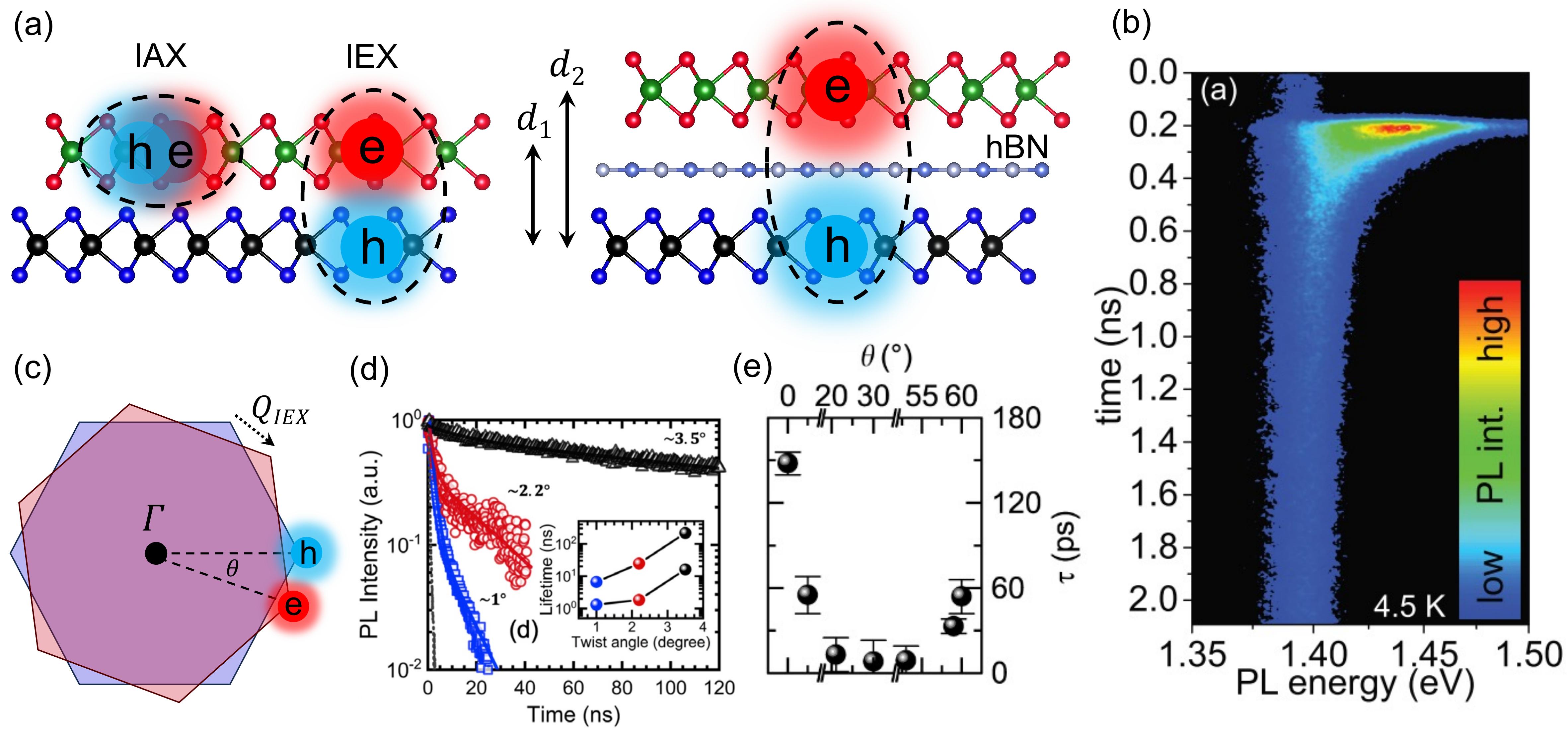} 
  \caption{(a) Illustration of a TMD heterostructure showing intralayer excitons (IAX), where the electron and hole wavefunctions strongly overlap.  For interlayer excitons (IEX), the overlap of the electron and hole wavefunction is reduced due to the spatial separation of the carriers. This can be further engineered by introducing additional hBN spacers between the constituent monolayers and further increase the distance between the electron and hole (right). Here, $ d_{1}<d_{2} $. (b) Time-resolved photoluminescence showing the long lifetimes (order of nanoseconds) of interlayer excitons, measured at T = 4.5 K (data taken from Ref  \cite{nagler2017interlayer}). This is much longer compared to IAX in monolayers, exhibiting picosecond lifetimes at the similar temperatures. (c) Schematic representation of the relative twist between the two Brillouin zones (red from top monolayer and blue from bottom monolayer) in the reciprocal space, introducing a finite momentum, $ Q_{IEX} $, in the center of mass momentum of IEXs. (d) Twist angle dependent IEX lifetimes showing longer decay time for larger twist angles. (e) Exciton decay time for different twist angles in twisted homobilayers due to varying interlayer hybridization. Figs. (b), (c,d) and (e) adapted respectively from Ref.  \cite{nagler2017interlayer}, \cite{choi2021twist} and \cite{merkl2020twist}.}
  \label{fig:B9}
\end{figure}

Now we examine in more detail the radiative lifetime of  interlayer excitons. The spatial separation between the electron and hole has important consequences in the decay rate of interlayer excitons but also in their valley depolarization rate compared to intralayer excitons in monolayers (Figure \ref{fig:B9}(a)). In previous sections we have seen the optical transition dipole moment dictating the optical selection rules in a semiconductor. This type of dipole moment characterizes the coupling between excitons and photons, given by the off-diagonal matrix element in an optical transition, $\braket{0|\hat{\mu}|X}$ (here, $\hat{\mu}$ is the dipole operator, $\ket{0}$ is the vacuum and $\ket{X}$ is the exciton) \cite{rivera2018interlayer}. The strength of the exciton-photon coupling is named \textit{oscillator strength} and is proportional to $|\braket{0|\hat{\mu}|X}|^{2}$. Calculations that consider the spatial separation between electrons and holes  \cite{yu2015anomalous,wu2018theory,ovesen2019interlayer,lu2019modulated}, estimate that the oscillator strength (linked to the transition dipole moment) for interlayer excitons is orders of magnitude smaller compared to intralayer excitons. Consequently, this results in much longer lifetimes for interlayer excitons, as confirmed experimentally by several groups (an early example is shown in Figure \ref{fig:B9}(b)  \cite{nagler2017interlayer}). For comparison, typical lifetimes for intralayer excitons at T = 4 K are on the order of one to a few picoseconds, whereas interlayer excitons extend into the nanosecond regime, even microseconds \cite{rivera2015observation,miller2017long,montblanch2021confinement}. Other reports claim that the intrinsic interlayer exciton lifetime is not on the order of nanoseconds but hundreds of picoseconds, opening the discussion for the role of the relaxation pathway before radiative recombination \cite{barre2022optical}. Besides the radiative lifetime, interlayer excitons exhibit long valley polarization lifetimes because their reduced oscillator strength is strongly linked to the exchange interaction, $ \Omega $, responsible for the valley depolarization \cite{glazov2014exciton}. Indeed, many recent reports demonstrate ultralong (tens of microseconds) valley lifetimes, rendering interlayer excitons as potential candidates for low-energy dissipation valleytronic devices. \cite{kim2017observation,jiang2018microsecond}. Note also that the decay time is affected by the presence of dark states \cite{Selig18}, with crucial impact on transport \cite{Beret22} (as we will see later), as well as by non-radiative mechanisms such as trapping in defects.

New degrees of freedom, such as the twist angle and the incorporation of ultrathin hBN spacers between the monolayers in van der Waals heterostructures, offer additional control over the lifetime of interlayer excitons.  In Figure \ref{fig:B9}(a), a schematic comparison between two heterostructures with different interlayer distances is shown. Introducing additional hBN layers between the two TMD monolayers, further reduces the spatial overlap between the electron-hole wavefunctions, offering the capability to engineer and further increase the lifetime of interlayer excitons \cite{zhou2020controlling}. Remarkably, the interlayer exciton lifetime drastically changes as a function of the twist angle (Figure\ref{fig:B9}(d)) \cite{choi2021twist}. Comparing samples between 1$ ^{o} $ and 3.5$ ^{o} $ twist angles, one order of magnitude longer lifetimes have been reported with increasing twist angle. This effect is attributed to an additional momentum mismatch ($ Q_{IEX} $, Figure\ref{fig:B9}(c)) that impacts the center of mass of the interlayer exciton, prolonging radiative recombination (Figure \ref{fig:B9}(d)). In the case of twisted homobilayers and hybridized excitons the lifetime is strongly impacted by changes in the hybridized wavefunctions in real space as a function of the twist angle \cite{merkl2020twist}(Figure \ref{fig:B9}(e)). We note that also optical cavities have been examined as tools to control the lifetime of interlayer excitons reaching tunability over two orders of magnitude due to the Purcell effect \cite{forg2019cavity}.
 
\subsubsection {Tuning of interlayer excitons by electric fields}\label{sec:Stark}

Besides the optical transition dipole moment, interlayer excitons also possess another type of dipole. It is named \textit{static electric dipole moment}, $\braket{X|\hat{\mu}|X}$, and represents the diagonal matrix element of the dipole operator $\hat{\mu}$ for an exciton state, \textit{X} \cite{rivera2018interlayer}. Aligned along the out-of-plane direction ($ \hat{z} $), this static dipole is proportional to the layer separation pointing from the negative charge to the positive one and its magnitude can be engineered by adding additional hBN spacers between the two monolayers \cite{zhou2020controlling} (see interlayer excitons with and without spacer in Figure \ref{fig:B9}(a)). Importantly, the static dipole moment can introduce energy shifts in the interlayer exciton state by applying vertical electric fields ($-E\cdot\braket{X|\hat{\mu}|X}$, Stark effect), a notable distinction compared to intralayer excitons that show negligible static dipole moments.  

\begin{figure}[t!]
  \centering
  \includegraphics[width=1\linewidth]{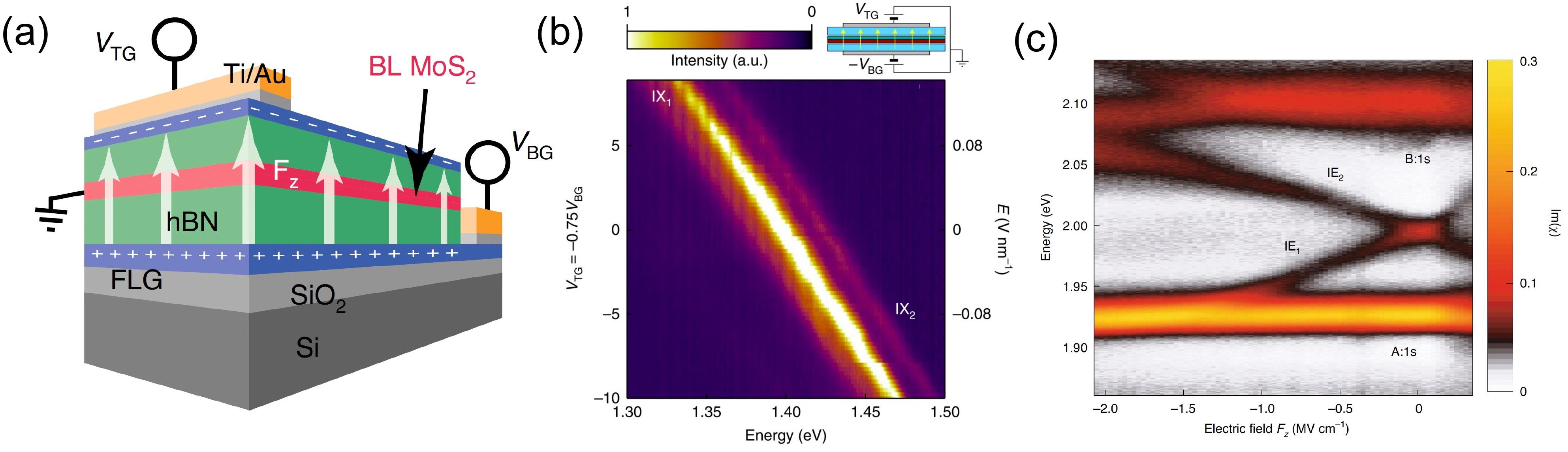} 
  \caption{(a) Illustration of a device designed to apply vertical electric fields in a van der Waals stack. Blue parts correspond to the few-layered graphene electrodes, green parts to dielectric hBN, while the TMD is presented with red color. (b) Map of photoluminescence emission showing a linear energy shift of the interlayer excitons as a function of the applied electric field. (c) Colour map of the absorption spectra as a function of the electric field $F_{Z}$ applied perpendicular to the stack, showing a clear Stark splitting of the two interlayer excitons. Figs. (a,c) and (b) adapted respectively from Ref.  \cite{leisgang2020giant} and  \cite{ciarrocchi2019polarization}.}
  \label{fig:B10}
\end{figure}

An illustration of a device to apply vertical electric fields using van der Waals materials is shown in Figure \ref{fig:B10}(a) \cite{leisgang2020giant}. Two graphene (or few layered graphene) flakes are used as a top and bottom electrodes (blue color) and the active material (here with red color, a homobilayer MoS$_2$ but it can be replaced by any heterostructure) is separated from the graphene electrodes by a dielectric material, typically hBN (green color). Applying a voltage between the top and bottom electrodes generates an electric field that shifts the energy levels of the interlayer exciton by $ \Delta E = - \mu_{z} \cdot F_{z} $, where $F_{z}$ is the applied electric field in the $ \hat{z} $ direction and $\mu_{z}$ is the static dipole moment.  The direction of the electric field can be conveniently controlled by switching the sign of the potential difference between the top and bottom electrode. Consequently, the static dipole moment can be experimentally measured by recording the interlayer exciton's energy shift as a function of the applied field. Moir\'e-trapped interlayer exciton-based quantum emitters in heterostructures, exhibit shifts on the order of 40 meV \cite{baek2020highly}, while broader singlet and triplet interlayer excitons show strong, linear energy shifts that exceed 100 meV \cite{ciarrocchi2019polarization} (see Figure \ref{fig:B10}(b)). The measured electric dipole moments of interlayer excitons typically fall within the range of 0.5 nm$\cdot q$, with $q$ representing the elementary charge, where 0.5 nm is roughly the interlayer distance of bilayers. The large tunability in the emission energy of interlayer exciton energy, combined with their extended lifetimes, has sparked considerable interest in developing excitonic transistors. These devices aim to control exciton transport through engineering of an electrostatic potential landscape \cite{unuchek2018room}.  

In contrast to heterobilayers, homobilayers accommodate two degenerate, hybridized interlayer exciton species with opposing electric dipole moments, along with strong oscillator strengths. Consequently, when subjected to $\hat{z}$-electric fields, absorption spectra reveal a distinct separation between the interlayer exciton species \cite{leisgang2020giant,peimyoo2021electrical} (compare Figures \ref{fig:B10}(b) with \ref{fig:B10}(c)). The magnitude of the Stark shift and the electric dipole moment values resemble those of interlayer excitons in heterostructures. Similar observations in the presence of electric fields have been reported for interlayer excitons that are both spatially and momentum indirect \cite{huang2022spatially,Tagarelli23}. The intriguing interaction and mixing between intralayer with interlayer exciton states has also been revealed by tuning their energies close to resonance using external electric fields \cite{sponfeldner2022capacitively} (for instance, see the avoided crossing between B-excitons and interlayer exciton, IE$_{2}$, in Figure \ref{fig:B10}(c)).
         
\subsubsection {Tuning of interlayer excitons by magnetic fields}

\begin{figure}[t!]
  \centering
  \includegraphics[width=1\linewidth]{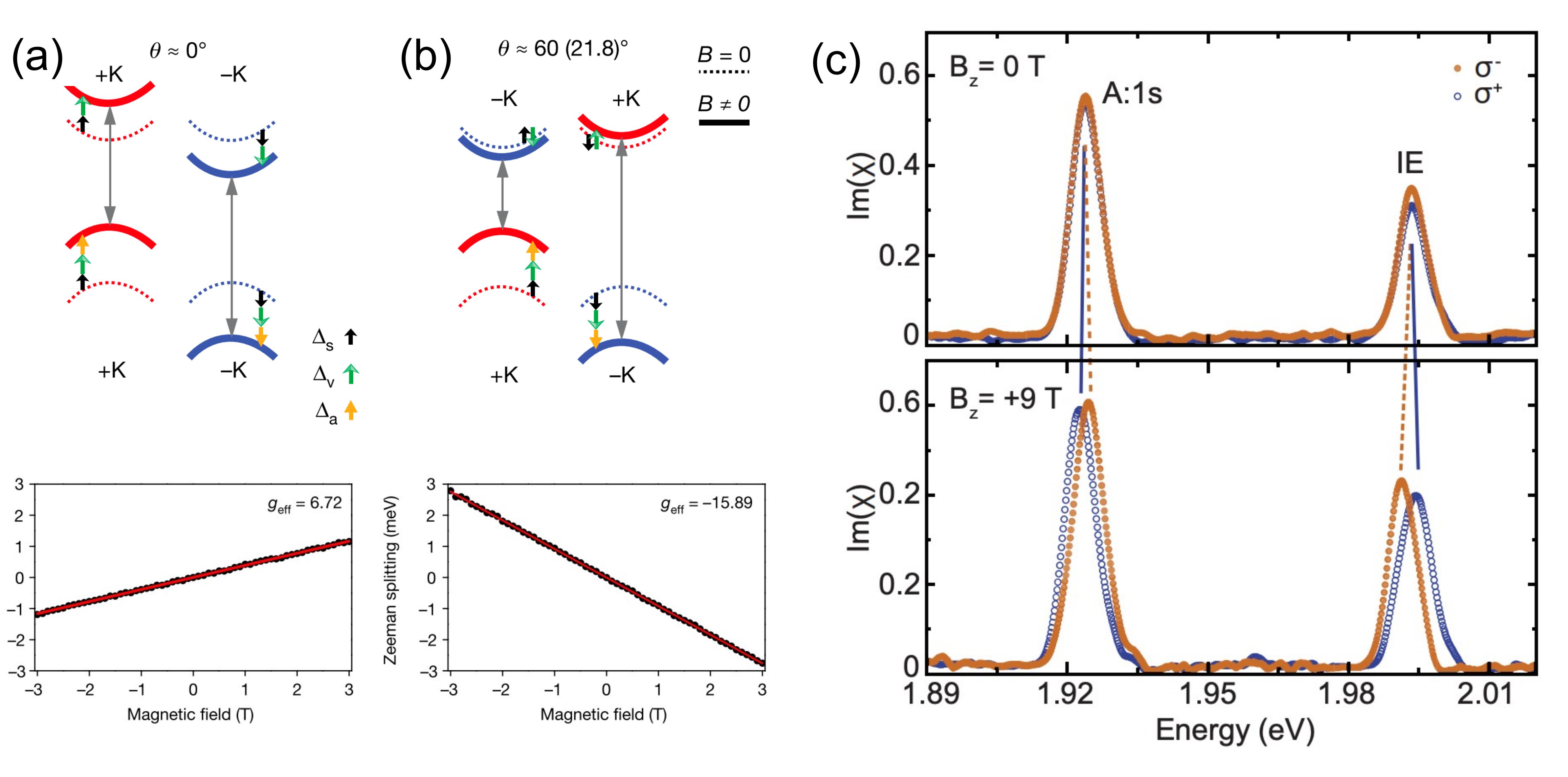} 
  \caption{Up: schematic showing the contributions to the interlayer exciton Zeeman shift from the electron spin ($ \Delta_{s} $, black), valley ($ \Delta_{v} $, green) and atomic orbital ($ \Delta_{a} $, orange) for heterobilayers with twist angles of (a) 0$^{o}$ and (b) and 60$^{o}$. Down: experimentally extracted interlayer exciton $g$-factors for twist angles of (a) 0$^{o}$ and (b) and 60$^{o}$. (c) Polarization-resolved absorption spectra of homobilayer MoS$_2$  at zero magnetic field (top)
and $B$ = 9 T (bottom). The orange and blue curves correspond to $\sigma^{-}$ and $\sigma^{+}$ polarization. Intra- and interlayer excitons show opposite Zeeman shifts. Figs. (a,b) and (c) adapted respectively from Ref.  \cite{seyler2019signatures} and  \cite{leisgang2020giant}.}
  \label{fig:B11}
\end{figure}

Magneto-optical experiments are routinely used to measure the effective interlayer exciton Land\'e $g$-factor. Similar to the electron $g$-factor, the effective Land\'e $g$-factor quantifies the response of an interlayer exciton to an external magnetic field. The effective Land\'e $g$-factor is determined through the measurement of the Zeeman energy splitting of an exciton, $ \Delta E = g \cdot \mu_{B} \cdot B$, where $\mu_{B}$  represents the electron’s Bohr magneton and $B$ denotes the applied external magnetic field that breaks the time inversion symmetry \cite{robert2021measurement}. $ \Delta E$ corresponds to the Zeeman splitting between right ($\sigma^{+}$) and left ($\sigma^{-}$) circularly-polarized light components. For instance, bright excitons in  TMD monolayers yield effective Land\'e $g$-factor values of $ g \approx -4 $. In general, semi-phenomenological models are employed to describe the exciton's $g$-factor. These models typically incorporate three main contributions: spin, valley, and orbital terms, which are added to determine the $g$-factor value \cite{koperski2018orbital} (see Figures\ref{fig:B11}(a,b)). Common optical methods employed for measuring the Land\'e $g$-factor of interlayer excitons in heterobilayers \cite{seyler2019signatures} and homobilayers \cite{slobodeniuk2019fine} include magneto-photoluminescence/absorption techniques where $\sigma^{+}$ and $\sigma^{-}$ components are measured in the presence of out-of-plane magnetic fields. 

It has been reported that the interlayer exciton $g$-factors differ between R-stacking ($\approx 7$ for triplet, $\approx -8$ for singlet \cite{ciarrocchi2019polarization,tran2020moire}) and H-stacking ($\approx 15$ for triplet, $\approx 10$ for singlet \cite{wang2019giant,tran2020moire}) depending on the local atomic registry where the interlayer exciton is localized and contributions from spin, valley, and orbital terms (see Figures \ref{fig:B11}(a,b)). It should be noted that strain can impact the local symmetry and create local potentials, thus affecting the value of the $g$-factor \cite{blundo2022strain}. Therefore, a uniform distribution of $g$-factors suggests that the structure is well retained over the exciton wavefunction \cite{seyler2019signatures}. Consequently, the interlayer exciton $g$-factor can assess the stacking order and effectively distinguish between singlet and triplet states. In homobilayers, a clear distinction between intra- and interlayer exciton transitions can be observed in magneto-optical experiments, where polarization-resolved differential reflectivity contrast reveals $g$-factors with opposite signs and amplitudes, approximately $-4$ for intralayer states and approximately $8$ for interlayer ones \cite{slobodeniuk2019fine} (see Figure \ref{fig:B11}(c) \cite{leisgang2020giant}). Notably, hybridized interlayer exciton states in homobilayers offer a strong manipulation of the exciton $g$-factor over a broad range, from $-4$ to $+14$, via mixing with intralayer states by simultaneously applying electric and magnetic fields \cite{feng2022highly}.


%

\section{Excitonic Transport}\label{sec:transport}

In previous sections we have discussed the remarkable excitonic landscape of TMD nanomaterials, which exhibit deeply-bound bright and dark excitons including spatially separated interlayer, hybrid and charge-transfer excitons in vertical and lateral TMD heterostructures. So far, we have mostly focused on exciton optics in these two-dimensional materials. Now we move our attention to exciton transport, which is particularly important for optoelectronics applications. 

\begin{figure}[t!]
    \centering
    \includegraphics[width=\linewidth]{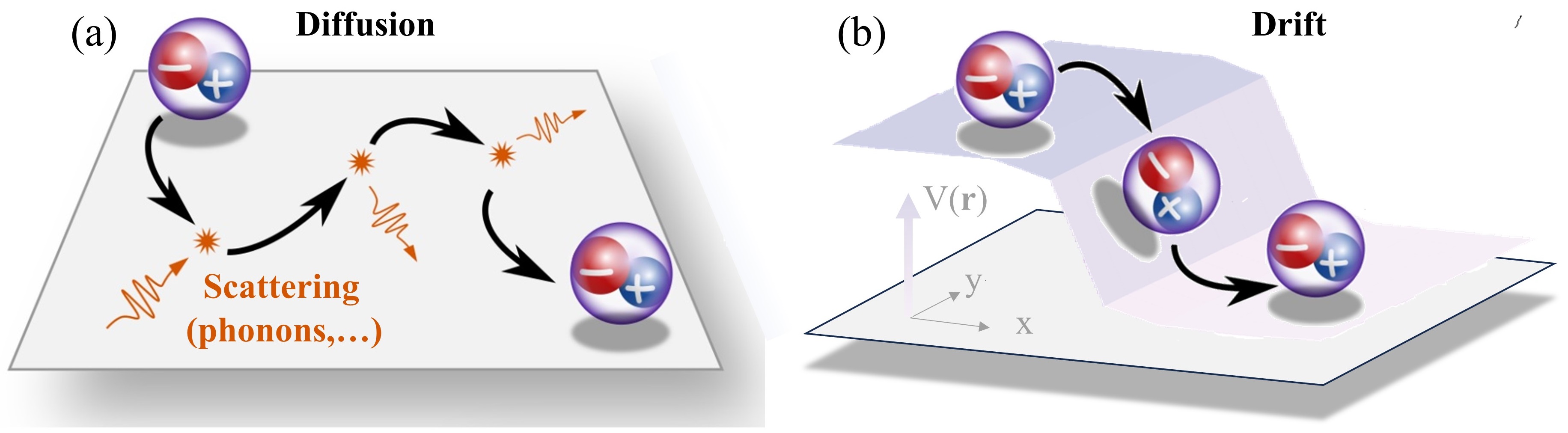}
    \caption{Sketch of  (a) exciton diffusion and (b) directional exciton drift. The diffusion is triggered by the excitonic center-of-mass momentum and partially counter-acted by scattering processes with phonons or other excitons. On average excitons move away from the excitation spot, resulting in a spatial broadening of the excited spatial profile. (b) Similarly to electronics, the directional drift of excitons can be obtained by in-plane potentials, whose origin has to go beyond regular electrostatic potential considering the neutral nature of excitons. Fig. (a) and portions of (b) adapted from Ref. \cite{Perea22b}}
    \label{tr1-sketch}
\end{figure}
All our electronic devices basically rely on electrons moving from one part to another. TMDs are characterized by excitons, which, luckily, also can move. The Coulomb interaction forces the bound electrons and holes to stay close to each other on a nanometer scale (corresponding to the exciton Bohr radius), but leaves free their center-of-mass position. 
This allows excitons to propagate through the crystal. The movement is driven by their center-of-mass momentum and slowed down by scattering mechanisms, cf. Fig. \ref{tr1-sketch}(a). Still on average excitons move away from the excitation spot where they were generated, resulting in an exciton diffusion which will be discussed in detail in Sec. \ref{sec:genDiffusion}. Here we will distinguish between conventional and transient diffusion, quantifying its speed via the diffusion coefficient and exploring which controlling mechanisms can speed up or slow down the diffusion (e.g. temperature, exciton density, strain, etc.).

On the other hand,  we can say that speed is nothing without control: Besides being fast, optoelectronics applications require excitons to drift in a given direction, cf. Fig. \ref{tr1-sketch}(b). In electronics, this can be easily obtained via electronic potentials.  However, excitons are charge neutral and hence can not be directly controlled with electric fields. In Sec. \ref{sec:Funneling}, we will review other efficient ways of controlling the excitonic drift. We will, in particular, discuss exciton funneling induced by inhomogeneous lattice strain profiles and explore additional opportunities offered by lateral TMD heterostructures via internal in-plane potentials. Despite the neutral nature of excitons we will show that even an electric control of exciton propagation is possible. We will microscopically investigate different mechanisms of exciton transport in TMD materials including strategies to tune and control the transport behavior.

\subsection{Excitonic Diffusion}\label{sec:genDiffusion}

A drop of wine spilled on a table starts expanding in the table cloth. Similarly, in two-dimensional semiconductors a confined optical excitation forms excitons, which subsequently spread in plane. 
As previously discussed, in TMD nanomaterials the excitonic landscape is particularly rich, including bright and dark excitons (cf. Fig. \ref{intro3}(c)), the corresponding charged excitons/trions \cite{Park21,Wagner23,Perea22} and even extending to interlayer, hybrid, moir\'e and charge-transfer excitons in heterostructures. cf. Fig. \ref{intro1}. The competition and interaction of such a multitude of species results in a remarkable excitonic transport, presenting multiple behaviors and controlling mechanisms, as we will explore in more details in the following.
\begin{figure}[t!]
    \centering
    \includegraphics[width=\linewidth]{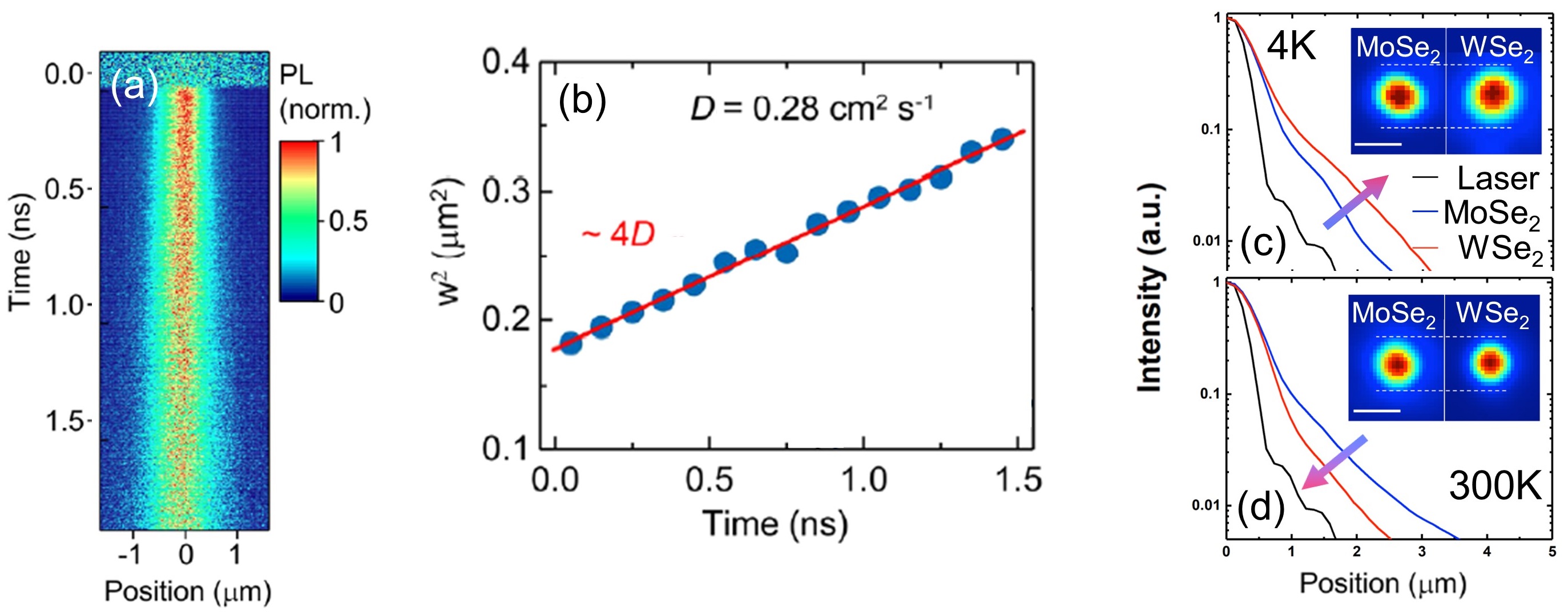}
    \caption{(a) Space- and time-resolved photoluminescence after confined pulsed excitation of a free-standing WS$_2$ monolayer at room temperature at room temperature and maximum exciton density of approximately 2$\times10^8$ cm$^{-2}$: Normalizing the intensity at each time step, it appears a spatial spreading with time  quantified by the profile squared width $w^2$ (b). This increases linearly with time and the slope is proportional to the diffusion coefficient $D$. (c,d) Space-resolved time-integrated PL on the MoSe$_2$ and the WSe$_2$ side of a hBN-encapsulated lateral MoSe$_2$-WSe$_2$ heterostructure at (c) 4 K and (d) 300 K. The different behaviour reflects the material-dependent variation of recombination time with temperature.  Figures (a,b) and (c,d) adapted from \cite{Kulig18} and \cite{Beret22}, respectively.}
    \label{tr2-conventional}
\end{figure}

How can transport be accessed experimentally? First, a spatially-inhomogeneous distribution is optically excited via a confined laser, most often via diffraction limited spots of several hundreds of nanometers \cite{Cadiz17,Kulig18,Cordovilla18,Sahoo18,Rosati21e,Harats20,Gelly22,Lee22,Choi23,Tagarelli23} but also via near-field techniques producing spots of tens of nanometers \cite{Beret22}. Second, the spatiotemporal dynamics  can be observed via multiple experimental techniques, including in particular photoluminescence (PL) \cite{Cadiz17,Kulig18,Unuchek18,Perea19,Cordovilla18,Cordovilla19,Harats20,Rosati21a,Rosati21a,Lee22,Beret22,Tagarelli23} and pump-probe techniques \cite{Yuan20,Choi23,Yuan23,Vazquez24}. These techniques track the excitonic spatial profile respectively by counting the number of photons emitted and by studying the absorption of a probe (typically resonant to the A-exciton peak, see Sec. \ref{sec:intro}) coming after an initial pulse.
Figure \ref{tr2-conventional}(a) shows a typical space- and time-resolved energy-integrated PL in WS$_2$ monolayers taken along one direction and normalized at each time step \cite{Kulig18}. 
The spatial profile  preserves its position while getting broader: This is a qualitative fingerprint of what we call diffusion. We now want to be more quantitative and  define the speed of diffusion, which will allow us to identify its different regimes and control strategies, as further discussed in Secs. \ref{sec:convDiff}, \ref{sec:transient} and \ref{sec:controlDiffusion}. 

\subsubsection{Conventional exciton Diffusion}\label{sec:convDiff}

Since the diffusion implies a spatial broadening, it is naturally quantified by the evolution with time $t$ of the profile area $w^2(t)$. For the more generic case, this quantity can be defined via the spatial variance of the excitonic spatial density $N(\textbf{R},t)$ as $w^2(t)=\left\langle \norm{\textbf{R} -\langle \textbf{R} \rangle}^2 \right\rangle$, where for a generic function  $f(\textbf{R})$ we introduced its average value  $\langle f(\textbf{R})\rangle=\int d\textbf{R} f(\textbf{R})N(\textbf{R},t)/n(t)$, with $n(t)=\int d\textbf{R} N(\textbf{R},t)$ being the total exciton population. For the typical case of quasi-Gaussian profiles the width $w$ can be directly extrapolated from the full-width half maximum FWHM via $w=$FWHM$/(2\sqrt{\log{2}})$ and it appears in the exponent of the Gaussian via $N(\textbf{R})\propto \text{Exp}[-\norm{\textbf{R}-\textbf{R}_0}^2/w^2]$, with $\textbf{R}_0\equiv \langle \textbf{R} \rangle$ being the position where $N$ is maximum.  

In the case of a free-standing WS$_2$ monolayer at room temperature and weak excitation this area increases linearly in time, cf. Fig. \ref{tr2-conventional}(b). This is a common behaviour \cite{Yuan17,Kumar14} and is called conventional diffusion, which takes place typically in the low-density regimes once excitons have reached a local quasi-equilibrium distribution in energy. 
The crucial figure of merit for the diffusion is the slope $4D$ of the $w^2(t)$ curve, with $D$ being the diffusion coefficient in cm$^2$/s. This quantity provides the speed of the diffusion, and its investigation is crucial to understand the efficiency of  exciton transport.

Conventional diffusion is driven by the gradient of the exciton density $N(\textbf{R},t)$, as excitons tend to occupy the whole space uniformly -- the same applies to the spilled wine. This can be described by  Fick's law
\begin{equation}\label{Fick}
    d_tN(\textbf{R},t)=D\Delta N(\textbf{R},t) \quad ,
\end{equation}
where the Laplacian $\Delta N(\textbf{R},t)$ indicates the important role of the density profile, which drives the diffusion proportionally to the diffusion coefficient $D$. For an initial Gaussian-type density $N(\textbf{R},t=0)=N_0 \exp{[-|\textbf{R}|^2/(2w_0^2)]}$, the solution of Eq. (\ref{Fick}) is a Gaussian profile with the area $w^2(t)=w_0^2+4Dt$, hence increasing linearly with time as shown in Fig. \ref{tr2-conventional}(b). The value of $D$ can be measured a posteriori from the observed profile, while a priori it can be microscopically predicted with the Wigner transport equation \cite{Hess96,Rosati20}. While allowing also the description of transient non-equilibrium effects beyond the conventional diffusion (cf. Sec. \ref{sec:transient}),  the Wigner equation approaches the Fick's law (\ref{Fick}) in the limit of local equilibrium, hence with Wigner distribution proportional to the thermalized Boltzmann distributions  \cite{Hess96,Rosati20}. 
In this way the diffusion constant for 2D excitons can be predicted as $D=\tau k_B T/M$, with $T$, $M$ and $\tau$ being temperature, mass and scattering time of excitons (approximated to be independent of the excitonic momentum or valley). 
This simple equation tells us that the diffusion coefficient is slower for more effective scattering, as the latter counteracts the exciton propagation, cf. Fig. \ref{tr1-sketch}(a). For a fixed $\tau$ the diffusion coefficient becomes higher for larger temperature or smaller masses, because in both cases states with higher group velocities are present. 

The spatial broadening, in fact, results from the co-existence of excitonic components having different group velocities, hence propagating with different speed. A counter-example is provided by electron transport in metallic carbon nanotubes, where only two group velocities are present (differing only by their direction along the nanotube): This results in a dispersionless propagation of two copies of the original peaks when the so-called backward scattering is switched off, in particular by excitations with energy smaller than half the one of the associated phonons \cite{Rosati15,Rosati15b}. In contrast the conventional electronic diffusion is observed in graphene, because here the whole continuum of directions is possible. Nevertheless huge diffusion coefficients of 200-400 cm$^2$/s are present, also thanks to the high (modulus of) the group velocity \cite{Jago19}. Interestingly, in TMDs linear bands with huge group velocities have been predicted via exchange potential for excitons \cite{Yu14,Wu17}, leading to interesting transport phenomena, in particular under uniaxial strain \cite{Thompson22}. In a nutshell, excitons in TMD nanomaterials typically show conventional diffusion, whose speed can be at first approximation predicted from lattice characteristics. Nevertheless, this simple picture can be extended and controlled in multiple ways, as we will discuss in the following.

Note that besides the diffusion coefficient, a second crucial figure of merit is the diffusion length $l=\sqrt{D\tau_{\text{dec}}}$, which provides the average distance travelled by excitons before decaying with the time constant $\tau_{\text{dec}}$. This length can be extrapolated also from time-integrated measurements, which are dominated by the conventional diffusion when the decay times are longer than the duration of transient diffusion, cf. Sec. \ref{sec:transient}. 
The decay processes can be non-radiative (for example via trapping in defects) or radiative, the latter being induced by bright excitons, while dark excitons act as an optically-inactive reservoir.
Since Mo- and W-based monolayers have a different dark exciton occupations, $\tau_{\text{dec}}$ is strongly material-dependent, including an opposite behaviour with increasing temperature in the two classes of materials \cite{Selig18}. This results in diffused profiles larger in WSe$_2$ compared to MoSe$_2$ at cryogenic temperatures, Fig. \ref{tr2-conventional}(c), while the opposite takes place at 300 K, Fig. \ref{tr2-conventional}(d). Here the measurements are taken at the two sides of a hBN-encapsulated lateral TMD heterostructures, where the two monolayers are integrated in the same plane. Compared to the one of excitons, a reduced oscillation strength leads to longer diffusion length  for trions \cite{Cadiz17}, interlayer excitons \cite{Unuchek18} and electrons, which can decay only via recombining with free holes (e.g. from positive doping) or via non-radiative trapping in defects, allowing the propagation of valley coherence for more than 10 $\mu$m \cite{Ren22}.

\subsubsection{Transient low-density exciton diffusion}\label{sec:transient}

\begin{figure}
    \centering
    \includegraphics[width=\linewidth]{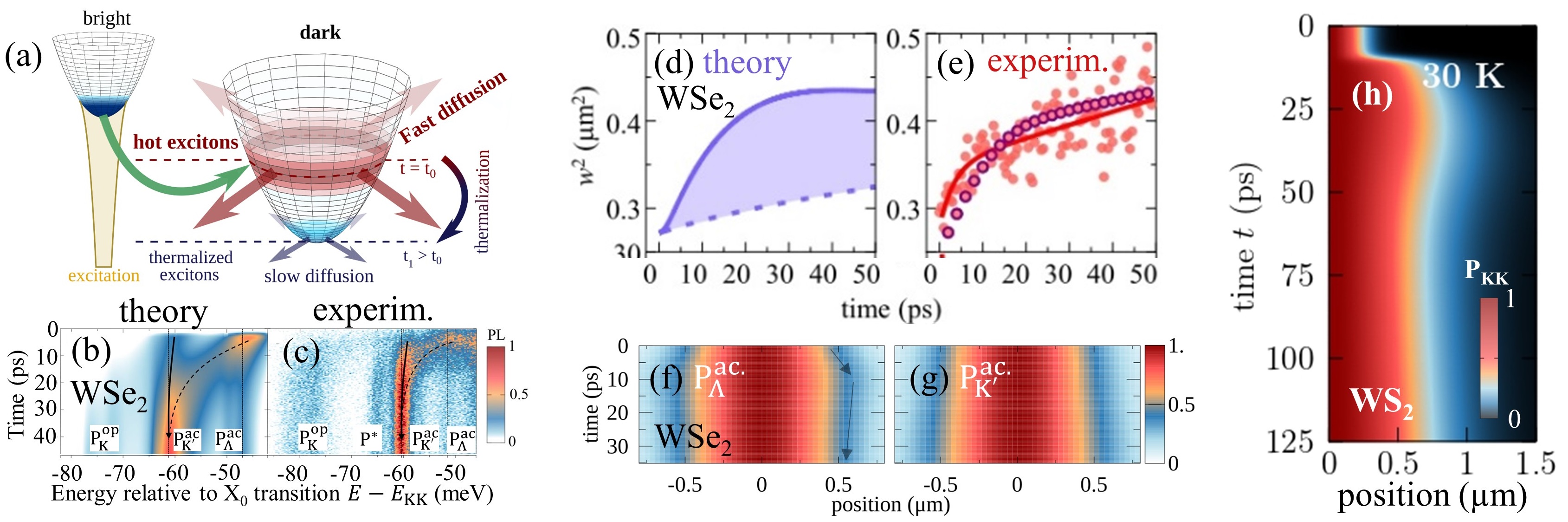}
    \caption{(a) Sketch of the direct excitation of KK excitons and the resulting formation of hot dark excitons via emission of phonons (green arrow). Before losing their excess energy dark excitons propagate fast due to their large group velocities. (b,c) Time-and energy-resolved cryogenic phonon-sidebands PL as predicted (b) and measured (c) in hBN-encapsulated WSe$_2$. The transient red-shift towards P$^{ac.}_{\text{KK}^\prime}$ (curved arrow) indicates the thermalization of hot KK$^\prime$ excitons. (d-g) Diffusion of (d,e) energy-integrated  and (f,g) energy-resolved  cryogenic phonon sidebands in WSe$_2$. A transient fast spatial broadening is observed with good theory-experiment agreement (d,e), while a small spatial shrinking is predicted at P$^{ac.}_{\Lambda}$ (f) or at the bright exciton peak (h) in hBN-encapsulated WS$_2$. Here an abrupt spatial broadening appears delayed by the time required to absorb a phonon and scatter from dark to bright excitons ($\approx$10 ps). Figs. (a,d-g), (b,c) and (h) adapted respectively from Refs. \cite{Rosati21c,Rosati20b,Rosati20}.}
    \label{tr3-transient}
\end{figure}

 So far we have only discussed the conventional diffusion and claimed it appears after the exciton thermalization is over. In order to 
 prove such a claim as well as to explore different regimes we now explore non-equilibrium processes where excitons are not yet thermalized, hence where the the energy distribution differs from the equilibrium (Boltzmann) one. The easiest way to do this is by decreasing the temperature, which crucially impacts the duration of the thermalization by altering the efficiency of exciton-phonon scattering \cite{Selig18,Brem18,Rosati20}.
Upon excitation, only the bright coherent excitons are directly excited (yellow in Fig. \ref{tr3-transient}(a)). In a second phase, a coherent-to-incoherent transfer occupies the states out of the light cone, in particular the momentum-dark valleys (cf. blue cloud in Fig. \ref{intro3}(c)) \cite{Selig18,Brem18}. This mechanisms is driven by phonons, resulting in a delayed occupation of the dark valleys observed, for example, in time-resolved angle-resolved photoemission spectroscopy (ARPES) \cite{Madeo20,Wallauer21,Schmitt22} or with infrared spectroscopies \cite{Merkl19}. The emitted phonons  have energies smaller than the energy difference between bright and dark excitons (green arrow in Fig. \ref{tr3-transient}(a)). As a consequence, hot dark excitons are initially formed. 
With respect to their valley energy minimum these hot excitons have an excess energy, which will be subsequently lost by interaction with phonons.
At room temperature, this energy thermalization takes place on a timescale of tens/few hundreds of femtoseconds \cite{Selig18,Rosati20}. In contrast, the timescale extends to 10 ps in W-based materials at cryogenic temperatures,
reflecting the drastically smaller efficiency of scattering processes with intravalley acoustic phonons, which decreases linearly with temperature \cite{Cadiz17b,Selig16} (cf. the linewidth in Fig. \ref{tr4-Temperature}(b)). 
Recently this loss of excess energy and thermalization have been directly observed via  time-resolved phonon-sidebands in PL spectra \cite{Rosati20b}, which appear below the A exciton resonance in W-based monolayers at cryogenic temperatures. Their origin is phonon-assisted scattering from dark excitons into a virtual state in the light cone followed by their radiative recombination \cite{Brem20}. As a consequence, these sidebands provide direct information about the dark states, as shown in Fig. \ref{tr3-transient}(b,c) after resonant excitation of hBN-encapsulated WSe$_2$ monolayer \cite{Rosati20b}, while the sidebands are absent in Mo-based monolayers due to the negligible role and occupation of dark excitons \cite{Brem20,Rosati20b}.  
A peak red-shifting by more than 10 meV toward the final energy P$^{ac.}_{\text{KK}^\prime}$ is observed- in an excellent agreement between theoretical predictions and experimental measurements, respectively Fig. \ref{tr3-transient}(b) and (c). 
 This peak is induced by KK$^\prime$ excitons and its red-shift directly tracks their loss of excess energy during the thermalization process.
 In addition, a transient higher-energy peak P$^{ac.}_{\text{K}\Lambda}$ is also predicted and observed. 
 While other phonon-sidebands or charged biexcitons at higher densities could also emit in similar energies \cite{He20,Ye18}, here P$^{ac.}_{\text{K}\Lambda}$ is predicted from K$\Lambda$ excitons.
Interestingly it would be absent at equilibrium, because K$\Lambda$ excitons have negligible occupation compared to the energetically-lowest  KK$^\prime$ states. Nevertheless this feature is visible in a transient phase thanks to a bottleneck retarding the thermalization from K$\Lambda$ towards  KK$^\prime$ states \cite{Rosati20b}.

Hot excitons have a larger group velocity, hence they propagate faster (red and blue arrows in Fig. \ref{tr3-transient}(a)). Their transient occupation results in a transient speed-up of the diffusion. This can be observed in Fig. \ref{tr3-transient}(d-e), where we investigate again hBN-encapsulated WSe$_2$ at cryogenic temperature and resonant excitation as in Fig. \ref{tr3-transient}(b-c), but now with confined excitation and space-resolving the PL. The spatial profile of the energy-integrated phonon sidebands shows an increase of $w^2$ differing importantly from the linear increase of the conventional diffusion, cf. Fig. \ref{tr2-conventional}(b). In contrast, the gedanken experiment of a thermalized exciton distribution would recover a slower and conventional diffusion, as shown by the dashed purple line in Fig. \ref{tr3-transient}(d), showing a weak and linear increase of the PL area. This difference between solid and dashed line  indicates qualitatively that the transient diffusion is much faster. To address this quantitatively,
we introduce a time-dependent effective diffusion coefficient $D(t)$, also called diffusivity. This is obtained from the derivative of $w^2(t)$ as $D(t)=\frac{1}{4}\frac{d w^2(t)}{dt}$, clearly recovering the regular diffusion coefficient $D$ for the conventional diffusion (where $w^2(t)=4Dt$). In the first 10 ps a transient diffusion coefficient of up to 50 cm$^2$/s has been measured, a value 10 times larger than the regular one displayed after thermalization \cite{Rosati21c}.  This speed-up by one order of magnitude is of great potential importance for technological applications. 

An even more involved exciton transport is predicted when further including the energy resolution, as shown in Fig. \ref{tr3-transient}(f,g) again for hBN-encapsulated WSe$_2$ monolayer \cite{Rosati21c}. 
The spatial profile of the signal at the higher-energy peak P$^{\text{ac}}_{\Lambda}$
shows a surprising behaviour, Fig. \ref{tr3-transient}(f): After ten picoseconds it shrinks, rather than expanding -- this unfortunately does not apply to spilled wine.
The just-introduced time-dependent effective diffusion coefficient becomes negative, $D(t)<0$, hence providing the so-called negative diffusion \cite{Rosati20}.
This remarkable shrinking of the profile typically stems from the competition between multiple species with different transient mobility \cite{Rosati20,Rosati21c,Beret23,Wietek24}.
Here it is due to the competition between hot KK$^\prime$- and  cold K$\Lambda$-excitons, the former being more mobile than the latter thanks to their high excess energy. In the first 10 ps these states have similar energies, hence emitting both around  P$^{\text{ac}}_{\Lambda}$, cf. Figs. \ref{tr3-transient}(b-c): In this transient phase the spatial profile of P$^{\text{ac}}_{\Lambda}$ is particularly broad thanks to the signal emitted by  KK$^\prime$ states, which have diffused more efficiently than the K$\Lambda$ ones in view of their excess energy. Similarly to Fig. \ref{tr3-transient}(b-c), also with localized excitation after 10 ps the KK$^\prime$ excitons stop emitting at P$^{\text{ac}}_{\Lambda}$, resulting in its apparent negative diffusion because the signal becomes dominated by the cold K$\Lambda$ states, which are spatially narrower than hot KK$^\prime$ excitons because less mobile.
Such a negative diffusion is hence apparent, as it does not directly involve a motion toward the excitation spot (as predicted in ZnSe \cite{Zhao03}).

Finally, Fig. \ref{tr3-transient}(h)  shows the predicted spatiotemporal dynamics at the A exciton in WS$_2$ monolayers at cryogenic temperatures. While its experimental detection is made difficult by the weaker occupation of the bright exciton states in W-based materials (due to the competition of dark states), it shows a peculiar behaviour, with an abrupt spatial broadening at about 10 ps followed by a negative diffusion, which is the spatial shrinking lasting up to 50 ps. 
The spatial broadening is induced by the fast propagation of hot dark excitons, which scatter back into KK valley via absorption of intervalley phonons. At cryogenic temperatures this scattering is a less efficient mechanism, hence a delay of 10 ps is required for its activation. The following spatial shrinking stems from the thermalization of KK excitons with all dark states including both hot and cold states. Here again as in Fig. \ref{tr3-transient}(f) the apparent negative diffusion stems from the competition between different exciton species. A qualitatively similar spatial shrinking has been observed in different material classes and under different conditions \cite{Ziegler20,Beret23,Wietek24}. While addressing the physical mechanisms behind each of them goes beyond the scope of this chapter, we stress that often the negative diffusion results from the competition between slower- and faster-diffusing exciton species.

The results discussed so far  refer to timescales longer than the  time of each scattering event (of about 1 ps for typical cryogenic scattering rates of the order of 1 meV, while being much faster at high temperatures). At shorter timescales, each exciton propagates according to its center-of-mass momentum. The resulting momentum integrated profile shows a squared width expanding quadratically with time \cite{Knorr98,Rosati14b}, the typical signature of ballistic transport \cite{Knorr98}. 
Interestingly, a long-lasting ballistic-like quadratic increase of $w^2$ with time has been reported as triggered by macroscopic inhomogeneities \cite{Su22}, such as dielectric or strain-induced bubbles \cite{Carmesin19,Blundo21}, while nanobubbles provide an involved transport and dynamics \cite{Rosati21d,Rosati18} via the locality of the carrier capture \cite{Glanemann05,Rosati17}. The ballistic regime shows a fast spreading, as excitons are free to  propagate without being slowed down by scattering with phonons. An even faster spatial broadening on a 100 fs timescale has been predicted for electrons  \cite{Rosati14b, Rosati13b}. The key for this are excited profiles reaching the quantum limit, in the sense of a product of the uncertainty in momentum and real space close to the Planck constant $\hbar$, as obtainable via near-field techniques \cite{Beret22,Lamsaadi2023}. Interestingly, this fast spatial broadening is promoted by  scattering with phonons, rather than being slowed-down by it as for the conventional diffusion. This happens because quantum effects make this scattering spatially non-local: Besides leading to a change in momentum, in this quantum limit the scattering can lead to a change in position \cite{Rosati14b}.
This is contrary to the case of regular diffusion, where the scattering is local, hence preserving the spatial position as described by the Boltzmann collision term withing the Wigner transport equation \cite{Hess96}. Nevertheless when a broader laser excitation or the evolution brings the product of the uncertainties well beyond $\hbar$, one enters the semiclassical limit, formally written as $\hbar \to 0$: Here it was shown that the non-local scattering superoperator becomes the local Boltzmann one \cite{Rosati14b}. 

In a nutshell, the transient exciton transport occurring initially after the optical excitation is faster than the conventional diffusion thanks to the large excess energy and the interplay between bright and dark excitons, in particular in tungsten-based monolayers. This speed-up of diffusion comes however with the drawback of its short duration. Luckily, TMD-based nanomaterials allow for other tuning knobs to speed-up the diffusion even at longer timescales, as it will be discussed in the following section.

\subsubsection{Controlling exciton diffusion}\label{sec:controlDiffusion}

Technological applications constantly strive for higher speeds
. 
A fast transport is crucial for optoelectronic devices. To this purpose we need to control the many-particle mechanisms behind the exciton propagation to be able to reach larger diffusion coefficients or even to go beyond the conventional diffusion, similarly to Sec. \ref{sec:transient} but potentially extending the temporal range. 
Here we present several experimentally accessible knobs to control exciton transport.

\paragraph{Temperature control}
\begin{figure}[t!]
    \centering
    \includegraphics[width=\linewidth]{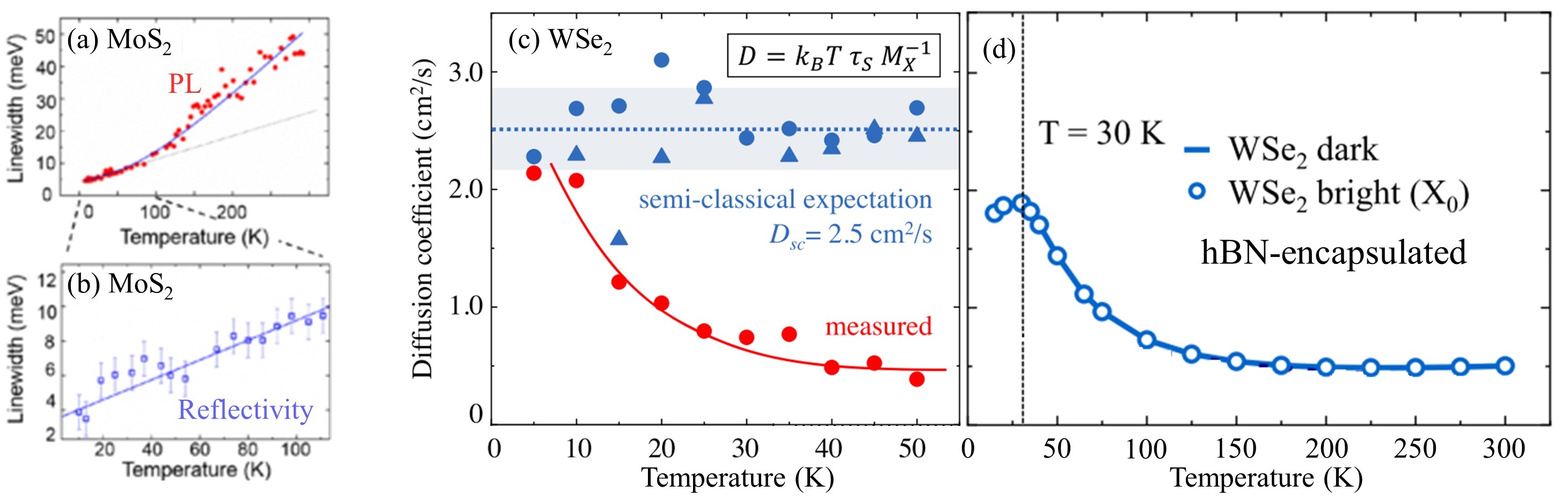}
    \caption{(a,b) Temperature dependent linewidth in hBN-encapsulated MoS$_2$ extracted from  (a) PL and (b) reflectance measurements, showing a linear increase induced by scattering with intravalley acoustic phonons before the onset at higher temperatures of scattering via absorption of optical phonons with finite energy. (c) Temperature-dependent diffusion coefficient measured in hBN-encapsulated WSe$_2$ (red-line), where due to quantum effects the diffusion slows down for increasing temperatures, contrary to the semiclassical expectation (blue line). (d) Predicted temperature-dependent diffusion in hBN-encapsulated WSe$_2$ and MoSe$_2$, showing a decrease at moderate temperature $T\gtrsim$ 50 K induced by thermal activation of intervalley scattering via phonon absorption. Figs. (a,b) and (c,d) adapted respectively from Ref. \cite{Cadiz17b} and \cite{Wagner21}.}
    \label{tr4-Temperature}
\end{figure}

Temperature modifies the exciton transport in many different ways. To see this we go back again to the conventional diffusion, which can be discussed by the time-independent diffusion coefficient  $D=\tau k_B T/M$, cf. Eq. (\ref{Fick}) and Sec. \ref{sec:convDiff}. Besides its explicit linear dependence on the temperature $T$, also the scattering time $\tau$ has a crucial dependence on temperature. This can be explored by investigating the exciton linewidth in a MoS$_2$ monolayer (Figs. \ref{tr4-Temperature}(a-b)), which is determined by the temperature-dependent phonon-mediated scattering rate of  bright excitons plus temperature-independent radiative recombination (and eventually residual inhomogeneous broadening). 
At small temperatures the linewidth linear increases with $T$ due to the scattering with intravalley acoustic phonons, whose efficiency increases linearly with temperature in all valleys, Figs. \ref{tr4-Temperature}(a-b). 
This implies a linear decrease of $\tau$ with temperature, suggesting that the product $\tau T$ and hence the expected diffusion coefficient $D$ are temperature-independent, cf. the blue line in Fig. \ref{tr4-Temperature}(c). Nevertheless, the direct experimental measurement of $D$ behaves very differently, showing a clear decrease for increasing temperatures, where  $D$ goes from about 2 cm$^2$/s at 10 K to about 0.5 cm$^2$/s at 50 K (cf. the red line in  Fig. \ref{tr4-Temperature}(c)). This decrease is induced by non-classical effects, namely a quantum interference inducing an effective exciton localization \cite{Glazov20,Wagner21}. Note that besides the decreasing/constant behaviour of $D$ with increasing temperature (cf. respectively  red   and blue in Fig. \ref{tr4-Temperature}(c)), an increase of $D$ with $T$ is expected in  presence of a co-existence of localized and extended states: Here temperature can promote the escape from trapped states \cite{Reiter07}, in qualitative agreement with the observation in twisted WS$_2$-WSe$_2$  \cite{Yuan20} and WS$_2$-hBN-WSe$_2$ heterostuctures \cite{Liu24}. Here the twist angle and the resulting moire potential can induce an excitonic localization \cite{Merkl20}, cf. Figs. \ref{intro1}(c) and \ref{fig:B3}. Such a localization crucially depends on the twist angle \cite{Merkl20}, cf. Eq. \ref{eq:moireperiod}, resulting in an angle-dependent exciton transport, cf. Ref. \cite{Knorr22}.

Further increasing the temperature toward several tens of Kelvin a second drop of the diffusion coefficient can appear, cf. Fig. \ref{tr4-Temperature}(d). This is induced by the opening of intervalley scattering via its thermal activation, as these mechanisms require the absorption of phonons of given energies of 12-15 meV \cite{Wagner21,Jin14}. This drop is particularly evident in WSe$_2$ (solid line), as here the temperature opens the  efficient scattering channel between KK$^\prime$ and K$\Lambda$ excitons. Note that the exact temperature at which this drops takes place is affected by small variations of the energy separation between these two valleys \cite{Wagner21}, in turn induced for example by disorder or strain \cite{Schmidt16,Saroj22}. A more moderate decrease is observed in MoSe$_2$ due to the activation of the less efficient scattering from KK to KK$^\prime$ excitons (dashed lines). 

\begin{figure}[t!]
    \centering
    \includegraphics[width=0.85\linewidth]{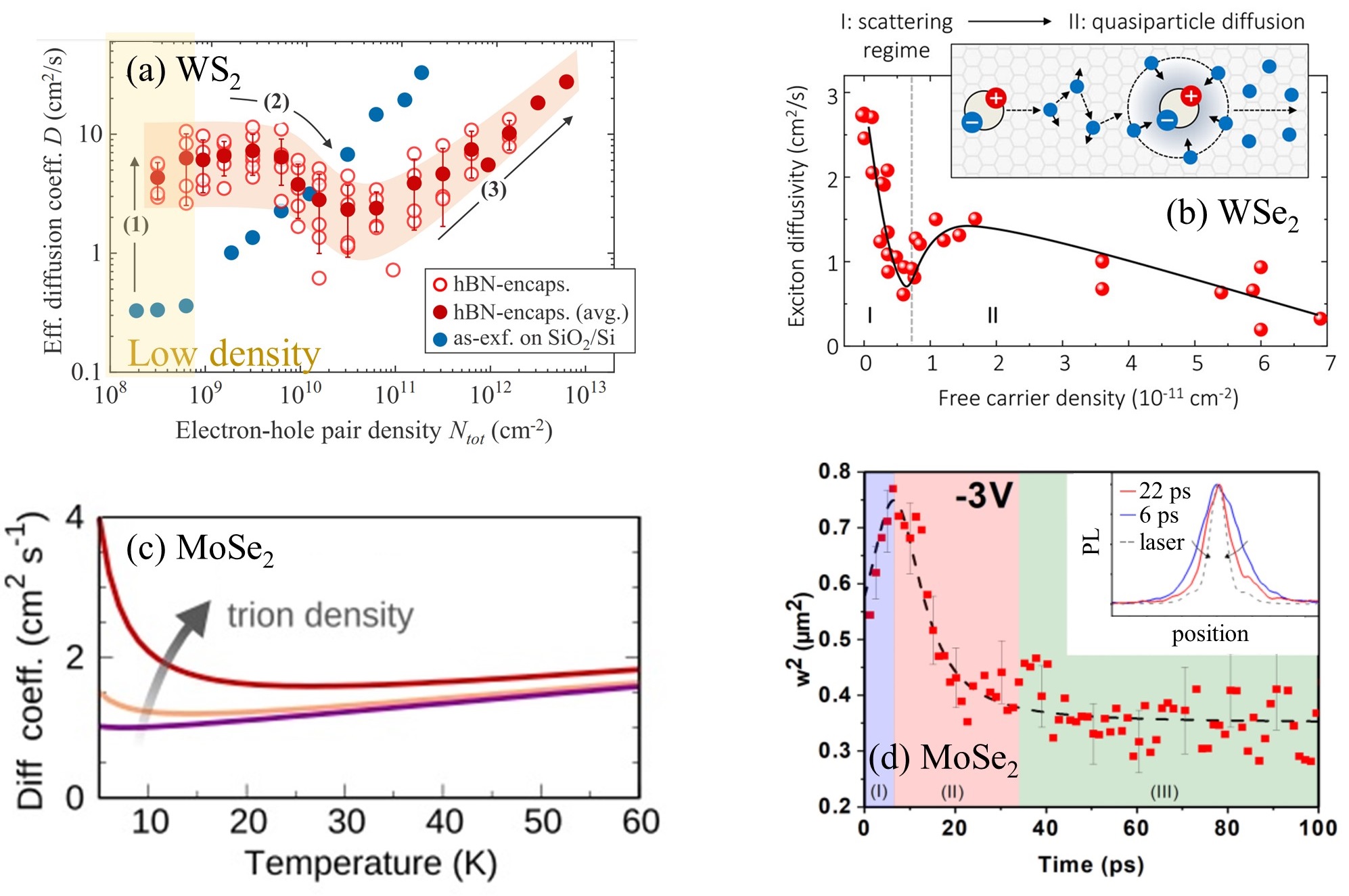}
    \caption{(a) Density-dependent room-temperature diffusion coefficient in WS$_2$ monolayer: In the low-density regime (yellow) a 10-fold speed-up of the diffusion is observed under hBN-encapsulation (red) via contribution of the continuum. (b) Doping-dependent diffusion coefficient in hBN-encapsulated WSe$_2$ at 5 K, showing a non-monotonic behaviour induced by the transition from exciton- to trion-dominated diffusion (sketch). (c) Temperature-dependent trionic diffusion in hBN-encapsulated MoSe$_2$, showing at low temperatures a high-density speed-up due to occupation of higher-energy states in the trionic Fermi-Dirac distribution. (d) Time-resolved squared width of the trionic profile in hBN-encapsulated MoSe$_2$ at 4K, with a negative diffusion (red area) indicating the competition between different species. Fig. (a), (b), (c)  and (d) adapted respectively from Ref. \cite{Zipfel20}, \cite{Wagner23}, \cite{Perea22} and \cite{Beret23}.}
    \label{tr7-charges}
\end{figure}

Finally, we emphasize that in some sense the temperature even affects the excitonic mass $M$ having direct impact on the diffusion coefficient. TMD materials show a remarkable excitonic landscape, with multiple excitonic valleys having different but yet comparable energies. While the mass of each valley is independent of temperature, a variation of thermal energy can vary the relative weight of different states. In thermalized WS$_2$ on typical substrate, for example, the most occupied excitonic valley is K$\Lambda$ at room temperature while KK$^\prime$ at  cryogenic temperatures, reflecting the competition between lower energy of KK$^\prime$ and the higher 3-fold degeneracy of K$\Lambda$ excitons. 
Similarly, high temperatures can allow a higher amount of unbound electron-hole pairs in comparison to the bound excitons, as it will be discussed in Fig. \ref{tr7-charges}(a). 
This already indicates how transport is affected by the competition of different species. This can be further triggered by high laser powers, which can promote the occupation of states with very high energy \cite{Kulig18,Perea19,Choi23} or even induce a Mott transition, where the excitons dissociate  into electron-hole plasma \cite{Chernikov15,Steinhoff17}. In next section we discuss how strong excitations affect the transport.

\paragraph{Density control}\label{sec:nonlinear}

The power of the laser dictates which excitonic densities are excited. Increasing the density, non-linear transport phenomena become important. These are induced by Coulomb interaction effects scaling quadratically with  the exciton density. These effects are negligible at low pump fluence (cf. Eq. (\ref{Fick})) but not under high excitation power. Different nonlinear transport mechanisms have been observed in TMD monolayers as well as in their vertical and lateral heterostructures, as we discuss in the following.
\begin{figure}[t!]
    \centering
    \includegraphics[width=\linewidth]{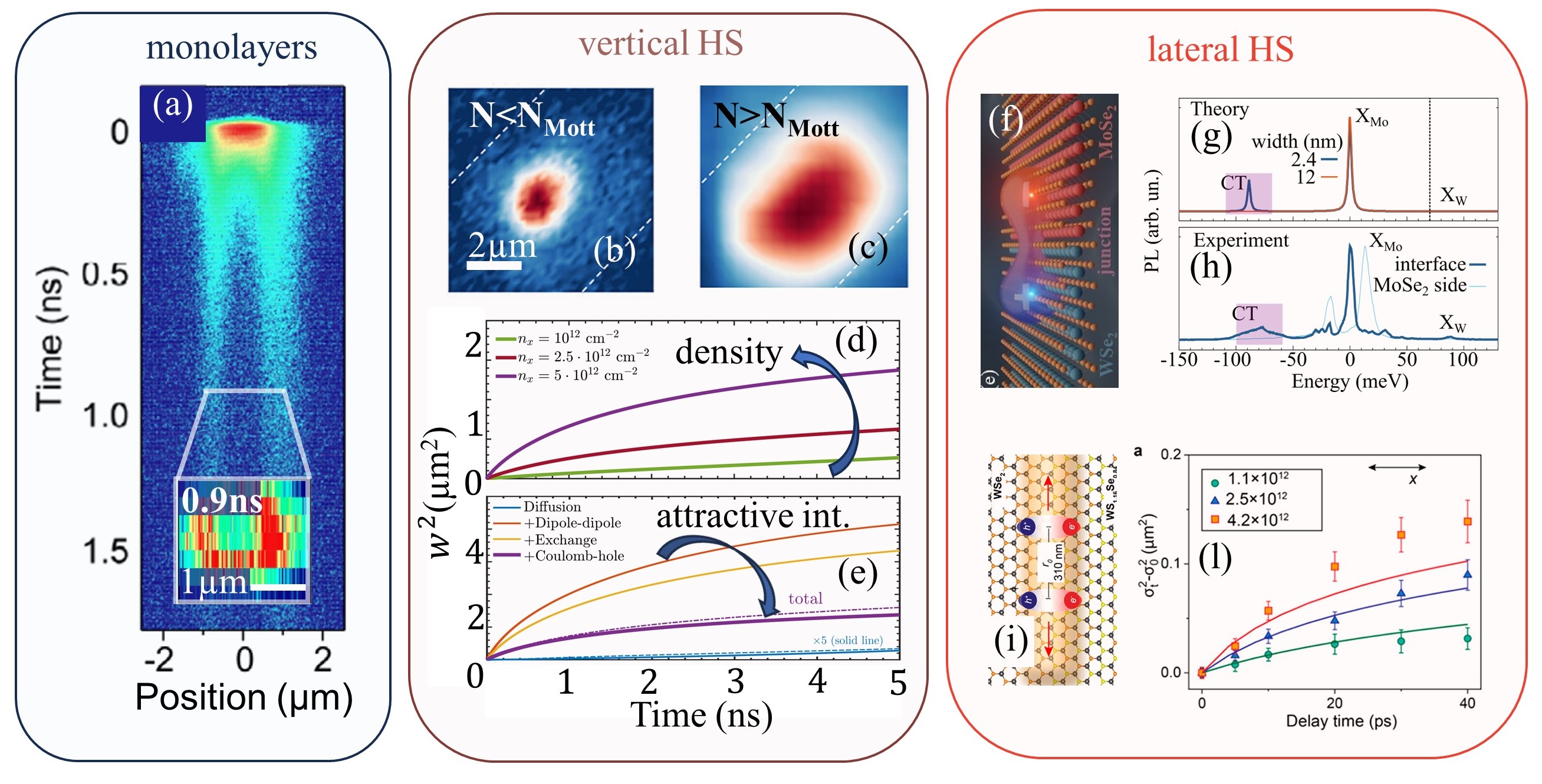}
    \caption{(a) Space- and time-resolved  photoluminescence in WS$_2$ on a SiO$_2$ substrate under a high laser fluence exhibiting the appearance of halos. (b,c) Diffused profiles in twisted vertical MoSe$_2$-WSe$_2$ heterostructure (HS) revealed in a pump-probe experiment after excitation  below (b) and above the Mott transition (c). The latter shows a much larger broadening due to Fermi pressure and Coulomb repulsion. (d) Anomalous diffusion in a WSe$_2$ homobilayer triggered by dipole-dipole repulsion of interlayer excitons (e), contrary to other non-linear mechanisms which slow down the diffusion. (f) Sketch of a lateral MoSe$_2$-WSe$_2$ heterostructure and the demonstration via photoluminescence of charge-transfer (CT) excitons (g,h), contrary to what predicted with larger widths (orange in (g)) or exciting away of the junction (thin in (h)). (i) Sketch of lateral WSe$_2$-WS$_{1.16}$Se$_{0.84}$ heterostructure and (l)  non-linear anomalous diffusion along the interface (e). Figs. (a), (b,c), (d,e), (g,h) and (i,l) adapted respectively from Refs. \cite{Kulig18}, \cite{Choi23}, \cite{Erkensten22}, \cite{Rosati23} and \cite{Yuan23}.}
    \label{tr5-nonLinear}
\end{figure}

Figure \ref{tr5-nonLinear}(a) shows
the PL spectrum measured in a strong-excitation regime of a WS$_2$ monolayer on a SiO$_2$ substrate \cite{Kulig18}. A qualitatively different diffusion behaviour is found compared to the conventional diffusion at low excitations (Fig. \ref{tr2-conventional}(a)). Two spatially-separated PL peaks splits appear: These propagate away from the excitation spot (Fig.  \ref{tr5-nonLinear}(a)), forming in the whole 2D plane a spatial ring (often denoted as a  halo, see inset) with a radius increasing with time \cite{Kulig18}. Before the formation of this halo at approximately $\approx$ 500 ps, the profile in Fig. \ref{tr5-nonLinear}(a) is much more extended than in Fig. \ref{tr2-conventional}(a), indicating a faster transport.  
The halos can be triggered by phonon wind and drag \cite{Glazov19} and/or be traced back to the efficient process of Auger recombination, a Coulomb mechanism where 1s excitons spontaneously recombine and their energy promotes another 1s exciton into a high-energy state \cite{Perea19}. Once this hot exciton relaxes down to lower energies, it emits a vast amount of phonons. This results in an off-equilibrium phonon distribution which can drive the halos via a large local gradient in the excitonic temperature and the associated Seebeck effect \cite{Perea19}. The Auger mechanisms strongly depends on the substrate, in particular being much smaller under hBN-encapsulation. This results in a suppression of halos in hBN-encapsulated WS$_2$ \cite{Zipfel20} or WSe$_2$ \cite{Cordovilla19}, while observed only for trions in hBN-encapsulated MoSe$_2$\cite{Park21}. 

Further increasing the excitation density, new phenomena can appear. A surprisingly fast propagation was observed in  MoS$_2$ under high excitation, with the pulse area expanding by 4x10$-4$ cm$^2$ in the first ps \cite{delAguila23} (corresponding to a diffusion coefficient $D\approx 10^6$ cm$^2$/s, which is 5-7 orders of magnitude larger than in Figs. \ref{tr2-conventional},\ref{tr4-Temperature}). This has been attributed to a collective exciton fluid behaviour \cite{delAguila23} and observed only for a low doping. At densities in the range of $\gtrsim 10^{13}$cm$^{-1}$, the Mott transition is reached, where  excitons break and build and electron-hole plasma \cite{Steinhoff17}, as observed in both TMD mono- \cite{Chernikov15,Sousa23} and bilayers \cite{Choi21,Wietek24}.  In van der Waals (vdW) heterostructures 
the diffused profile below the Mott transition is much smaller than the one above it, as revealed by pump-probe measurements and shown respectively Fig. \ref{tr5-nonLinear}(b) and \ref{tr5-nonLinear}(c). This reflects a more efficient diffusion of an electron-hole plasma induced by Fermi pressure and Coulomb repulsion, as the high density of electrons leads to a Fermi-Dirac distribution with a high fermionic temperature \cite{Choi23}. The competition between transient phases with high and low mobility  above and below the Mott transition can also give rise to a negative diffusion \cite{Wietek24}. 

Even remaining below the Mott transition, vertical  heterostructures show a peculiar non-linear transport at increasing exciton densities. 
These heterostructures in fact host interlayer excitons, which are composed by electrons and holes in different layers, cf. Fig. \ref{intro1}(b).
This electron-hole separation forms a spatial dipole of about 6 $\AA$, corresponding to their interlayer separation, or even more if enhanced  e.g. by adding hBN spacers between the two TMD layers \cite{sun22,Latini17,Erkensten22}, as in Figs. \ref{tr5-nonLinear}(d,e).
At high exciton density, dipolar excitons are close enough to repel each other via dipole-dipole interaction. In optics this has been observed via a blueshift of the excitonic peak for increasing laser power \cite{Yuan20,Unuchek19,Nagler17}. Such a blueshift becomes particularly important in transport: Here localized densities induce a spatially-dependent blueshift, which in turn acts as a  potential pushing the excitons away of the excitation spot.
This leads to the behaviour in Fig. \ref{tr5-nonLinear}(d), where the increase of the squared width clearly deviates from the linear behaviour of the conventional exciton diffusion on a nanosecond timescale \cite{Erkensten22}.
Importantly, in view of the fluency-dependent dipole-dipole repulsion, the diffusion is faster for higher excitation densities, in full agreement with experiments \cite{Sun2022,Yuan20}.
Besides the  repulsive dipole-dipole interaction, which enhances the exciton propagation, there are also  attractive contributions of Coulomb exchange and bandgap renormalizations \cite{Erkensten22,Steinhoff23} which counteract partially the dipolar repulsion, cf. red and purple line in Fig. \ref{tr5-nonLinear}(e). Nevertheless the diffusion becomes overall faster than without any nonlinear effects, cf. purple and blue in Fig. \ref{tr5-nonLinear}(e), because the magnitude of the dipole-dipole repulsion is stronger than the attractive interactions \cite{Erkensten22}. 

Particularly large dipoles can be obtained in  lateral TMD heterostructures (Fig. \ref{tr5-nonLinear}(f)). These structures are composed by two different TMD monolayers grown one beside each other and covalently bound together \cite{Huang14,Duan14,Li15,Xie18,Sahoo18}. At the interface they form a type-II heterostructure, hence potentially allowing the formation of charge-transfer excitons, with Coulomb-bound electrons and hole separated from each other at two different sides of the junction \cite{Lau18,Rosati23,Yuan23}, cf. the sketch in Fig. \ref{tr5-nonLinear}(f). These CT excitons have only recently been observed in lateral MoSe$_2$-WSe$_2$ \cite{Rosati23} and WSe$_2$-WS$_{1.16}$Se$_{0.84}$ heterostructures \cite{Yuan23}. The microscopic study of these quasi-particles revealed the crucial role played by the interface width, which only when comparable to the exciton Bohr radius allows  the formation of stable CT excitons and their optical observation  \cite{Rosati23}. 
Larger junction widths in fact lead to a reduction of occupation, binding energy and
 oscillator strengths of CT excitons 
 \cite{Rosati23}. 
To the purpose of optical observation it is hence crucial to have samples with a reduced amount of alloying at the interface, as allowed by the recent developments in CVD growth techniques \cite{Li15,Xie18,Rosati23,Beret22,Najafidehaghani21,Ichinose22}. The appearance of CT excitons is demonstrated in Fig. \ref{tr5-nonLinear}(g), where the spectrally-resolved PL after excitation at the interface is shown for junction widths of 2.4 and 12 nm. Only in the first case, a new peak appears almost 90 meV below the A exciton from MoSe$_2$  (purple box). This prediction is in full agreement with the experimental observation, showing a peak at similar energies -  in contrast to the case when  exciting away of the interface (Fig. \ref{tr5-nonLinear}(h)). Besides the interface width, the microscopic study revealed the importance of an environment with a large dielectric screening, as it is the case for the hBN encapsulation shown in Fig. \ref{tr5-nonLinear}(g,h). In contrast, free-standing samples would not allow the observation of CT excitons due to a reduced energy separation between the monolayer and CT exciton, resulting in the weak occupation of the latter \cite{Rosati23}. 

CT excitons are technologically interesting as they exhibit dipoles of several nanometers \cite{Lau18,Rosati23,Yuan23}, which are one order of magnitude larger than in vertical heterostructures. The reason is that  here the electron-hole separation is not affected by any geometrical restriction, contrary to the case of vertical heterostructures, where it is limited by the distance between layers, although partially increasable via insertion of hBN spacers \cite{sun22,Latini17,Erkensten22}. 
The large dipoles make CT excitons highly interesting to study. First, they lead to a non-linear blue-shift of CT excitons \cite{Rosati23}. Second, they can be the reason for the recently reported anomalous diffusion along the interface of a WS$_{1.16}$Se$_{0.84}$ lateral heterostructure, Fig. \ref{tr5-nonLinear}(i), as shown in Fig. \ref{tr5-nonLinear}(l). Here the profile variance along the interface is shown, with a non linear increase  with time becoming more effective at larger densities. This is similar to the anomalous diffusion observed in vertical heterostructures \cite{Erkensten22,Tagarelli23}, Fig. \ref{tr5-nonLinear}(d), suggesting an efficient repulsion of excitons trapped at the interface. Note that effective non-linear speed-up of the excitonic diffusion were also observed in coupled quantum wells building one-dimensional channels  via engineering of the environment \cite{Voegele09}.  This makes use of a gate field, which can also allow gate-induced homojunctions \cite{Pospischil14,Baugher14,Ross14} potentially resulting in  bound excitons for p-i-n junctions confined to few tens of nanometers \cite{Thureja22}.
In a nut-shell, these results indicate that lateral heterostructures can lead to an interesting transport along  the interface, in addition to the remarkable transport across the interface which will be further discussed in Sec. \ref{sec:lateralDrift}. 

\paragraph{Cavity control}\label{sec:cavity}

\begin{figure}[t!]
    \centering
    \includegraphics[width=\linewidth]{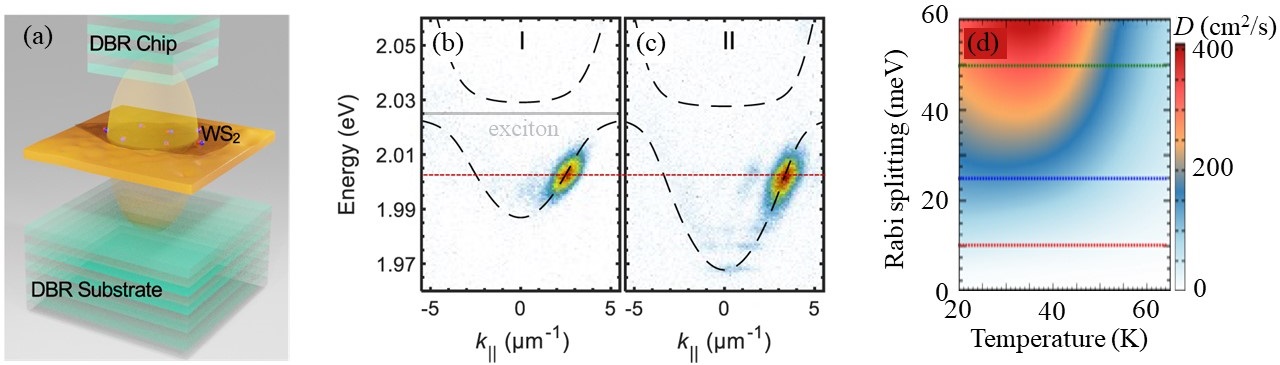}
    \caption{Sketch of the integration of a TMD nanomaterial within a planar cavity, resulting in a localization of the photon modes around the monolayer (yellow cloud), hence increasing the coupling with excitons. (b,c) Angle/momentum resolved PL from a WS$_2$ monolayer embedded in a cavity and collected from spatial positions approximately 7 $\mu$m and 14 $\mu$m at the right of the excitation spot. The signal preserves its excitation energy and has contributions only from positive momenta, reflecting a ballistic propagation toward the collection spot located at the right. (d) Diffusion coefficients in MoSe$_2$ as a function of temperature and Rabi splitting, which reflects the cavity-induced photon confinement. The diffusion is 2-3 orders of magnitudes larger than without the cavity. Figs. (a-c) and (d) adapted from Ref. \cite{Wurdack21} and \cite{Ferreira22}.}
    \label{tr6-cavity}
\end{figure}

Due to their remarkably strong light-matter interaction  TMD nanomaterials are suitable for reaching the strong-coupling regime in optical cavities (Fig. \ref{tr6-cavity}(a)). This confines photons around the 2D material, hence increasing the exciton-photon interaction beyond the exciton and the cavity decay rate. Here, new quasi-particles are formed, namely the exciton polaritons, which are hybrid particles combining properties of excitons and photons. They are observable  via  an avoided crossing between the exciton and photon dispersion giving rise to a Rabi splitting, cf. the dashed lines in Fig. \ref{tr6-cavity}(b). Optical studies allow to resolve the energy of polaritons as well as their in-plane momentum, which is determined by the propagation direction of the absorbed/emitted photons out of the cavity. The interaction between (the photonic part of) polaritons and out-of-the-cavity photons in fact preserves the momentum, whose ratio between inplane and out-of-plane component determines the direction at which the photon propagate out of the cavity. 

Exciton polaritons combine the high mobility of photons with the control offered by excitons via scattering with phonons. The huge group velocity allows important ballistic propagation, as shown in Fig. \ref{tr6-cavity}(b,c), where the PL collected respectively 7 and 14 $\mu$m away of the excitation spot is shown. This shows a distribution centered in the excitation energy and in positive momenta, reflecting that the detection spot is located in real space at the right side of the excited regions (while measurements at the left side show only negative momenta \cite{Wurdack21}). 
While the diffusion is characterized by an isotropic distribution centered at low-energies \cite{Rosati20}, the strong anisotropy of this PL indicates that a large fraction of polaritons can propagate ballistically for tens of micrometer according to their energy and momentum direction. 
A qualitatively similar behaviour has been predicted for bright excitons at cryogenic temperatures without strong coupling, however limited to few picosecond and few hundreds of nanometers due to  the considerably smaller group velocities \cite{Rosati20}.

On the other hand, the scattering of polaritons with phonons can be observed in  optical spectroscopy measurements, with  the absorption showing step-like increases in linewidth as well as in intensity when new scattering channels are opened \cite{Ferreira23,Ferreira24}. Crucial for the increase in absorption intensity is the critical coupling \cite{Fitzgerald22}, allowing maximum absorption when the phonon-induced polariton decay matches the cavity decay induced by the mirror quality. 
The polariton optics is particularly affected by the scattering with dark excitons, which are crucial for circumventing a polariton relaxation bottleneck \cite{Fitzgerald24}
while the scattering of lower polaritons with intravalley acoustic modes is suppressed \cite{Ferreira22}, contrary to the case of regular excitons \cite{Selig16} (cf. Fig. \ref{tr4-Temperature}(a)). 
The suppression of scattering with intravalley acoustic modes together with the huge polariton group velocities leads to diffusion coefficients 2-3 orders of magnitudes larger than for excitons, cf. Fig. \ref{tr6-cavity}(d). The faster polariton diffusion, already observed in other materials \cite{Bley98,Zaitsev15}, and the important ballistic propagation make TMD-based exciton polaritons particularly interesting for transport purposes. In a nutshell, the strong interaction with photons can significantly speed up the excitonic transport and even allow propagations for tens of micrometers. 

\paragraph{Gating control}\label{sec:charge}

Excitons dominate the response of TMD materials due to their huge binding energies. This however does not mean that one can forget about charges carriers, in particular when discussing about transport. They have typically a higher mobility \cite{Jin14} and a slower decay time, potentially resulting in faster diffusion \cite{Zipfel18} and longer diffusion length \cite{Ren22}. However, tackling their contribution to transport is often complicated, as generally interconnected  with the excitonic transport -- in contrast to what happens in optics, where continuum, trions and excitonic result in  spectral features well separated in energy. Here we discuss the effects on transport induced by electron-hole continuum as well as charge doping and how this can be controlled by dielectric engineering and gating.

Optical excitations create electron-hole pairs, both bound as excitons or unbound in the continuum above the bandgap according to the laser energy. Once formed, these pairs  relax down in energy toward the equilibrium \cite{Brem18}, which results in a large fraction of pairs being 1s excitons thanks to their low energy.
Nevertheless, the unbound electron-hole pairs have larger density of states than the bound excitons (forming respectively a four and two-dimensional space).
At room temperature and/or with small binding energies this can lead to a finite amount of unbound electron-hole pairs, in turn having
a drastic impact on transport. In the low-density regime the conventional diffusion of WS$_2$ shows a 10-fold speed-up under hBN-encapsulation compared to the case of deposition on SiO$_2$-air \cite{Zipfel20}, Fig. \ref{tr7-charges}(a), or the free-standing one \cite{Kulig18}. This has been attributed to the higher dielectric screening induced by hBN-encapsulation, which results in lower binding energies and hence in a much larger fraction of electron-hole plasma compared to the SiO$_2$-air or freestanding surroundings. This eventually results in the increased diffusion via the higher mobilities of electron and holes \cite{Zipfel20}.

Besides showing intrinsic doping, in TMD monolayers the amount of free charges can be extrinsically controlled by means of gating via metallic contacts \cite{He20}. This results in a non-trivial and non-monotonic variation of the diffusion coefficient, as shown in Fig. \ref{tr7-charges}(b) exploring hBN-encapsulated WSe$_2$ at cryogenic temperatures, where the electron-hole plasma plays no role. 
A general decrease of diffusion coefficient with higher doping is induced by an increased scattering with free charges (as revealed by the linewidth \cite{Wagner23}). 
This is interrupted at a doping of approximately 10$^{11}$cm$^2$ by a transition from exciton- to trion/polaron-dominated system (sketch), resulting in the non-monotonic behaviour of diffusion with doping \cite{Wagner23}), Fig. \ref{tr7-charges}(b).

We now proceed discussing more directly the diffusion of trions. First we consider the regular conventional diffusion, whose predictions are shown in Fig. \ref{tr7-charges}(c) for MoSe$_2$ as a function of temperature \cite{Perea22}. Here the energetically-lowest trion states are composed by electron and holes all living in K or K$^ \prime$ valley, leading to results almost independent of the charge sign (here negative, with two electrons and a hole forming the trion). In the low-density limit (purple) the diffusion becomes faster with increasing  temperatures due to the smaller scattering of high-energy state \cite{Perea22}.
When increasing the density of trions a drastic speed-up of diffusion appears, up to a factor of four at low temperatures (cf. red and purple in Fig. \ref{tr7-charges}(c)). This happens because trions behaves as fermions, hence increasing their density leads to state-filling in particular at low $T$, resulting in the occupation of states with higher energy and hence larger mobility in the Fermi-Dirac distribution. 

The diffusion of trions becomes even more interesting looking at its transient behaviour, Fig. \ref{tr7-charges}(d), which clearly shows a spatial shrinking of the profile width appearing between approximately 5 and 35 ps (red area). This remarkable negative diffusion indicates the competition between  different species, cf. Fig. \ref{tr3-transient}.
In contrast, a trivial linear increase of $w^2$ is observed for the spin-dark trions \cite{Beret23}. In addition this mechanism can be controlled by gating \cite{Beret23}, similarly to Fig. \ref{tr7-charges}(b) but now affecting the  transient diffusion. In a nutshell, all these results show how the free charges provide an additional turning knob to control the exciton and trion transport in TMD monolayers.

\paragraph{Strain control}\label{sec:strain}

\begin{figure}[t!]
    \centering
    \includegraphics[width=\linewidth]{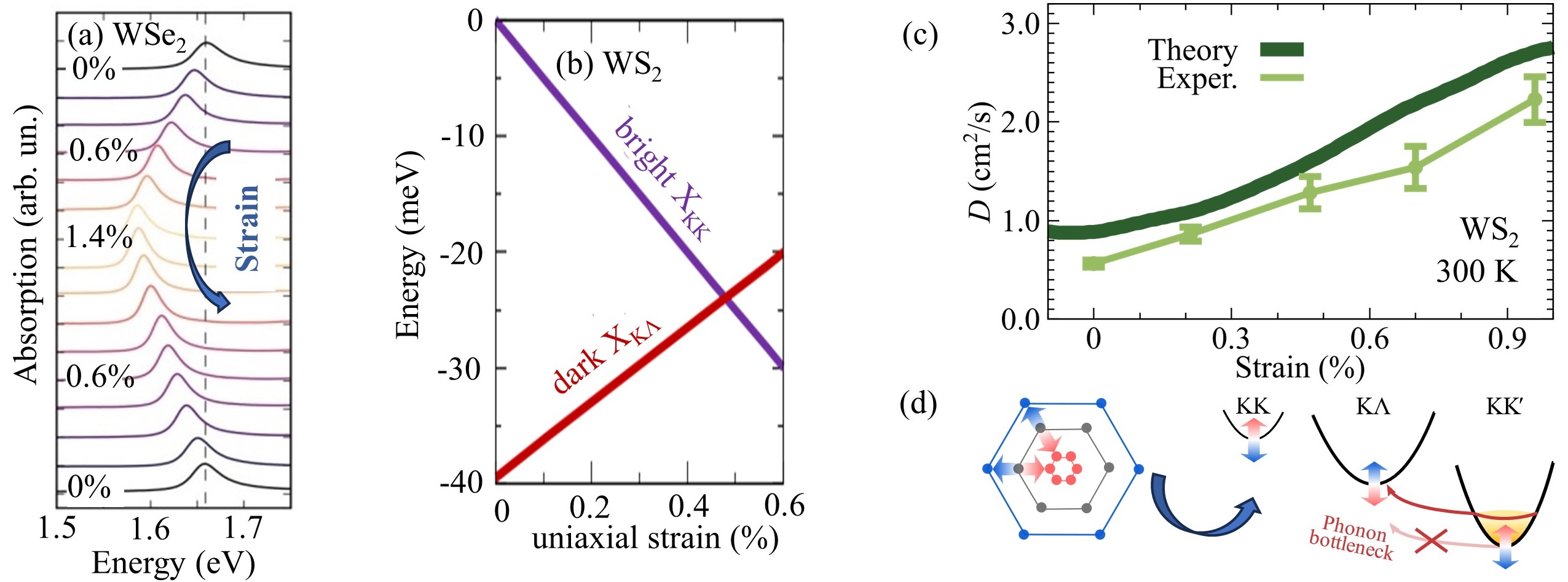}
    \caption{(a) Room-temperature absorption spectra of WSe$_2$ for varying uniaxial strain, which induces a red-shift of the A-exciton peak. (b) Strain-induced variation of the energy of bright KK and dark K$\Lambda$ excitons, showing an opposite red- and blue-shift. (c) Room-temperature diffusion constant in WS$_2$ as a function of uniaxial strain, showing a 3-fold increase with an excellent theory-experiment agreement. This is induced by the suppression of interlayer scattering between the dark KK$^\prime$ and K$\Lambda$ valleys when these become non-degenerate thanks to the tensile strain sketched in (d). Fig. (a), (b) and (c,d) adapted from Ref. \cite{Schmidt16}, \cite{Rosati21e} and \cite{Rosati21a}.}
    \label{tr8-strainDiff}
\end{figure}

Finally,  we discuss  a mechanism  able to control exciton diffusion but also to induce a directional exciton funneling, hence building a bridge between this and next section. The mechanism is based on strain engineering, which provides a powerful and valley-dependent tool to control excitons and their transport \cite{Schmidt16,Khatibi18,Rosati21a,Rosati21e,Kumar23}. Strain is a lattice deformation  obtained for example via a substrate bending, electrostatic forces, piezoelectric or 
patterned substrates \cite{Schmidt16,Cordovilla18,Harats20,Moon20,Lee22,Gelly22,Kumar23}. 
While allowed by the flexibility of TMD nanomaterials, which can sustain up to 10\% strain, these lattice deformations crucially affect the optoelectronic properties of TMDs.
The strain affects both phonon and exciton energies. The former can be revealed by Raman spectroscopy, 
which allow an accurate read-out of strain despite minor variations of  phonon energies in the range of 0.1-0.4 meV/\% uniaxial strain \cite{Dadgar18}. In contrast, the strain-induced variations of excitonic energies are macroscopic, of the order of  100 meV/\% biaxial strain (50 meV/\% uniaxial strain), leading to a drastic impact on the optical and transport response of TMDs. This is shown in the strain-dependent absorption spectra in Fig. \ref{tr8-strainDiff}(a), with the A peak red-shifting by approximately 50 meV/\% uniaxial strain \cite{Schmidt16}. This reflects the gauge factor of bright excitons, in strong contrast with the blue-shift of momentum-dark K$\Lambda$ excitons, cf.  Fig. \ref{tr8-strainDiff}(b). This stems from the strain-dependent single-particle energies, with the K and $\Lambda$ valley in the conduction band red-shifting with a respectively larger and smaller gauge factor than the K point in the valence band\cite{Chen22}. This results in strain-induced modifications of the single-particle bandgap \cite{Ghorbani13,Plechinger15,Feierabend17b,Khatibi18,Blundo22,Li23} which dominate the excitonic gauge factor, with small residual effects from the variations of the binding energies, smaller than 10 meV/\% biaxial strain for KK excitons in free-standing WS$_2$ \cite{Kumar23}.
The excitonic energies vary quasi-linearly with typical strain as large as 3\% biaxial \cite{Feierabend17b}, while deviations from a linear gauge factor can appear at larger lattice deformations \cite{Chang13}.
Such co-existence
and competition of valleys behaving differently with strain allows for a remarkable control of exciton transport. Here we discuss the impact of strain on exciton diffusion, whereas in Sec. \ref{sec:Funneling} we discuss its impact on exciton funneling.

The conventional diffusion is drastically enhanced by strain in tungsten-based materials \cite{Rosati21a,Uddin22}. In WS$_2$ the diffusion coefficient increases by a factor of 3 upon application of 1\% of uniaxial strain, as shown in Fig. \ref{tr8-strainDiff}(c) -  in an excellent theory-experiment agreement \cite{Rosati21a}.
This happens because in this material the diffusion is dominated by two dark exciton valleys, K$\Lambda$ and KK$^\prime$, which are quasi-degenerate in unstrained WS$_2$ but behave very differently with strain, sketch in Fig. \ref{tr8-strainDiff}(d). While K$\Lambda$ blue-shifts,  KK$^\prime$ red-shifts with a similar gauge factor as KK (cf. energy separation between spin-dark and bright excitons varying by 6.4 meV/\% uniaxial strain \cite{Dirnberger21}). As a result, strain brings K$\Lambda$ and KK$^\prime$ out of resonance (Fig. \ref{tr8-strainDiff}(d)), hence suppressing the otherwise very effective intervalley scattering between these two valleys (responsible also for the drop of diffusion in WSe$_2$ of Fig. \ref{tr4-Temperature}(d)). This phonon bottleneck results in a considerable speed-up of the diffusion via a decrease of the phonon-mediated scattering with  the most populated excitons. Such a strain-induced bottleneck involves only dark excitons, while the scattering of the bright excitons is not affected in this strain range (although in general the 
linewidth and its asymmetric shape can be affected by strain \cite{Niehues18}): This shows how transport allows the observation of physical mechanisms linked to states which are not directly optically active. 

\begin{figure}[t!]
    \centering
    \includegraphics[width=\linewidth]{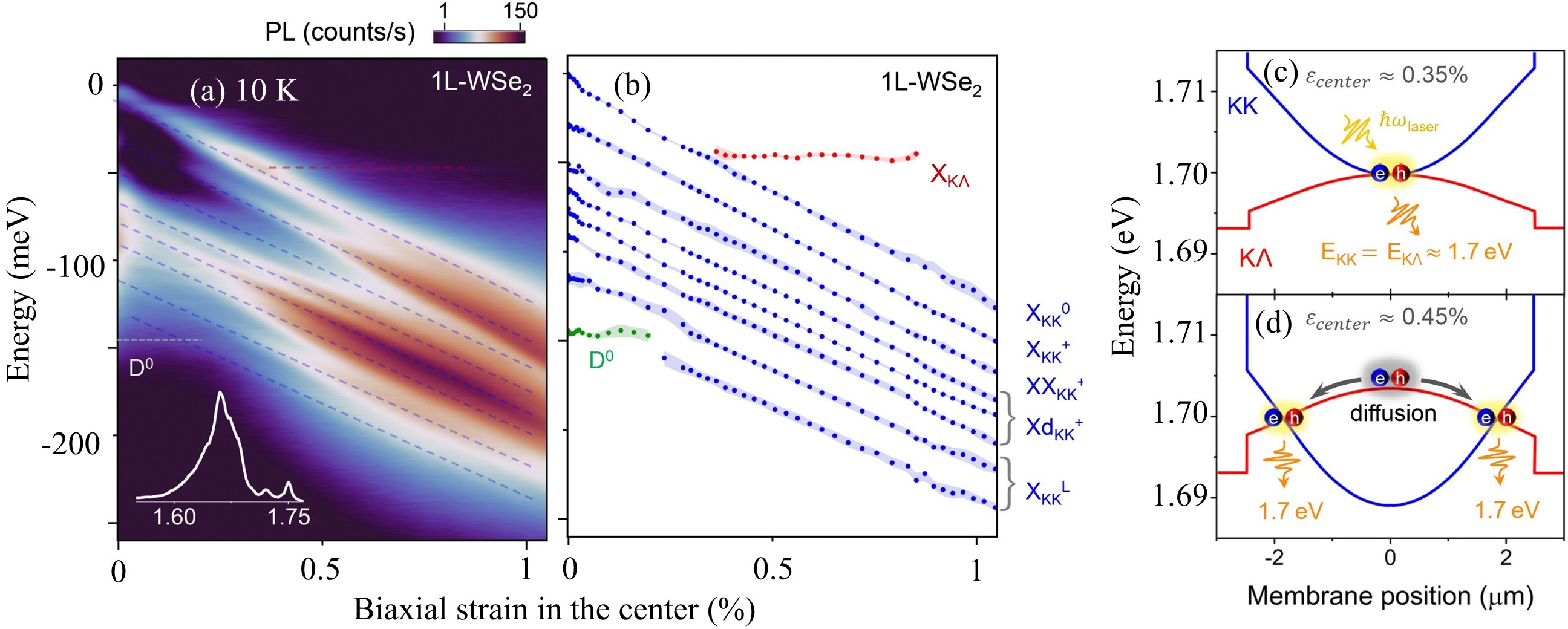}
    \caption{Cryogenic photoluminescence spectra in freestanding WSe$_2$ (b) and energy position of its peaks (b) as a function of strain.
    The peak X$_{\text{K}\Lambda}$ appearing at fixed energy and in a given strain range is induced by the hybridization between KK and K$\Lambda$ excitons, which takes place always at the same energy and only for central strain larger than approximately 0.35\% (c). Larger central strains implies an hybridization happening away from the membrane center (d): When this spatial separation becomes larger than the diffused profile, peak X$_{\text{K}\Lambda}$ disappears.  Fig. adapted from Ref. \cite{Kumar23}.}
    \label{tr9-strainFingerprinting}
\end{figure}

More recently the valley-dependent gauge factors have allowed to fingerprint the excitons behind the cryogenic photoluminescence \cite{Kumar23}, Figs. \ref{tr9-strainFingerprinting}(a). 
Such spectra show a multitude of peaks in both WS$_2$ and WSe$_2$, Figs. \ref{tr9-strainFingerprinting}(a), whose energy change with strain, cf. Fig. \ref{tr9-strainFingerprinting}(b) for WSe$_2$. Most of the peaks redshifts with gauge factors of approximately 110 meV/\%, blue lines in Figs. \ref{tr9-strainFingerprinting}(a-b): These are emitted by quasiparticles related to KK excitons, including trions and phonon replicas. However, two very different peaks appear, peak D$^0$ induced by defect \cite{Lopez22} (green) and peak X$_{K\Lambda}$ induced by the hybridization between the bright excitons with the dark K$\Lambda$ ones. Such a peak shows a peculiar behaviour, with a constant energy and its presence observable only in a finite range of strain between 0.3 and 0.7 \%. The reason stems from the hybridization,
which takes place only when K$\Lambda$ and KK excitons are almost resonant. 
In view of the slightly inhomogeneous strain, the energy profile of KK and K$\Lambda$ excitons are weakly space-dependent. The electrostatic potential can change the height of the strain profile while changing its shape \cite{Kumar23}.
A minimum height of the strain profile is required for the hybridization, Fig. \ref{tr9-strainFingerprinting}(c), while too large heights would push such crossing point too far away from the membrane center, where the laser spot is positioned.  
When the diffused excitonic spatial density is narrower in space than the distance between crossing point and the center of the membrane, no hybridization can take place, resulting in the given range of strain allowing the observation of X$_{\text{K}\Lambda}$. In addition to  these features, in WS$_2$ a new peak with a moderate redshift of only 60 meV/\% is observed \cite{Kumar23}. This is induced by $\Gamma\Lambda$ excitons, as microscopically predicted from their redshift \cite{Kumar23} and allowed in freestanding WS$_2$ due to a small energy separation from the bright excitons, contrary to the case of WSe$_2$ and of supported/encapsulated WS$_2$. Finally, peaks from KK, K$\Lambda$ and $\Gamma \Lambda$ have also been directly observed in homobilayer WSe$_2$ and assessed via their strain-induced energy shifts \cite{Kumar23}. In addition, peaks emitted from intralayer K$\Lambda$ excitons have been observed in sequentially-stacked vertical heterostructures, interestingly emitted only from the top layer likely due to compressive strain associated with the stacking procedure \cite{Sebait23}.
In a nutshell, these results provide finally the direct visualization of the remarkable interplay between different excitons in W-based nanomaterials.

So far, the cases of spatially homogeneous lattice deformation or excitations at the maximum of a weakly inhomogeneous strain profile have been considered, respectively Fig. \ref{tr8-strainDiff} and  \ref{tr9-strainFingerprinting}. The situation becomes even more interesting combining localized strain (as obtainable via patterned substrates, tips, electrostatic potentials of surface acoustic waves, etc) with excitation away of its maximum, where the bright excitons have minimum energy. 
Can excitons drift toward these points? This question is  treated in the next section, where we focus on  directional exciton transport, i.e. exciton drift and how it can be achieved.

\subsection{Exciton drift}

The principle of minimum energy states that the energy of physical systems decreases toward its minimum value. This general rule applies to transport as well. Charge carriers move toward spatial regions, where the external potential is minimum. This can be expressed by extending  Fick's law of Eq. \ref{Fick} to a drift-diffusion equation
\begin{equation}\label{drift-diffusion}
    d_tN(X,t)=\nabla D(\textbf{R}) \cdot \nabla N(\textbf{R},t) +\nabla( \mu(\textbf{R})  N(\textbf{R},t) \nabla V(\textbf{R}, t )) \quad .
\end{equation}
The first term describes the regular diffusion discussed already in the last section, but now  with  a space-dependent diffusion coefficient $D$ describing also the case of localized strain profiles $s\equiv s(\textbf{R})$. The second term of Eq. (\ref{drift-diffusion}) describes the drift driven by an external potential $V(\textbf{R},t)$ and proportional to the mobility $\mu$, which can be typically written via the Einstein relation $\mu=D/k_BT$ with potential modification from traps \cite{li23c}. In electronics, the external potential is the electrostatic one induced by metallic contacts. Unfortunately, the same mechanism can not be applied to excitons due to  their neutral character. Nevertheless, different mechanisms exist to bypass this limitations, as we proceed discussing in the following.

\subsubsection{Strain-induced funneling}\label{sec:Funneling}

In Sec. \ref{sec:strain} we learned how strain drastically modifies the exciton energies. For homogeneous lattice deformation, this impacts transport via a 3-fold faster diffusion upon application of 1\% uniaxial strain (Fig. \ref{tr8-strainDiff}(c)). Now we consider localized strain profiles, which can be obtained via patterned substrates \cite{Tonndorf15,Branny17,Palacios17,Rosati21e} or spontaneously via formation of bubbles\cite{Blundo21,Carmesin19,Rosati21d} and wrinkles\cite{Castellanos13}. 
\begin{figure}[t!]
    \centering
    \includegraphics[width=0.9\linewidth]{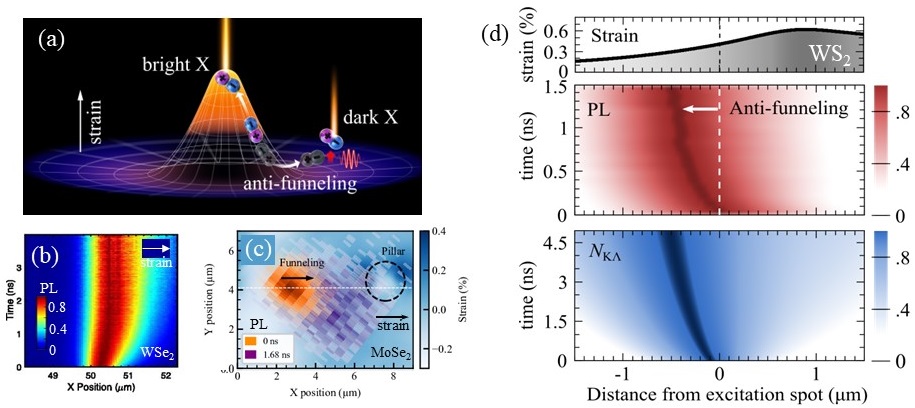}
    \caption{(a) Sketch of a spatially-localized strain profile and its impact on funneling and anti-funneling of respectively bright and dark excitons. (b,c) Observation of funneling toward high-strain regions in WSe$_2$ (b) and MoSe$_2$(c). (d) Time- and space-resolved photoluminescence (center) along one direction after exciting WS$_2$ in points with inhomogeneous strain (top). The anti-funneling toward low-strain regions is observed and induced by dark excitons (low). Figs. (a,c,d) and (b) adapted from Ref. \cite{Rosati21e} and \cite{Cordovilla18}.}
    \label{tr8-funneling}
\end{figure}

Considering the gauge factors of Fig. \ref{tr8-strainDiff}(b), local strain results in space-dependent exciton energies, which in turn act as a potential for these excitons. Bright excitons have their energy minimum in regions with high strain, Fig. \ref{tr8-funneling}(a), driving the exciton toward these regions in a  regular  so-called funneling. This was observed first in the wrinkles of bilayer MoS$_2$ \cite{Castellanos13} and then in monolayers \cite{Cordovilla18,Moon20,Harats20,Rosati21e}, as shown in Fig. \ref{tr8-funneling}(b-c)  for WSe$_2$ and MoSe$_2$. Here the time- and space-resolved PL reveals the drift of a localized exciton profile toward spatial regions with high strain, where the energy of bright excitons is lower. 

More recently, the  opposite scenario of anti-funneling has been observed, defined as a motion  toward low-strain regions as observed in WS$_2$, cf. Fig. \ref{tr8-funneling}(d). Here a high-strain profile (top panel) has been investigated varying the laser position. The resulting space- and time-resolved PL taken along a direction  shows anti-funneling toward low-strain regions, central panel. The effectiveness of this anti-funneling decreases with time, with the central position drifting by 200 nm in the first 300 ps (dark red in central panel). In addition, varying the laser positions reveals a stronger and faster anti-funneling exciting in regions where the strain profile is steeper, with drifts in the first nanosecond ranging from few tens to 300 nanometers as in Fig. \ref{tr8-funneling}(d) \cite{Rosati21e}. 

Similarly to the case of strain-induced increase of the diffusion coefficient, the reason behind the anti-funneling lies again in the competition between different valleys. 
To show this, Fig.  \ref{tr8-funneling}(d) provides the microscopically predicted evolution of the K$\Lambda$ excitons. Interestingly, this shows a behaviour very similar to the experimental one. First of all, the K$\Lambda$ density anti-funnels toward low strain regions, exactly as the PL signal in experiments. This happens because at high strain the energy of K$\Lambda$ states is maximum, hence these excitons drift toward low strain regions, despite the competition of the  phonon-driven intervalley thermalization from K$\Lambda$ to KK and KK$^\prime$ excitons. 
In full analogy with the experiments, the antifunneling of K$\Lambda$ is predicted to be initially faster before saturating after few hundreds of picosecond. This is mainly due to the strain-dependent diffusion coefficients of Sec. \ref{sec:strain} and Fig. \ref{tr8-strainDiff}(c). While anti-funneling the excitons reach regions with lower strain, where the diffusion coefficient is smaller \cite{Rosati21e}. The theory-agreement experiment remain also varying the laser position, further indicating that the K$\Lambda$ excitons are the reason for the experimentally-observed anti-funneling. This is further confirmed by the analogous experiment in MoSe$_2$, Fig. \ref{tr8-funneling}(c). Here regular funneling toward high-strain regions is observed, because in this material K$\Lambda$ excitons play a negligible role as they are higher in energy compared to the bright states. Interestingly, in MoSe$_2$ funneling of 1 $\mu$m in the first nanosecond has been observed \cite{Rosati21e}.

The K$\Lambda$ excitons can also drive the transport toward dielectric inhomogeneities, in particular in TMD bilayers \cite{Su22}, as will be discussed in Sec. \ref{sec:dielectric} together with other dielectric controls.  
Besides bright and K$\Lambda$ excitons, other species may be involved in the directional funneling.
The funneling of spin-dark excitons has been observed via tip-enhanced measurements \cite{Gelly22}, while the conversion of excitons into trions could also affect the transport \cite{Harats20}. 
Strain profiles can also be changed dynamically via electrostatic potentials \cite{Lopez22}, tips \cite{Koo21,Harats20,Gelly22,Lee22} and surface acoustic waves, which have allowed a controllable drift of excitons in particular in TMD bilayers \cite{Datta22}. These studies exemplify how exciton transport in TMD-based devices can be
engineered with strain.

\subsubsection{Dielectric control}\label{sec:dielectric}

\begin{figure}[t!]
    \centering
    \includegraphics[width=\linewidth]{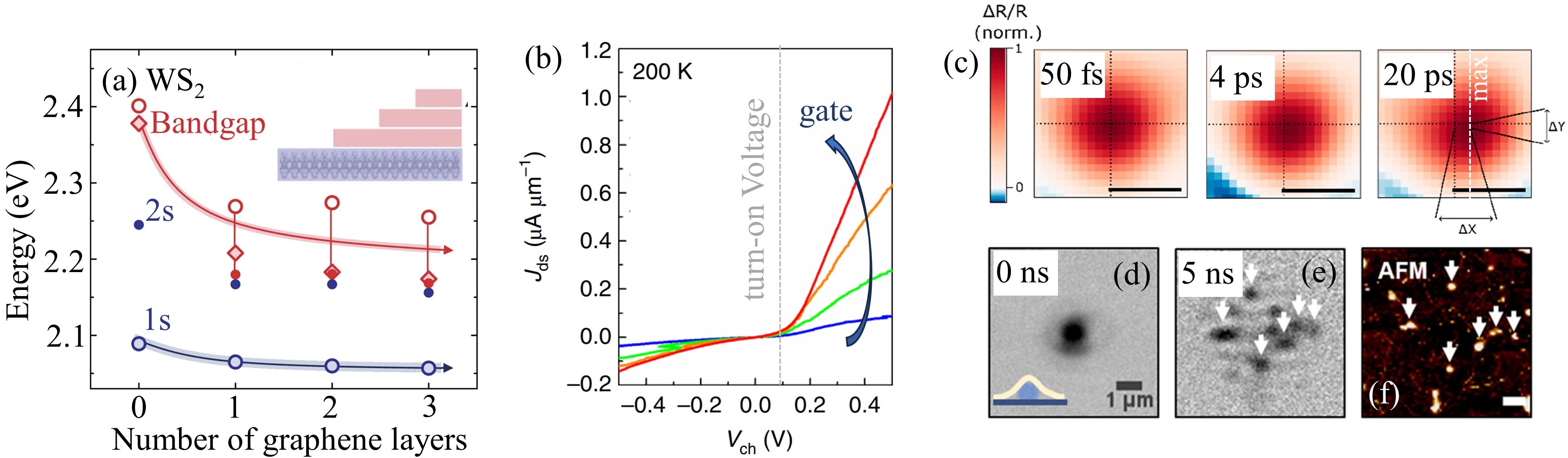}
    \caption{(a) Relevant energies as a function of number of graphene layers between a WS$_2$ monolayer and the SiO$_2$ substrate (sketch): Both the measured 1s and 2s excitons (blue) and the bandgap estimated from their separation $\Delta_{12}$ (red) show a non-linear decrease with graphite thickness. (b) Rectification of electric current in a MoS$_2$ monolayer on a dielectric lateral heterostructure: The onset potential of 90 meV (vertical line) reflects the conduction band offset. (c) Space- and time-resolved pump-probe measurements after non-resonant excitation of a WSe$_2$ monolayer in a dielectric lateral heterostructure, showing a drift/funneling toward the side with lower dielectric constant. (d,e) Two-dimensional excitonic distribution at 0 and 5 ns measured by stroboSCAT pump-probe microscopy, revealing a drift toward the dielectric nanobubbles, the latter observed also in AFM (f) and distinguished from the strain nanobubbles via lack of redshift in phototoluminescence. Figs. (a), (b), (c) and (d) adapted from Ref. \cite{Raja17}, \cite{Utama19}, \cite{Gauriot24} and \cite{Su22}.}
    \label{tr10-dielectric}
\end{figure}

The remarkable funneling and anti-funneling discussed in Sec. \ref{sec:Funneling} relies on strain via its property of modifying the excitonic energies in TMD monolayers. Other mechanisms can change the excitonic energies, in particular the so-called \textit{dielectric} control, obtained by varying the dielectric constants of the surroundings of the TMD crystal. Although this control present both quantitative and qualitative differences compared to  strain, see Fig. \ref{tr10-dielectric}(a), it can also lead to drift and antifunneling, as we discuss in the following and show in Fig. \ref{tr10-dielectric}(b-d).

Fig. \ref{tr10-dielectric}(a) show how key energies depend on the dielectric constant, here changed experimentally by depositing a monolayer WS$_2$ on a varying amount of graphene layers on top of SiO$_2$ (sketch). The results show the direct measurements of the energy of 1s and 2s excitons (blue) as well as the direct bandgap estimated from the measured energy separation $\Delta_{12}$ between 1s and 2s states (red). Increasing the number of graphene layers all the energies decrease but saturating to negligible variations for a high number of layer: Contrary to the case of strain, the energies vary non-linearly with dielectric surrounding \cite{Raja17,Waldecker19}, compare Fig. \ref{tr8-strainDiff}(b) and Fig. \ref{tr10-dielectric}(a). A second qualitative difference comes from comparing the magnitude of the shift of three  species of electron-hole pairs: While strain affects in similar way all the excitonic states from 1s and 2s to the continuum, the dielectric control becomes more efficient for decreasing binding energies. 
This happens because the binding energies are drastically affected by the dielectric surrounding via the Coulomb screening \cite{Chernikov14,Wang18,Mueller18}, whereas they are almost independent of strain (except for tiny variations induced by the residual variation of the effective mass on strain \cite{Feierabend17b,Kumar23}). As a result, in the dielectric control there is the competition of two mechanisms: Increasing the dielectric constant both the bandgap and the binding energies decrease, leading respectively to a red- and blueshift of the energy of 1s excitons. This leads to a partial cancellation, leading to a key quantitative  difference between strain and dielectric control when applied to 1s excitons: While the first can easily modify the 1s energy by (more than) 100 meV \cite{Schmidt16,Kumar23}, the second can typically induce variations of few tens of meV (becoming several tens of meV when starting from free-standing samples). This competition and cancellation become weaker for smaller binding energies: As a result, the dielectric control of the single-particle bandgap is more effective than the excitonic one. As a consequence, we expect a particularly interesting dielectric control of transport involving free charges, as we proceed discussing.

First, in Fig. \ref{tr10-dielectric}(b) the dielectric control of pure electron transport is shown. Here a so-called dielectric-defined lateral heterostructure is considered, with a bare MoS$_2$ monolayer deposited of a substrate composed of Cytop and hBN at the left and right side of the sample (sketch). A drastic increase of the electronic current across the dielectric lateral heterostructure is obtained increasing the in-plane bias V$_{\text{ch}}$ above the value of 90 meV (vertical line). This value is correlated to the electronic band offset between the two sides of the monolayer as induced by the two different substrates. More recently, a similar dielectric lateral heterostructure has been obtained depositing a WSe$_2$ on a homogeneous SiO$_2$ substrate, and then capping only half of it with HfO$_2$. A pump-probe measurement reveals a drift toward the side with higher dielectric constant of several tens of nm in the first 5 ps \cite{Gauriot24}, Fig. \ref{tr10-dielectric}(c). Here the pump is strongly non-resonant, hence a large component of free charges is still expected to be present before the phonon-driven energy relaxation toward 1s excitons \cite{Brem18}: This could contribute to the very effective drift.

Finally, in Fig. \ref{tr10-dielectric}(d,e) 
 the excitonic distribution after localized non-resonant excitation of a homobilayer WSe$_2$-hBN heterostructure is shown at two different instants of 0 and 5 ns. Such a distribution is measured via stroboscopic scattering microscopy (stroboSCAT), a pump-probe technique hence allowing to track excitons independently of their bright/dark nature. The initially localized but isotropic distribution of Fig. \ref{tr10-dielectric}(d) quickly spreads in a very anisotropic distribution, Figs. \ref{tr10-dielectric}(e), in particular localizing in the vicinity of the so-called dielectric nanobubbles, arrows in Fig. \ref{tr10-dielectric}(e,f). These are found via parallel AFM measurement, Fig. \ref{tr10-dielectric}(f), and induced by a separation between WSe$_2$ and hBN. Contrary to the strain nanobubbles, the dielectric ones do not show redshift of the bright-exciton peak, cf. Fig. \ref{tr8-strainDiff}, but can nevertheless alter the dark-exciton energy,
  as deduced from the increase of lifetime measured in correspondence of these impurities. This results in the trapping of these states \cite{Su22}, Fig. \ref{tr10-dielectric}(e), which dominate the response of the homobilayer WSe$_2$ \cite{Kumar23}.
 In a nutshell, these results indicate how the dielectric control can trigger directional transport, potentially showing interplay between free carriers and bound excitons.
 

\subsubsection{Electric control of drift}\label{sec:electricDrift}

Excitons are neutral, hence they cannot be influenced by using electric fields in a strainghtforward way, contrary to the case of regular electronics. Nevertheless, TMD-based nanomaterials host a rich landscape of quasi-particles. In particular, trions and interlayer excitons exhibiting a dipole can be influenced by electrical fields.  
\begin{figure}[t!]
    \centering
    \includegraphics[width=\linewidth]{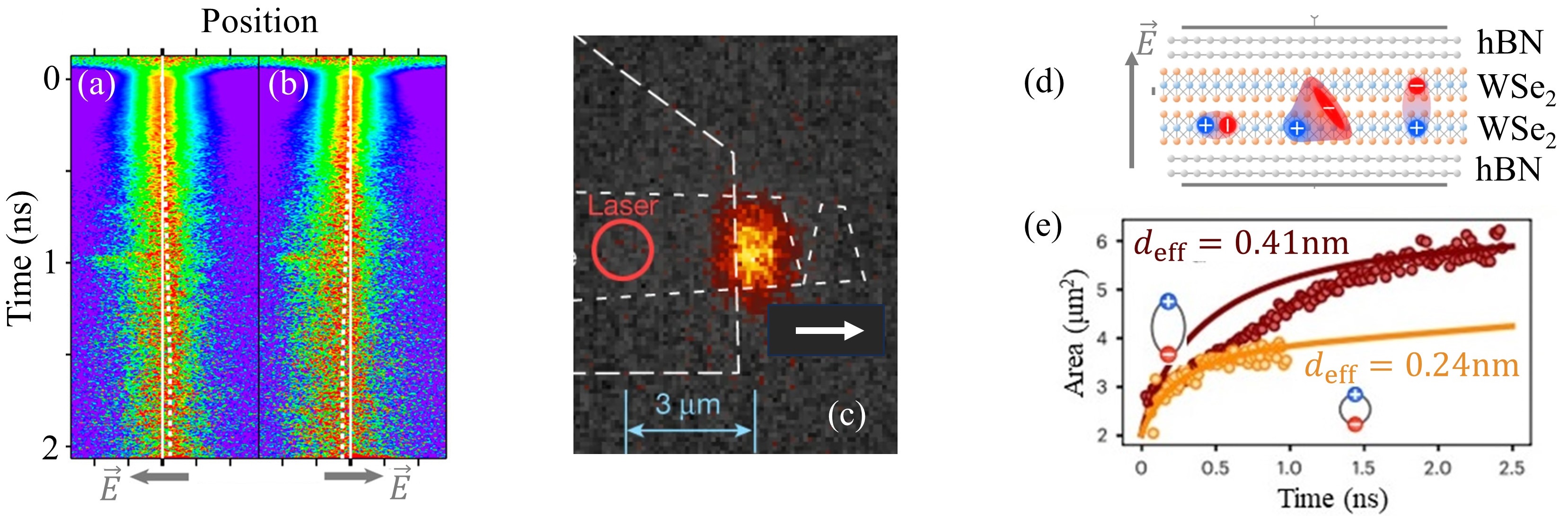}
    \caption{(a,b) Space- and time-resolved photoluminescence in WS$_2$ with an in-plane electric field pointing left and right, showing a drift according to the field direction. (c) Space-resolved PL after excitation of a MoS$_2$-WSe$_2$ vertical heterostructures, showing a drift of 3 $\mu$m with respect to the laser position thanks to the presence of multiple metallic contacts creating a space-dependent gate field. (c) Sketch of intralayer, interlayer and hybridized excitons in a homobilayer WSe$_2$ inserted in two metallic contacts: These create an electric field able to promote the interlayer component of hybrid excitons. This results in an electric control of anomalous diffusion (e) via an effective dipole depending on the field (sketch). Figs. (a,b), (c)  and (d,e) adapted respectively from Ref. \cite{Cheng21}, \cite{Unuchek18} and \cite{Tagarelli23}.}
    \label{tr12-drift}
\end{figure}

Trions are charged, hence allowing an electrical control. This is shown in Fig. \ref{tr12-drift}(a,b), which shows the normalized PL after localized excitation of WS$_2$ in the presence of a bias of respectively $V_{\text{bias}}=-60V$ and $V_{\text{bias}}=60V$ applied to two electrodes separated by 6.7$\mu$m, predicted to result in inplane electric fields of approximately 4 V/$\mu$m \cite{Cheng21}. The field drives the trion propagation of about 74 nm/ns toward left or right according to the polarity  (a,b), while leading to no drift in the absence of a field \cite{Cheng21}. 
While this drift is smaller than the one found via strain control (with funneling and anti-funneling in the first ns of more than respectively 1$\mu$m and 300 nm in MoSe$_2$ and WSe$_2$\cite{Rosati21e}), it indicates a technologically important way to electrically control the directional transport. Note that comparable in-plane electric fields of the order of few V/$\mu$ could give rise to the dissociation of 1s exciton via tunneling ionization \cite{Massicotte18}, also potentially affecting the transport in the presence of bias potentials.

In Sec. \ref{sec:Stark} we discussed how out-of-plane electric fields $E_z$ 
couple directly with the dipole $d$ of excitons, resulting in an electrostatic potential $V=-E_zd$ inducing a shift of the emission energy \cite{Jauregui19,Kiemle20,Ciarrocchi19,Leisgang20}. It follows that an inhomogeneous potential $V(\textbf{R})$ would be present, if such electric field $V=-E_z(\textbf{R})$ is localized via small electrodes. This results in an electrical control of the drift as shown in Fig. \ref{tr12-drift}(c). Here a gate-induced negative electrostatic potential allowed the directional collection of of photoluminescence 3$\mu$m away of the emission spot. 
Crucial for this purpose is varying this potential along the plane, as obtained via multiple metallic contacts connected to different gates. In addition the mechanism makes use of the long lifetimes of interlayer excitons to obtain this long-range micrometric transport, cf. Sec. \ref{sec:lifetime}.

This remarkable electric control does not require high-density excitations, although out-of-plane electric fields can trigger also intriguing non-linear transport. This happens in particular thanks to the hybrid excitons and their quantum mechanical components.
The hybrid excitons are Coulomb-bound electron-hole pairs, where either the electron or the hole is located in both layers to a certain percentage thanks to an efficient interlayer tunneling. Hybrid excitons are hence partially inter- and partially intralayer excitons. An out-of-plane electric field obtainable by gating  increases the interlayer component of hybrid excitons, Fig. \ref{tr12-drift}(d), hence inducing a larger effective dipole, cf. sketch in Fig. \ref{tr12-drift}(e).
This in turn leads to a speed-up of the nonlinear diffusion, which as shown in \ref{tr12-drift}(e) is anomalous. Such a behaviour is again triggered by dipole-dipole repulsion as in Fig. \ref{tr5-nonLinear}, but here controlling the dipole electrically.  This electrical control of diffusion has great technological implications, in particular given that also the exciton drift can be controlled electrically, Fig. \ref{tr12-drift}(c).

\subsubsection{Directional transport in lateral heterostructures}\label{sec:lateralDrift}

\begin{figure}[t!]
    \centering
    \includegraphics[width=\linewidth]{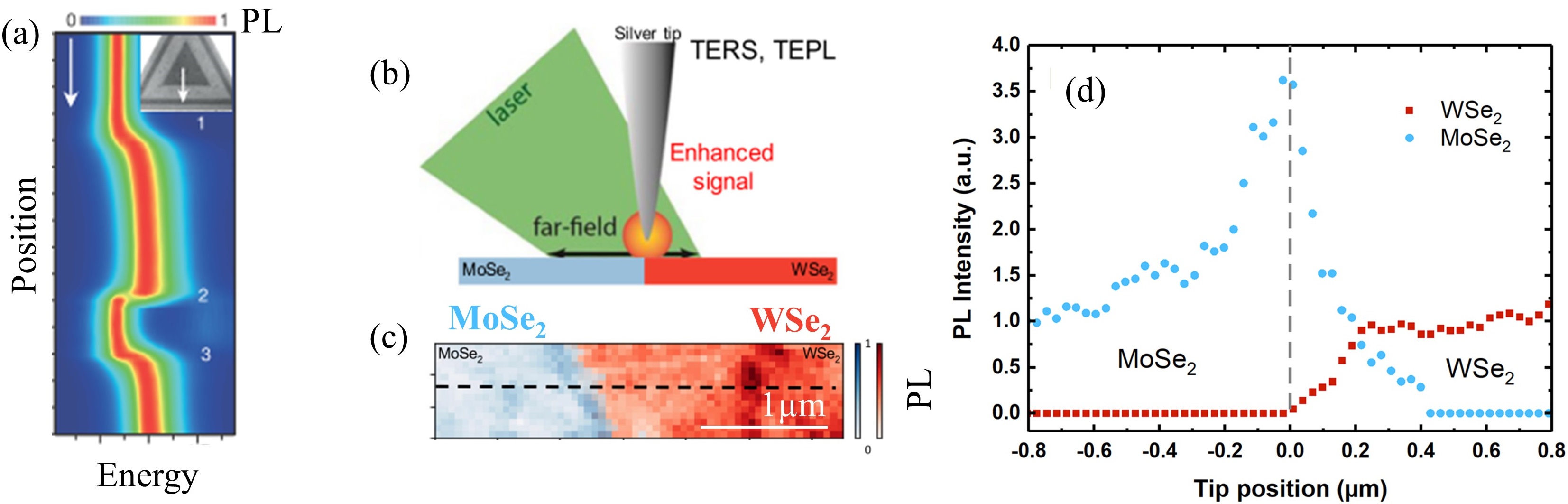}
    \caption{(a) Raster scanning of a lateral MoSe$_2$-WSe$_2$ heterostructure at room temperature. According to the position of the laser, the photoluminescence recovers the typical MoSe$_2$ or WSe$_2$ spectrum. (b) Sketch of a near-field raster scanning and (c) space-resolved photoluminescence intensity, indicating respectively in blue and red the regions with emission in the MoSe$_2$ and WSe$_2$ spectral range, respectively. A more quantitative analysis cutting along the dashed line (d) reveals how the MoSe$_2$ spectrum dominates even when the tip is placed 200 nm into the WSe$_2$(d). This happens due to unidirectional transport, leading WSe$_2$ excitons to drift toward the MoSe$_2$ before recombining into light. Figs. (a) and (b-d) adapted from Ref. \cite{Sahoo18} and \cite{Beret22}.}
    \label{tr13-drift-lateral}
\end{figure}

In Sec. \ref{sec:nonlinear} it has been shown how the lateral heterostructure can provide interesting transport along the interface \cite{Yuan23} making use of the presence of CT excitons allowed by narrow interface \cite{Rosati23} with small residual alloying. 
Lateral TMD  heterostructures provide also one of the most natural ways to apply external potentials in TMD-based 2D materials. Here, the two materials lying at the two sides of the interface have different excitonic energies. 
While conceptually similar to the dielectric lateral heterostructure (cf. Fig. \ref{tr10-dielectric}(b,c)), here the excitonic offset can be one order of magnitude higher and reaching few hundreds of meV \cite{Beret22,Yuan23,Sahoo18,Bellus18}. Such excitonic offset  can be directly shown by raster scanning the sample with a localized laser and then spectrally-resolving the PL \cite{Sahoo18,Beret22}. As shown in Fig. \ref{tr13-drift-lateral}(a), a change in the laser position implies a change in shape and energy position of the resulting PL peak. An additional low-energy peak due to charge-transfer excitons can be observed \cite{Rosati23,Yuan23}, in particular for low temperatures and crucially for sharp interfaces and high dielectric constants \cite{Rosati23}, cf. Fig. \ref{tr5-nonLinear}(d). The naive interpretation of the raster scanning is that the PL reflects the intrinsic excitonic energies of the spatial points directly excited by the laser. Such an interpretation, however,  can be further enriched by the inclusion  of transport.

This is shown by a more recent near-field raster scanning of a lateral MoSe$_2$-WSe$_2$ heterostructure, cf. Fig. \ref{tr13-drift-lateral}(b). 
Qualitatively the signal far away of the interface at its left and right side is still dominated by the MoSe$_2$ and WSe$_2$, respectively (Fig. \ref{tr13-drift-lateral}(c)). On a more quantitative level a peculiar behaviour is found close to the interface, as shown in Fig. \ref{tr13-drift-lateral}(d) scanning along the dashed line of Fig. \ref{tr13-drift-lateral}(c).
Only moving the tip deep into the WSe$_2$ side a finite WSe$_2$ peak appears, whereas exciting at the interface only the MoSe$_2$ optical feature is present. 
The MoSe$_2$ peak is still visible even when exciting 400 nm into the WSe$_2$ side, a length one order of magnitude larger than the dimension of the tip.
This happens because before recombining, excitons excited in the WSe$_2$ have time to reach the spatial points where their energy is minimal, which means the MoSe$_2$ side (with an offset of about 90 meV). Such an excitonic diode-like unidirectional transport is triggered by the offset in the excitonic energies, similarly to what has been  observed in other lateral heterostructures \cite{Beret22,Lamsaadi2023,Bellus18,Yuan23,Shimasaki22}. This shows the remarkable opportunities provided by lateral heterostructures for transport, which could be additionally enriched by the role of charge-transfer excitons, cf. Fig. \ref{tr13-drift-lateral}.

\section{Conclusions}

In last decade atomically thin 2D materials have attracted much
attention in research and their vast technological potential has been realized. In this chapter, we have in particular focused on one prominent material class, namely transition metal dichalcogenides (TMD). They exhibit tightly bound excitons with much larger binding energies than in typical III-V or II-VI quantum wells, hence dominating the their response independently of the temperature. In addition, the excitonic landscape shows a remarkable coexistence of states, ranging from  bright and dark excitons in TMD monolayers to the spatially separated interlayer, hybrid, moir\'e and charge-transfer excitons in TMD heterostructures. 
In this chapter  we have focused on exciton transport and optics phenomena.
The versatile exciton landscape results in an intriguing propagation behaviour including diffusion and drift effects and how they can be controlled via temperature, gating, varying excitation conditions, strain and dielectrics engineering, integration into cavities etc. Furthermore, twisted TMD homo- and heterobilayers can exhibit deep moire potentials confining excitons and resulting in quantum emission of light. Here, atomic reconstruction effects play also a major role at low twist angles considerably changing the properties of these materials. Overall, this book chapter gives an overview on the still ongoing research on the fascinating exciton  physics in atomically thin nanomaterials.

\bibliographystyle{unsrt}
\bibliography{rosatiBib_short.bib,optics/Optics_biblio}

\end{document}